\documentclass[twocolumn]{aastex631}

\usepackage[utf8]{inputenc}

\DeclareRobustCommand{\ion}[2]{%
\relax\ifmmode
\ifx\testbx\f@series
{\mathbf{#1\,\mathsc{#2}}}\else
{\mathrm{#1\,\mathsc{#2}}}\fi
\else\textup{#1\,{\mdseries\textsc{#2}}}%
\fi}

\begin{document}

\title{Gemini High-resolution Optical SpecTrograph (GHOST) at Gemini-South:\\ Instrument performance and integration, first science, and next steps 
}

\author[0000-0002-4641-2532]{V. M. Kalari}
\affiliation{Gemini Observatory/NSF’s NOIRLab, Casilla 603, La Serena, Chile}
\author[0000-0001-9716-5335]{R. J. Diaz}
\affiliation{Gemini Observatory/NSF’s NOIRLab, Casilla 603, La Serena, Chile}
\author[0000-0001-5528-7801]{G. Robertson}
\affiliation{Australian Astronomical Optics, Macquarie University, 105 Delhi Rd, North Ryde NSW 2113, Australia}
\affiliation{Sydney Institute for Astronomy, School of Physics, University of Sydney, NSW 2006, Australia}
\author[0000-0003-4666-6564]{A. McConnachie}
\affiliation{NRC Herzberg Astronomy and Astrophysics Research Centre, 5071 West Saanich Road, Victoria, B.C., Canada, V9E\,2E7}
\author[0000-0002-6194-043X]{M. Ireland}
\affiliation{Research School of Astronomy and Astrophysics, College of Science, Australian National University, Canberra 2611, Australia}
\author{R. Salinas}
\affiliation{Gemini Observatory/NSF’s NOIRLab, Casilla 603, La Serena, Chile}
\affiliation{Departamento de Astronom\'ia, Universidad de La Serena, Av. Juan Cisternas 1200, La Serena, Chile}
\author{P. Young}
\affiliation{Research School of Astronomy and Astrophysics, College of Science, Australian National University, Canberra 2611, Australia}
\author{C. Simpson}
\affiliation{Gemini Observatory/NSF’s NOIRLab, 670 N. A’Ohoku Pl., Hilo, HI 96720, USA}
\author{C. Hayes}
\affiliation{NRC Herzberg Astronomy and Astrophysics Research Centre, 5071 West Saanich Road, Victoria, B.C., Canada, V9E\,2E7}
\affiliation{Space Telescope Science Institute, 3700 San Martin Drive, Baltimore, MD 21218, USA}
 \author{J. Nielsen}
\affiliation{Research School of Astronomy and Astrophysics, College of Science, Australian National University, Canberra 2611, Australia}
\author{G. Burley}
\affiliation{NRC Herzberg Astronomy and Astrophysics Research Centre, 5071 West Saanich Road, Victoria, B.C., Canada, V9E\,2E7}
\author{J. Pazder}
\affiliation{NRC Herzberg Astronomy and Astrophysics Research Centre, 5071 West Saanich Road, Victoria, B.C., Canada, V9E\,2E7}
\author{M. Gomez-Jimenez}
\affiliation{Gemini Observatory/NSF’s NOIRLab, Casilla 603, La Serena, Chile}
\author[0000-0002-5084-168X]{E. Martioli}
\affiliation{Laborat\'{o}rio Nacional de Astrof\'{i}sica, Rua Estados Unidos 154, 37504-364, Itajub\'{a} MG, Brazil}
\author{S. B. Howell}
\affiliation{NASA Ames Research Center, Moffett Field, CA 94035, USA}
\author{M. Jeong}
\affiliation{Department of Astronomy, Space Science, and Geology, Chungnam National University, Daejeon 34134, Republic of Korea}
\affiliation{Gemini Observatory/NSF’s NOIRLab, Casilla 603, La Serena, Chile}
\author{S. Juneau}
\affiliation{NSF’s NOIRLab, 950 N. Cherry Avenue, Tucson, AZ 85719, USA }
\author{R. Ruiz-Carmona}
\affiliation{Gemini Observatory/NSF’s NOIRLab, Casilla 603, La Serena, Chile}
\author{S. Margheim}
\affiliation{Vera C. Rubin Observatory/NSF’s NOIRLab, Casilla 603, La Serena, Chile}
\author{A. Sheinis}
\affiliation{Canada-France-Hawaii Telescope, 65-1238 Mamalahoa Highway, Kamuela, HI 96743, USA}
\author{A. Anthony}
\affiliation{NRC Herzberg Astronomy and Astrophysics Research Centre, 5071 West Saanich Road, Victoria, B.C., Canada, V9E\,2E7}
\author{G. Baker}
\affiliation{Australian Astronomical Optics, Macquarie University, 105 Delhi Rd, North Ryde NSW 2113, Australia}
\author{T. A. M. Berg}
\affiliation{NRC Herzberg Astronomy and Astrophysics Research Centre, 5071 West Saanich Road, Victoria, B.C., Canada, V9E\,2E7}
\author{T. Cao}
\affiliation{Department of Astronomy, Xiamen University, 422 Siming South Road, Xiamen 361005, People's Republic of China}
\author{E. Chapin}
\affiliation{NRC Herzberg Astronomy and Astrophysics Research Centre, 5071 West Saanich Road, Victoria, B.C., Canada, V9E\,2E7}
\author{T. Chin}
\affiliation{Australian Astronomical Optics, Macquarie University, 105 Delhi Rd, North Ryde NSW 2113, Australia}
\author{K. Chiboucas}
\affiliation{Gemini Observatory/NSF’s NOIRLab, 670 N. A’Ohoku Pl., Hilo, HI 96720, USA}
\author{V. Churilov}
\affiliation{Australian Astronomical Optics, Macquarie University, 105 Delhi Rd, North Ryde NSW 2113, Australia}
\author[0000-0001-9796-2158]{E. Deibert}
\affiliation{Gemini Observatory/NSF’s NOIRLab, Casilla 603, La Serena, Chile}
\author{A. Densmore}
\affiliation{NRC Herzberg Astronomy and Astrophysics Research Centre, 5071 West Saanich Road, Victoria, B.C., Canada, V9E\,2E7}
\author{J. Dunn}
\affiliation{NRC Herzberg Astronomy and Astrophysics Research Centre, 5071 West Saanich Road, Victoria, B.C., Canada, V9E\,2E7}
\author{M. L. Edgar}
\affiliation{Australian Astronomical Observatory}
\author[0000-0003-2530-3000]{J. Heo}
\affiliation{Gemini Observatory/NSF’s NOIRLab, Casilla 603, La Serena, Chile}
\author{D. Henderson}
\affiliation{Gemini Observatory/NSF’s NOIRLab, 670 N. A’Ohoku Pl., Hilo, HI 96720, USA}
\author{T. Farrell}
\affiliation{Australian Astronomical Optics, Macquarie University, 105 Delhi Rd, North Ryde NSW 2113, Australia}
\author{J. Font}
\affiliation{Gemini Observatory/NSF’s NOIRLab, Casilla 603, La Serena, Chile}
\author{V. Firpo}
\affiliation{Gemini Observatory/NSF’s NOIRLab, Casilla 603, La Serena, Chile}
\author{J. Fuentes}
\affiliation{Gemini Observatory/NSF’s NOIRLab, Casilla 603, La Serena, Chile}
\affiliation{ESO, Alonso de Cordova 3107, Vitacura, Santiago de Chile, Chile}
\author{K. Labrie}
\affiliation{Gemini Observatory/NSF’s NOIRLab, 670 N. A’Ohoku Pl., Hilo, HI 96720, USA}
\author{S. Lambert}
\affiliation{NRC Herzberg Astronomy and Astrophysics Research Centre, 5071 West Saanich Road, Victoria, B.C., Canada, V9E\,2E7}
\author{J. Lawrence}
\affiliation{Australian Astronomical Optics, Macquarie University, 105 Delhi Rd, North Ryde NSW 2113, Australia}
\author{J. Lothrop}
\affiliation{NRC Herzberg Astronomy and Astrophysics Research Centre, 5071 West Saanich Road, Victoria, B.C., Canada, V9E\,2E7}
\author{R. McDermid}
\affiliation{Australian Astronomical Optics, Macquarie University, 105 Delhi Rd, North Ryde NSW 2113, Australia}
\author{B. W. Miller}
\affiliation{Gemini Observatory/NSF’s NOIRLab, Casilla 603, La Serena, Chile}
\author{G. Perez}
\affiliation{Gemini Observatory/NSF’s NOIRLab, Casilla 603, La Serena, Chile}
\author[0000-0003-4479-1265]{V. M. Placco}
\affiliation{NSF’s NOIRLab, Tucson, AZ 85719, USA}
\author{P. Prado}
\affiliation{Gemini Observatory/NSF’s NOIRLab, Casilla 603, La Serena, Chile}
\author{C. Quiroz}
\affiliation{Gemini Observatory/NSF’s NOIRLab, Casilla 603, La Serena, Chile}
\author{F. Ramos}
\affiliation{Gemini Observatory/NSF’s NOIRLab, Casilla 603, La Serena, Chile}
\author{R. Rutten}
\affiliation{Gemini Observatory/NSF’s NOIRLab, Casilla 603, La Serena, Chile}
\author{K. M. G. Silva}
\affiliation{Gemini Observatory/NSF’s NOIRLab, Casilla 603, La Serena, Chile}
\author{J. Thomas-Osip}
\affiliation{Gemini Observatory/NSF’s NOIRLab, Casilla 603, La Serena, Chile}
\author{C. Urrutia}
\affiliation{Gemini Observatory/NSF’s NOIRLab, Casilla 603, La Serena, Chile}
\author{W. D. Vacca}
\affiliation{Gemini Observatory/NSF’s NOIRLab, 670 N. A’Ohoku Pl., Hilo, HI 96720, USA}
\author{K. Venn}
\affiliation{NRC Herzberg Astronomy and Astrophysics Research Centre, 5071 West Saanich Road, Victoria, B.C., Canada, V9E\,2E7}
\author{F. Waller}
\affiliation{NRC Herzberg Astronomy and Astrophysics Research Centre, 5071 West Saanich Road, Victoria, B.C., Canada, V9E\,2E7}
\author{L. Waller}
\affiliation{Australian Astronomical Optics, Macquarie University, 105 Delhi Rd, North Ryde NSW 2113, Australia}
\author{M. White}
\affiliation{Research School of Astronomy and Astrophysics, College of Science, Australian National University, Canberra 2611, Australia}
\author{S. Xu}
\affiliation{Gemini Observatory/NSF’s NOIRLab, 670 N. A’Ohoku Pl., Hilo, HI 96720, USA}
\author{R. Zhelem}
\affiliation{Australian Astronomical Optics, Macquarie University, 105 Delhi Rd, North Ryde NSW 2113, Australia}


\begin{abstract}

The Gemini South telescope is now equipped with a new high-resolution spectrograph called GHOST (the Gemini High-resolution Optical SpecTrograph). This instrument provides high-efficiency, high-resolution spectra covering 347-1060\,nm in a single exposure of either one or two targets simultaneously, along with precision radial velocity spectroscopy utilizing an internal calibration source. It can operate at a spectral element resolving power of either 76\,000 or 56\,000, and can reach a SNR $\sim$5 in a 1\,hr exposure on a $V\sim$20.8\,mag target in median site seeing, and dark skies (per resolution element). GHOST was installed on-site in June 2022, and we report performance after full integration to queue operations in November 2023, in addition to scientific results enabled by the integration observing runs. These results demonstrate the ability to observe a wide variety of bright and faint targets with high efficiency and precision. With GHOST, new avenues to explore high-resolution spectroscopy have opened up to the astronomical community. These are described, along with the planned and potential upgrades to the instrument. 

\end{abstract}



\section{Introduction}

High-resolution spectroscopy has enabled incredible discoveries spanning diverse fields of astrophysics. The detection of extrasolar planets based on the displacement of spectral lines due to the presence of orbiting planets \citep{peg51}, constraints on the fine structure constant and the proton-to-electron mass ratio using intervening quasar absorption systems \citep{qso}, and studying elemental abundance patterns of stellar populations to estimate the yields of {\it r} and {\it s}-process elements \citep{frebel}, are all examples of astrophysical disciplines that owe their existence to the high-resolution spectrograph (having resolving power, $R$=$\lambda$/$\Delta\lambda\geq$\,10\,000). With the increase in spectral resolution, the total incident flux is split into an ever-increasing number of individual spectral elements, and thereby an ever-decreasing number of photons hits a detector at higher resolving powers. For a fixed (desired) signal-to-noise ratio (SNR), the area of the telescope must increase as the resolution increases. Therefore, to increase the scientific potential of high-resolution spectroscopy and reach the needed SNRs in a given intergration time for a variety of science cases, an 8-meter class of telescope is essential.

In September of 2010, the Gemini Observatory initiated a new facility instrument project in response to the Gemini community's request for a high-resolution workhorse spectrograph with improved blue throughput \citep{toll10,boccas}. The proposed design needed to address the multiple challenges of providing wide-wavelength coverage, high efficiency, and high-resolution for a telescope without a Nasmyth/Coud{\'e} focus and a red-sensitive primary mirror. The final design was thus motivated not only by science requirements, but also by restrictions imposed by the telescope. In 2011, the conceptual design phase began and was competed by three different instrumentation groups. A collaboration led by the Australian Astronomical Optics (AAO) and the Australian National University (ANU) was awarded the project to build the Gemini High-resolution Optical SpecTrograph (GHOST; \citealt{ireland12}) in 2013. The GHOST design was based on a bench spectrograph fed by fibers from a focal plane unit mounted on the Cassegrain focus of the Gemini South 8.1\,m Telescope at the Cerro Pach{\'o}n site in Chile. In 2014, the National Research Council of Canada (NRC) at the Herzberg Research Center was subcontracted to design and build the main body of the spectrograph. 

Following this, the preliminary design review (PDR) was conducted in 2014, with the final design review (FDR) approved by Gemini in late 2015. This began the process of building and integration along with lab testing, with delivery to the Gemini-South site in Cerro Pach{\'o}n in early 2020. The final integration and commissioning were conducted between June 2022 and November 2023. Table\,1 lists the contributions of the institutions involved. A recent overview of the spectrograph and its performance during the commissioning runs in June and September 2022 are given in McConnachie et al. (2024).  

\begin{deluxetable*}{lll}
\tablenum{1}
\tablecaption{Summary of institutional contributions \label{}}
\tablewidth{0pt}
\tablehead{
\colhead{Institution } & \colhead{ Reference} & \colhead{Project scientists}}
\startdata
{\bf Australian Astronomical Optics} & \cite{sheinis16} & G. Robertson \\
\multicolumn{3}{l}{Cassegrain unit and fiber link}\\
\multicolumn{3}{l}{Acquisition and Guide unit}\\
\multicolumn{3}{l}{Slit-viewing unit}\\
\hline
{\bf Australian National University} &  &  \\
{Data reduction pipeline} & \cite{ireland14} & M. Ireland \\
{Instrument control software} & & \\
\hline
{\bf Gemini Observatory} & This paper & V. Kalari and R. Diaz \\
\multicolumn{3}{l}{Systems integration and}\\
\multicolumn{3}{l}{commissioning}\\
\hline
{\bf National Research Council Herzberg} & \cite{mcconn24} & A. McConnachie \\
\multicolumn{3}{l}{Bench spectrograph}\\
  \enddata
   \tablecomments{$^{1}$ Significant contributions to the data reduction pipeline were also made by Gemini Observatory (C. Simpson), and NRC Herzberg (C. Hayes).   }
\end{deluxetable*}

\begin{figure}
\includegraphics[width=1\columnwidth]{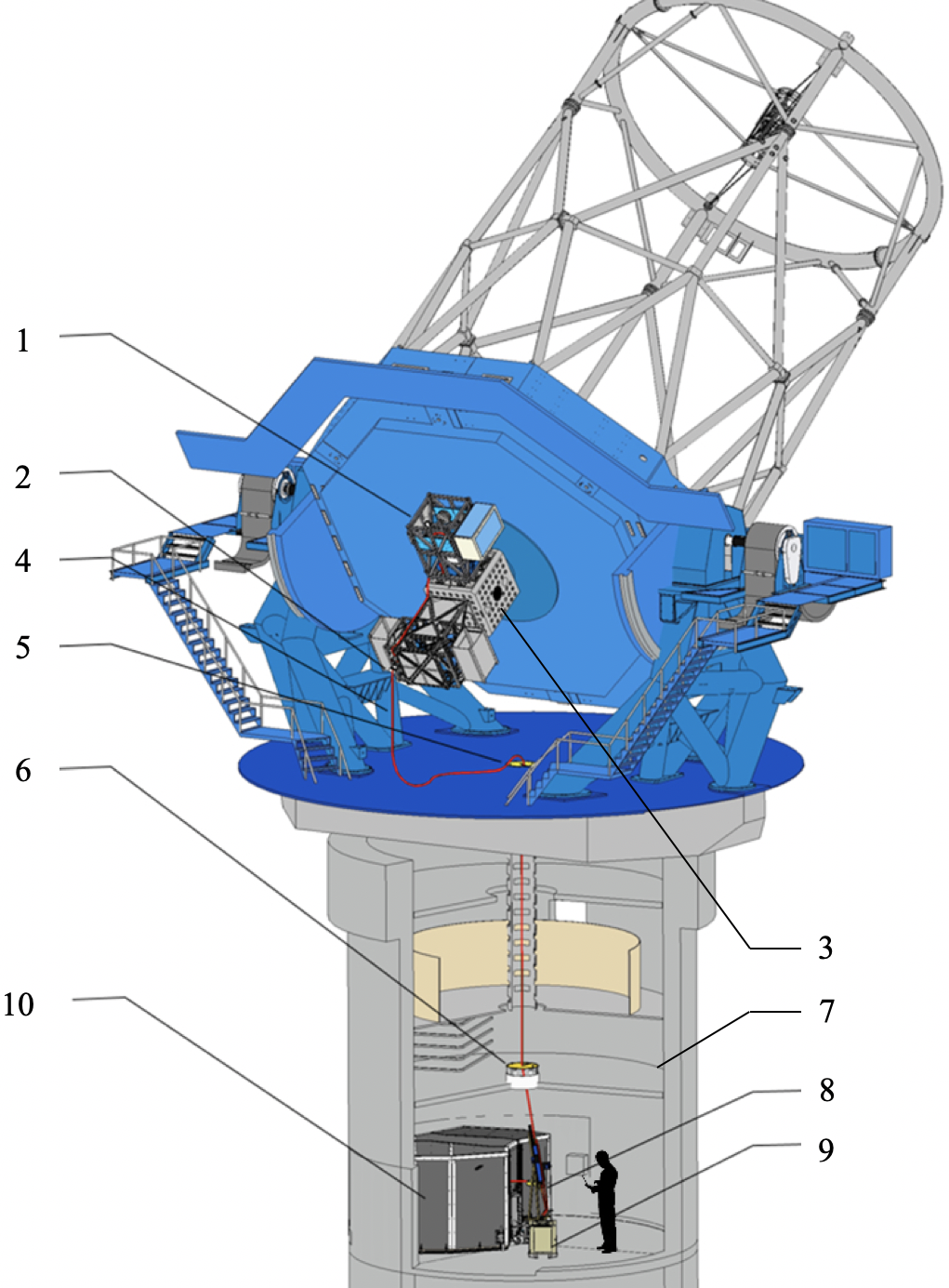}
\caption{Overview of major systems within the telescope (adapted from \citealt{sheinis16}). 1 and 2 represent the Cassegrain Unit mounted either on the side, or up-looking port respectively with the ballast frame also shown. 3 indicates the instrument support structure. 4 marks the fiber cable, and 5 shows the cable passing through the anchor and to the pier lab, and 6 marks the location of the hole in the Gemini South Pier Lab allowing for the fiber cable to pass through landing in the Pier lab indicated by 7. Additionally, the location of the fiber agitator and the acquisition and guiding unit (which includes the internal Th-Xe lamp) outside the bench thermal enclosure is shown by 8, and 9 respectively. 10 marks the outer enclosure, which is thermally enclosed, and a pressure-monitored chamber containing the bench spectrograph inside the inner enclosure. A human being is shown for scale.
\label{schematic}}
\end{figure}

This paper’s aims are to provide examples of early science enabled by GHOST in the context of the key science drivers and present the results of the instrument performance and complete integration into the Gemini observing queue. An overview of the key science cases are outlined in Section 2, with the instrument design described in Section 3, including a summary of the available modes. In Section 4 we present the results of on-sky performance during the instrument integration and commissioning period. The first science results are presented in Section 5. Future upgrades to the instrument to improve performance, and broaden science cases are discussed in Section 6. Our final conclusions are presented in Section 7.

\section{Key science drivers}

GHOST was designed as a workhorse instrument, that is aimed to address a wide range of science requirements. These requirements were set by the Gemini community after an open solicitation for science cases\footnote{https://www.gemini.edu/instrumentation/ghost/other-documentation}. These include:
\begin{enumerate}

\item {\it Measuring the elemental abundances of faint, or metal-poor stars.} Chemical abundances of faint (16\,$<${\it V}\,$<$\,20\,mag), or metal-poor ([Fe/H]$\leq-$1) stars are critical scientific laboratories to study stellar nucleosynthesis, and galactic archaeology \citep{abundances, nearcosmology}. 

\item {\it Studies of nearby dwarf galaxies.} Dwarf galaxies are the most common type of galaxy and are the local analogs of the dark matter mini-halos proposed in $\lambda$CDM model as the building blocks of larger galaxies in the early universe \citep{dwarfs, dwarfs2}. High-resolution spectroscopy is needed to obtain stellar kinematics and abundances in such galaxies.  

\item {\it Follow-up of low resolution or photometrically identified targets:} With the advent of imaging and spectroscopic surveys (e.g. such as the \citealt{gaia} survey), multiple science cases may require follow-up of individual objects using high resolution spectrographs at large telescopes.

\item {\it Exoplanetary studies:} During the early design phase, the ability to acquire high precision radial velocity observations (1\,m\,s$^{-1}-$few 10s\,m\,s$^{-1}$) to identify exo-planets \citep{Mayor1995} became an added driver for the instrument. In addition, transmission spectroscopy of planetary transits \citep{transmission} was included as an additional case for exo-planetary science. 

 \item {\it High-resolution observations of extra-galactic systems:} studying abundance patterns in extra-galactic absorption systems seen towards quasars may be key to understanding the early circum-galactic medium \citep{quasarigm}, or deriving high precision values of fundamental constants \citep{quasarfund}. Such studies require not only high-resolution spectroscopy on large telescopes, but also long-term instrument stability.

 \item {\it Transients:} Gemini's long-term strategic goals in the era of Multi-Messenger Astronomy\footnote{https://www.gemini.edu/about/gemini-era-multi-messenger-astronomy} to follow-up exciting new discoveries (for e.g. the recent binary neutron star merger seen by \citealt{nsmerger}) were thought to complement the broad wavelength coverage and high efficiency of the planned instrument. Coupled with the ability to acquire and guide on faint targets in a variety of weather conditions, the spectrograph was envisioned to follow-up transient objects. 
\end{enumerate}

These science cases necessitated the construction of stable, high-efficiency high-resolution ($R\sim50000-75000$) spectrograph covering the entire visible wavelength range. The absolute radial velocity precision, and stability also was necessitated to be in the range of m\,s$^{-1}$, and few 10s of m\,s$^{-1}$, respectively. In addition, for chemical abundance studies, especially of stars multiplexing was considered attractive. Based on these science, and aforementioned telescope requirements, the GHOST instrument was designed and constructed.

\begin{deluxetable*}{lll}
\tablenum{2}
\tablecaption{Instrumentation modes \label{}}
\tablewidth{0pt}
\tablehead{
\colhead{Mode } & \colhead{Standard resolution} & \colhead{High resolution}} 
\startdata
Spectral coverage$^1$  &  347--\,1060\,nm (simultaneous) & 347--\,1060\,nm (simultaneous)   \\
Spectral resolving power  & 56\,000 & 76\,000  \\
Binning modes & 2,4,8 (spatial); 2, 4 (spectral) & 2,4,8 (spatial); 2, 4 (spectral)  \\
Radial velocity precision  & 150\,m\,s$^{-1}$ & 50\,m\,s$^{-1}$  \\
Multiplexing$^2$  & Two targets (minimum separation 102$\arcsec$) &  Single target \\
Field of view  &  7.34$\arcmin$ (overlap of two IFU's 16$\arcsec$) &  7.34$\arcmin$   \\
IFU aperture  & 0.94\,$arcsec^2$ &  0.92\,$arcsec^2$ \\
Sky aperture & 0.4\,$arcsec^2$ & 0.34\,$arcsec^2$ \\
Microlens configuration  &  7 microlenses in each object IFU & 19 microlenses in object IFU \\
&  3 microlenses in sky IFU & 7 microlenses in sky IFU \\
Calibration source   & GCAL (ThAr)   & GCAL (ThAr), and Internal ThXe lamp\\
Limiting magnitude$^3$ & $V\sim$\,20.8\,mag & $V\sim$\,19.6\,mag \\
  \enddata
  \tablecomments{$^{1}$Represents the complete wavelength coverage of GHOST. The wavelength range across which  the throughput is above 2\% is 383--1000\,nm. $^2$ The instrument multiplexing capabilities are described in the design in \cite{sheinis16}. $^3$ Defined at the magnitude at which a SNR of 5 per resolution element is achieved in 1\,hr under clear skies with no moon, in median site seeing.   }

\end{deluxetable*}

\section{Instrument design} \label{design}

The GHOST instrument concept consists of three main components: (i) a high-resolution efficient spectrometer with a wide wavelength coverage coupled with a Cassegrain focus usable over a wide range of conditions. (ii) Two integral field units (IFUs) with Atmospheric Dispersion Corrector (ADCs) positioned at the Cassegrain focal plane capture the target. These images are reformatted into a pseudo-slit image, thereby increasing resolution and efficiency. (iii) This final image reaches a bench spectrograph located below the telescope via optical fibers thus providing stability and reliability. The input beam is reformatted before entering an echelle spectrograph, which provides simultaneous coverage between 347-1060\,nm. GHOST's design saves both size and weight at the focal plane while improving throughput, resolution, and wavelength stability (i.e. avoiding flexure issues) in the final spectra. 

A schematic of the light path is given in Fig.\,\ref{schematic}. After light exits the telescope focus, it enters the Cassegrain unit \citep{zhelem18} which transmits light from either one or two individual integral field units (IFUs), each equipped with an internal atmospheric dispersion corrector (ADC). This light is transferred to the Gemini pier lab via a optical fiber relay \citep{zhelem20,churilov18}, where the bench spectrograph resides \citep{pazder22,pazder20}. The light from the optical cables is reformatted to create a pseudo-slit and increase the resolution, either by a factor of 5 (high resolution; HR), or a factor of 3 (standard resolution; SR) in comparison to a slit of equivalent width. This light then enters the spectrograph, split by a dichroic at 530\,nm into red and blue arms, from where the final echellogram is captured. During calibrations, a mirror is inserted to send light from the Gemini calibration unit (GCAL). An internal ThXe lamp exists to provide on-sky simultaneous calibration for precision radial velocity measurements. A detailed description of each individual component design is to be found in the associated references, with the instrument concept described in \cite{ireland14}, \cite{sheinis16}, and \cite{mccon22}. The relevant references for each subsystem are to be found in Table\,1, and the available configurations for Gemini users are described in Table\,2. In this section, we describe briefly all the systems relevant to science performance and describe in detail the IFU and slit unit, and integration and data flow. A short summary is provided in Table\,3 (telescope interface components), and Table\,4 (bench spectrograph).

\subsection{Telescope interface and fiber link}

\begin{deluxetable}{ll}
\tablenum{3}
\tablecaption{Specifications of GHOST telescope interface components}
\tablewidth{0pt}
\tablehead{
 \colhead{} \\[-1cm]
}
\startdata
\multicolumn{2}{l}{{\bf On-telescope unit}}\\
Integral field units & Two (total of 62\,microlenses)\\
IFU1 configuration &  7 SR + 3 SR Sky microlenses \\
 &  19 HR microlenses \\
IFU2 configuration & 7 SR + 3 HR Sky microlenses \\
Scale & 1.641$\arcsec$\,mm$^{-1}$ \\
\hline
\multicolumn{2}{l}{{\bf Acquisition \& Guide unit}}\\
Limiting magnitude & $V$=\,19\,mag$^{1}$ \\
Guiding frequency & $\geq$10\,Hz \\
Guide fibers & 6 microlenses per IFU\\
Extended objects & Yes$^{2}$ \\
Maximum guide star &  6.75$\arcmin$ for dual target mode\\
 distance &  9.85$\arcmin$ for single target mode\\
\hline
\multicolumn{2}{l}{{\bf Fiber unit}}\\
Fibers & Circular \\
Fiber core & 53$\mu$m\\
No. of cables & 62\\
Agitator frequency & 0.5--1\,Hz\\
\hline
\multicolumn{2}{l}{{\bf Slit unit}}\\
Slit-viewer filter bandpasses & \\
\hspace{10pt}Blue & 420--600\,nm\\
\hspace{10pt}Red  & 600--760\,nm\\
Internal calibration & Th-Xe lamp \\
SR pseudo-slit width & 0.4$\arcsec$ \\
HR pseudo-slit width & 0.24$\arcsec$\\
\hline
\enddata
\tablecomments{$^1$ Refers to the limiting magnitude for direct acquisition and guiding in median site conditions for SR resolution mode. $^{2}$ guiding on extended objects depends on source morphology, and they can be observed unguided with fine guiding, using only peripheral telescope guiding.}
\end{deluxetable}

\begin{figure}
\includegraphics[width=1\columnwidth]{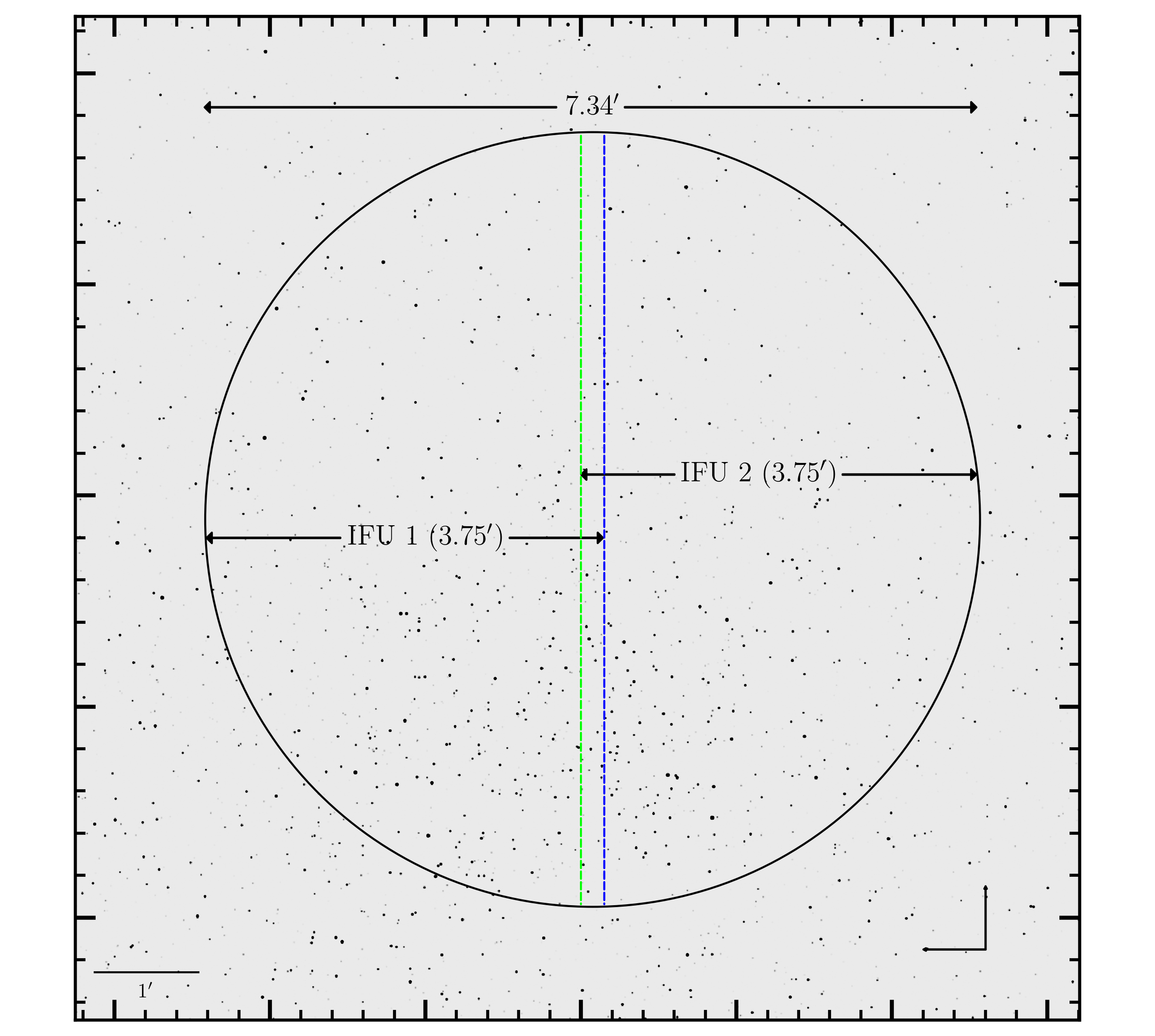}
\caption{Field of view of the Cassegrain unit overlaid on a simulated sky image, with the dimensions marked. The total diameter subtended on the sky is 7.34$\arcmin$, beyond which the probes suffer from vignetting. The field of view of each IFU semi-circle is marked, with the overlapping region also shown. The labels are offset from the center for display. The closest separation between the two IFUs (not shown) is 102$\arcsec$. \label{fov}}
\end{figure}

At the Cassegrain unit, the Instrument Support Structure (ISS) includes two robotic positioners, each carrying an IFU head. IFU 1 contains three individual IFUs, and IFU 2 contains two individual IFUs.

The field of view of each IFU positioner is a semi-circle (with the overlapping areas shown in Fig.\,\ref{fov}). The limiting value is based on vignetting of the IFUs by the Gemini wavefront sensor, and is smaller than the design expectation of 7.5$\arcmin$ \citep{sheinis16}. Further details on the specifications of the individual IFUs are provided in Table\,2.


\begin{figure}
\epsscale{1.1}
\includegraphics[width=1\columnwidth]{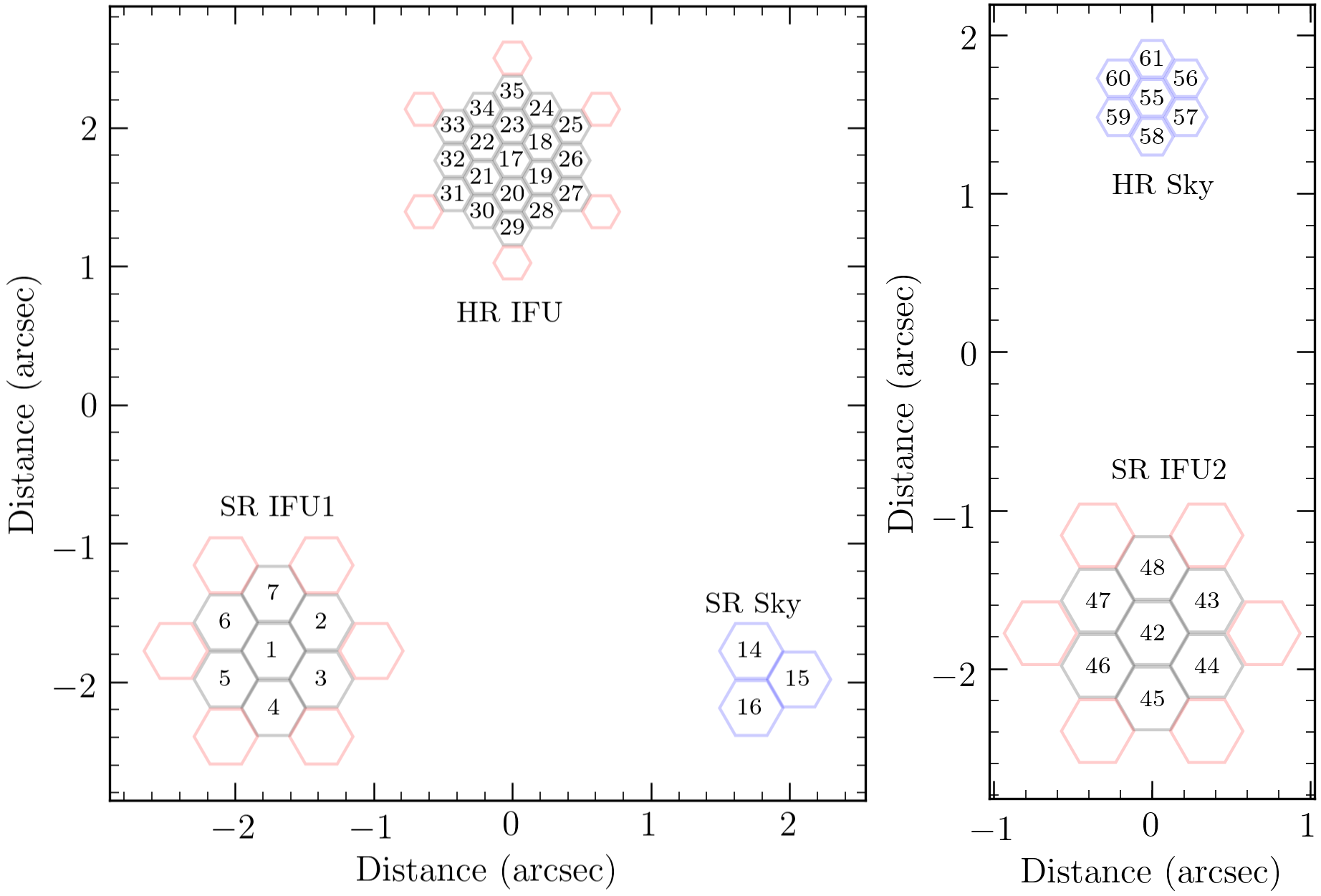}
\caption{The microlenses in the two probe arms are labeled and shown in the instrument's focal plane, with their distance translated to the on-sky scale. The microlenses for scientific purposes are indicated in black, those for guiding in red, and those for capturing sky information in blue. These labeled microlenses can be matched with the pseudo-slit image and echellogram. To maintain clarity, the guide microlenses are not labelled.
\label{micro}}
\end{figure}

Each IFU is shown in Fig.\,\ref{micro}, showing the configuration of the microlenses comprising. The units given are {\it arcsec}, as subtended on sky by the instrument. This presents the final configuration of the IFU units as seen on sky accounting for filling-factors between the individual microlenses based on the microlens metrology, which is slightly larger than design expectations noted in \cite{sheinis16} and \cite{mcconn24}. The complete Cassegrain unit can be rotated to any given position angle. Each science IFU subtends 1.2$\arcsec$ on sky, and is surrounded by guide microlenses directing light to a guide unit, responsible for centering and guiding along with the telescope guiding. The description of acquisition and guiding is given in Section 4.1.

\begin{figure}
\centering
\includegraphics[width=1\columnwidth]{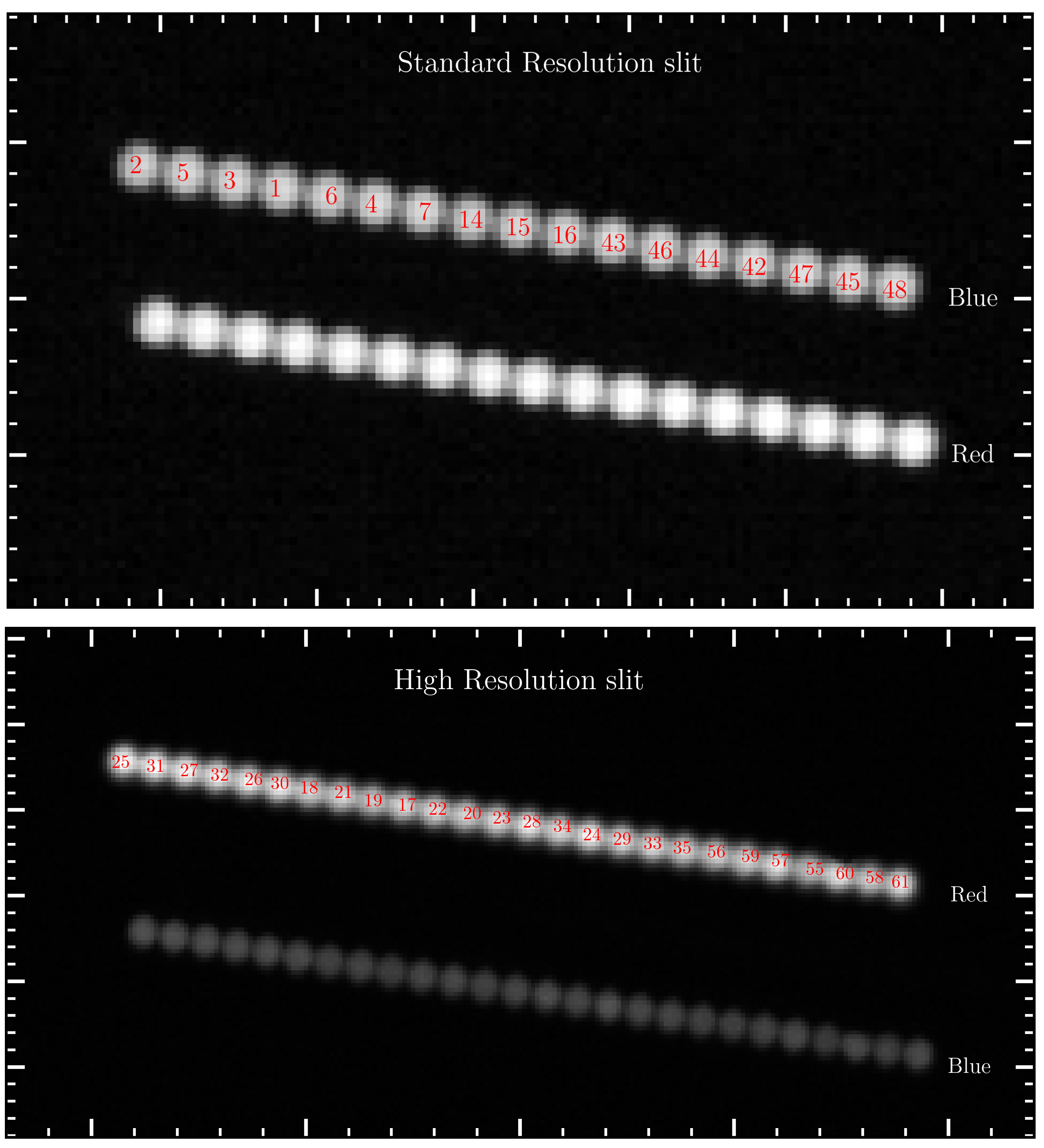}
\caption{{\it Top}: Cutout of a standard resolution slit unit flat from the dual IFU (SRIFU1 and SRIFU2). The microlenses are arranged horizontally, with seven IFU1 microlenses, three sky microlenses, and seven IFU2 microlenses from left to right. The pseudo-slit is captured in the slit viewing unit (receiving 1\% of the light entering the spectrograph), where it is split into a red, and blue pseudo-slit. The filter bandpasses of the slit-viewing camera are given in Table\,3. The top reconstructed slit is seen through a blue filter and the bottom one through a red filter.
{\it Bottom}: Cutout of high-resolution slit unit flat from a single high-resolution IFU is shown. The slits are reversed in the vertical direction for display purposes, and there are 19 IFU1 microlenses and seven sky microlenses visible. Microlens 62 is the final internal arc lamp microlens (not shown) and is separated from microlens 61 by an empty microlens. The final psuedo-slit image is tilted by 11\,degrees. Microlens nomenclature is the same as Fig.\,3.
\label{slit}}
\end{figure}

The optical path from the telescope to the IFUs in the two heads passes through a field lens and an ADC for each individual IFU, which is a conventional Risley Prism pair that rotates to allow a variable amount of dispersion along a chosen position angle axis, counteracting atmospheric dispersion up to a maximum zenith angle of 60$^\circ$ by orienting the prisms so that the dispersion is along the parallactic angle. This mini-ADC design allows for thinner prisms, with less loss especially in the flint component. This allows for a glass combination matched better to the ambient atmospheric dispersion curve at the Cerro Pach{\'o}n site.

To reduce modal noise in the extremely high signal-to-noise regime, the fibers are provided with agitators. Focal ratio degradation (FRD) effects have been shown to be minimal. A more detailed description of further design aspects of the on-telescope Cassegrain unit may be found in \cite{zhelem18}.

\subsection{Slit unit}

Images at the focal surface are reformatted after exiting the fiber cable system and emerge at a linear pseudo-slit (shown in Fig.\,\ref{slit}), increasing the spectral resolution. This is the pseudo-slit seen by the spectrograph. The injection and extraction optics between the on-sky unit to the slit unit by the fiber are the same for the two standard resolution IFUs ($f$/2.8 injection to the fibers), but different for the high-resolution mode ($f$/4.7 injection to the fibers). The arrangement of the microlenses from the focal plane onto the slit unit is chosen to maintain the best output for faint targets, and also for the precision radial velocity mode. For faint objects, this arrangement of the pseudo-slit ensures that fibers that are far apart at the input end, are not adjacent at the slit end. For precision radial velocity mode, this arrangement ensures the final image entering the slit unit is scrambled in the radial direction, and in the azimuthal direction. This setup allows the resultant image to be well scrambled in the azimuthal direction, resulting in a uniformly illuminated final slit profile. 

As light is injected separately into the spectrograph, the final spectra spectral stability has little dependence on FRD variations. The final pseudo-slit image is captured in the slit unit, which receives 1\% of the light entering the spectrograph. This light further passes through a blue and red filter. The image, therefore, consists of a blue and red pseudo-slit containing the slit illumination information at each observing epoch. The slit viewing camera obtains simultaneous exposures along with the spectrograph detectors and can be configured separately.

\begin{deluxetable}{ll}
\tablenum{4}
\tablecaption{Specifications of GHOST spectrograph components}
\tablewidth{0pt}
\tablehead{
 \colhead{} \\[-1cm]
}
\startdata
\multicolumn{2}{l}{{\bf Bench spectrograph}}\\
Dichroic split & 530\,nm \\
Blue orders & 35\,\,(64--98) \\
Red orders & 33\,\,(33--65) \\
\vspace{-0.3cm}\\
\multicolumn{2}{l}{{\it Standard resolution}}\\
Pseudo slit width & 0.4$\arcsec$\\
Spectral resolving power$^1$ & 56\,000\,\,(47\,000/\,23\,000) \\
Radial velocity precision & 150\,m\,s$^{-1}$\\
Sampling & 3.5\,pixels\\
SNR=100 & $V=15$\,mag$^2$ \\
SNR=10 & $V=19.9$\,mag$^3$  \\
\vspace{-0.3cm}\\
\multicolumn{2}{l}{{\it High resolution}}\\
Pseudo slit width & 0.24$\arcsec$\\
Spectral resolving power$^1$ & 76\,000\,\,(56\,000/\,31\,000) \\
Radial velocity precision & 50\,m\,s$^{-1}$\\
Sampling & 2.1\,pixels\\
SNR=100 & $V=$$14$\,mag$^2$ \\
SNR=10 & $V=19.3$\,mag$^3$ \\
\hline
\multicolumn{2}{l}{{\bf Detectors}}\\
Pixel size & 15$\mu$m\\
Read rate & 10, 5, 2\,$\mu$s\,pix$^{-1}$\\
Blue format$^4$ & 4096$\times$4112 $e2v$ \\
Red format$^5$ & 6144$\times$6160 $e2v$ \\
Dark current & $\sim$1\,$e^{-1}$pix\,$hr^{-1}$\\
Read noise & 2.1\,$e^{-1}$\\
Gain & 0.6\,$e$\,DN$^{-1}$ \\
\enddata
\tablecomments{$^{1}$Parentheses refer to a binning in the spectral direction of 2, and 4 respectively. $^{2}$SNR per pixel based on spatial binned (8), spectrally unbinned data in clear skies and median seeing, and for single object in 1\,hr in the Vegamag, at 550\,nm. $^{3}$SNR per resolution element based on data spatially binned (8), and spectrally binned (2) data taken in clear, dark skies and median seeing, and for single object in 1\,hr in Vegamag, at 550\,nm. $^4$\url{http://www.e2v.com/resources/account/download-datasheet/1364} $^5$\url{https://www.teledyne-e2v.com/en-us/Solutions_/Documents/datasheets/ccd231-c6.pdf} }
\end{deluxetable}

\subsection{Bench spectrograph}

The spectrograph is mounted on an aluminum bench with aluminum mounts (to reduce thermal effects) and resides in a thermal enclosure in the pier lab under the Gemini South telescope. It is designed as a box-in-a-box, with active temperature control for all interior surfaces. The enclosure is pressure monitored (but not controlled) and is not vacuum enclosed, with comprehensive details found in \cite{lothrop}. The resulting stability of the radial velocity measurements is described in Section\,4.5. 

The spectrograph has two arms and uses an echelle white-pupil design employing Volume Phase Holographic (VPH) gratings for cross-dispersion. The incoming beam is split into the red and blue VPH gratings. Further details on spectrograph design choices, and final assembly are provided in \cite{pazder20,pazder22}.

\subsection{Data flow and integration}

The instrument control and data archiving are integrated into the Gemini software systems. In practice, this means that all processes, from proposing for observations to the archiving of data products take place within the Gemini software architecture, allowing one to track the progress, and quality of observations at each level while allowing maximum flexibility in scheduling in Gemini's queue observing process \citep{bryan, puxley}.

Observations including targets, instrument configuration, and observing sequences, are defined in the Gemini Observing Tool \citep{nunez}. The observing sequence steps are passed to the Sequence Executor which sends commands to the instrument software (described in \citealt{young16, nielsen}) via the Gemini Instrument Application Programming Interface (GIAPI). An overview of the process is described in \cite{miller20}.

\begin{figure}
\plotone{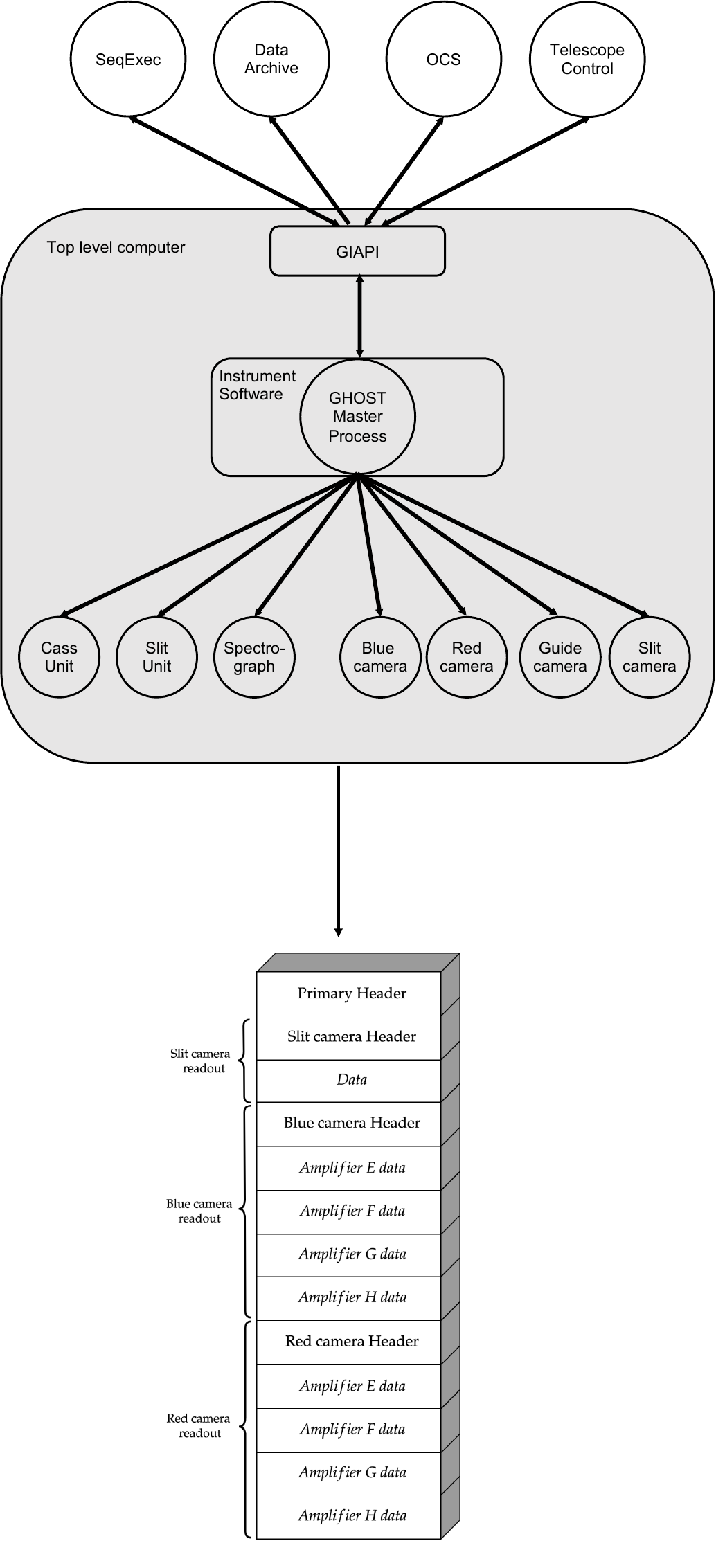}
\caption{ Simplified block diagram, depicting the data flow architecture during GHOST observations at Gemini South. The top panel shows how the various software systems coordinate to send commands to the top level computer (TLC). The TLC then controls the various instrument sub-systems to produce a multi-extension FITS (MEF) file. The bottom panel shows the structure of the MEF file outputted.  \label{data}}
\end{figure}

Once an observation sequence is executed via the Gemini Observatory Control System (OCS) and GIAPI, commands are transferred to the GHOST instrument software. The instrument software combines the control of four main optomechanical systems (the red and blue spectrograph cameras, the slit viewing, and the guide cameras), the mechanical systems and their controllers (spectrograph, slit unit, and Cassegrain unit) via a top-level computer. Each sub-system is controlled by its own computer, with the master processes being controlled by the TLC (top-level computer) via the GHOST Master process. The primary product is a multi-extension FITS (Flexible Image Transport System; \citealt{fits}) file (MEF), with each image taken from the spectrograph, and slit-viewing cameras written following the structure shown in Fig.\,\ref{data}. The guide camera images are stored independently on the top level computer and used only for monitoring performance. These files are then transferred from the TLC via the data handling system. 

\begin{deluxetable}{ll}
\tablenum{5}
\tablecaption{Overheads during queue observations}
\tablewidth{0pt}
\tablehead{
\colhead{Overhead} & \colhead{Time (s)} }
\startdata
Change of instrument & -- \\
Acquisition (direct) & 300 \\
Acquisition (blind offset/spiral search) & 900 \\
Readout & 5--100$^1$\\
NFS$^2$ write & 15 \\
\enddata
\tablecomments{$^{1}$ Readout limits are for high binning fast read, and no binning slow read $^{2}$ Network file system}
\end{deluxetable}

These completed observations are then ingested permanently into the Gemini Observatory Archive \citep{archive}. The ingestion stage involves updating the FITS headers of the final data products. Limited parameters, such as on-sky seeing are measured during the night from the pseudo-slit images to assess data quality. Multiple high-precision temperature and pressure sensors are located on the optical bench and enclosures, and the information taken from them is stored in the Gemini instrument monitoring system. This information is also written in the final MEF file. In addition, a quick visual assessment of the echellogram using a dedicated tool is employed by nighttime observers to assess whether the spectra meet any further requirements requested by the user (e.g. checking whether an emission line is visible; pseudo-slit counts). The final raw data products are written to the Gemini archive in almost real-time, allowing fast and easy access for users to the raw data products. A schematic of the whole data flow process described is given in Fig.\,\ref{data}. 

Integration into the Gemini software systems allows the observer to rapidly change between instruments during the night, and permits setting up, recording observation history, and archiving relatively easily. A short overview of the overheads involved in this process is given in Table\,5. From Table\,5, it is evident that the overheads involved in acquiring even a faint target, and writing a file to the archive are minimal such that one can gain access to an important target of opportunity observations rapidly once triggered during the night (excluding observing time).

Processing from the raw to reduced data products is handled by the user, via a dedicated pipeline written in Python \citep{ireland16, ireland18, hayes22}, and integrated into the DRAGONS data reduction platform \citep{dragons}. Users are able to conduct data processing sequentially. The pipeline performs basic CCD reductions such as bias subtraction and flat-fielding, along with additional corrections such as rejection of cosmic rays, and bad pixel masking before applying a wavelength calibration, including the barycentric correction. Flux calibration is also possible if a spectrophotometric standard star is taken with the same instrument configuration. Finally, the user has multiple options to perform optimal extraction of the reduced 2-D echellogram, with the general use case for optimal extraction of stellar sources utilizing the weighting function determined from the pseudo-slit illumination. However, depending on the on-sky extent of each target, and the sky subtraction choice the users can extract spectra without using the weighted slit image. A future publication is planned that will describe the complete data reduction process, including for the planned precision radial velocity mode.

 \begin{figure*}
\epsscale{1.2}
\plotone{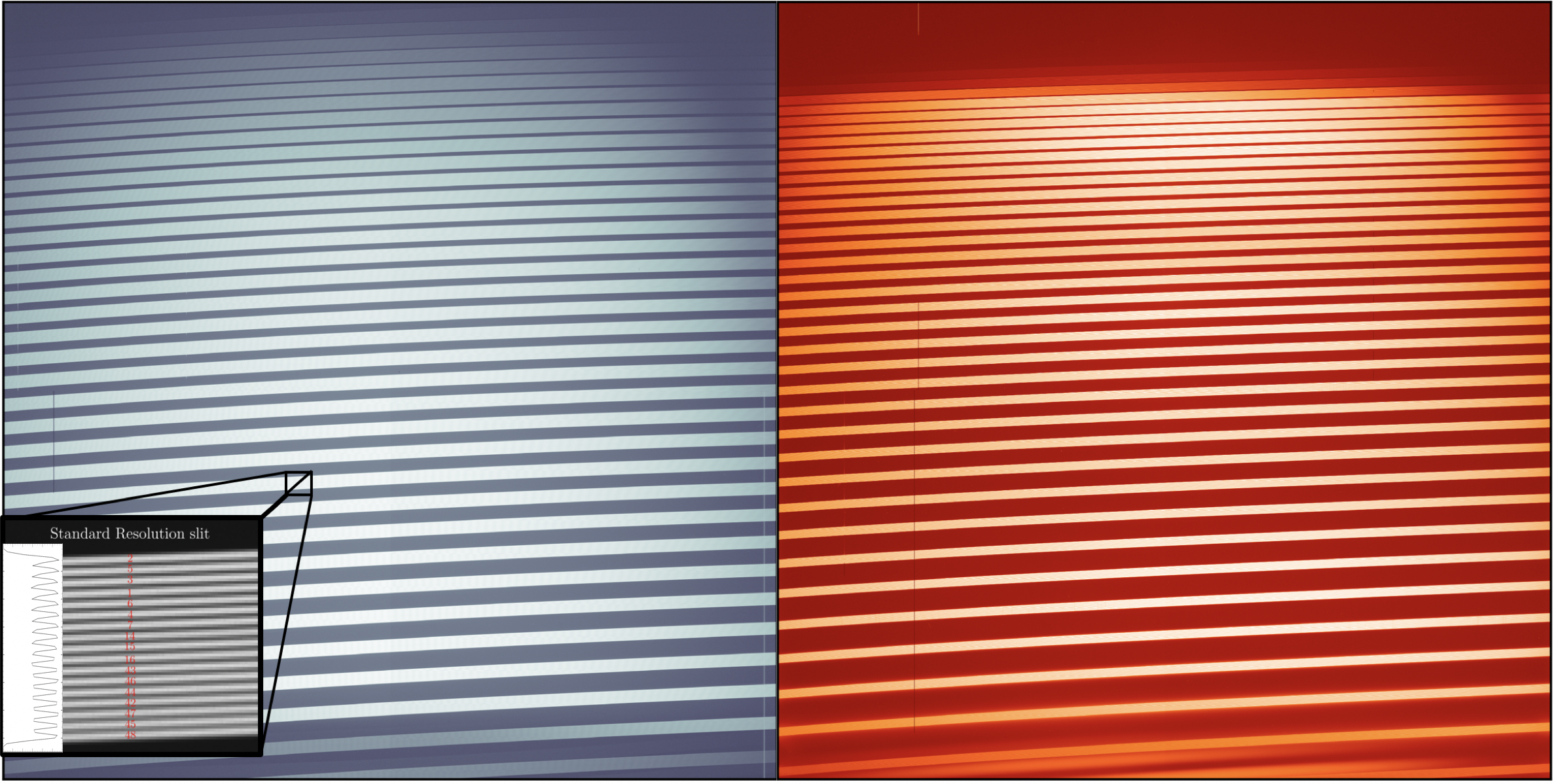}
\caption{Flat fields of the blue (left) and red (right) spectrograph detectors in standard resolution mode, showing the orders from top to bottom. The orders are not numbered for clarity. The zoom-in to the bottom left shows the individual fibers, resolved in order 72, and numbered. A cut across the order is also shown, displaying the profiles of the individual fibers revealing the image quality of the final spectrum. The fiber numbers within each order have the same convention and can be traced to the pseudo-slit image (Fig.\,\ref{slit}), and to the initial light entrance in the IFU as seen in Fig.\,\ref{micro}. Also visible is the extremely low inter-order background, and scattered light in the final echellogram. \label{format}}
\end{figure*}

\section{Performance} \label{perfm}

\subsection{Acquisition and guiding} \label{perfm}

The GHOST Cassegrain Unit lacks an on-instrument wavefront sensor. This can cause flexure to shift the position of the unit, which depends on the gravity vector and is influenced by the Cassegrain rotator angle and telescope zenith distance. Additionally, the telescope peripheral wavefront sensor has modest flexure and uncertainties in its probe mapping, both of which can mimic flexure effects in GHOST.

To address these issues, each science IFU has six guide fibers surrounding it, falling onto a 4.5$\micron$ pixel Peltier-cooled detector, with peak throughput corresponding to around 600\,nm, although the wavelength range is from 400--1000\,nm ($\gtrsim$80\% till 800\,nm) covering nearly the whole spectral range of the instrument and allowing sensitivity to a variety of objects. These fibers center the targets and correct for centering errors or flexure against the Peripheral Wave Front Sensor (PWFS) before and during exposure. During the exposure, the fibers are continuously monitored for error signals that are used to make small positioner corrections to keep the star optimally centered. The software employs a Proportional–integral–derivative (PID) controller loop to effect corrections based on the calculated centroid. The parameters of the PID loop are same for all the science IFUs, and the tolerance beyond which corrections are applied is currently set at 0.05$\arcsec$. Details on design of the guiding system is to be found in \citep{nielsen}.

The GHOST guiding loop also provides slow corrections to the IFU positioners with respect to the telescope tracking at low frequencies ($<$10 Hz), essential for observing faint targets for prolonged exposures (e.g. exposures longer than 1200 seconds), but also centering bright sources ensuring minimal slit loss. The system also ensures no conflicts with high-frequency peripheral wavefront sensor tracking at frequencies larger than 50 Hz. GHOST can additionally track slow-moving objects without on-telescope guiding for short durations if PWFS loses guiding for a few minutes.

In addition, the two IFUs cannot move on the curved Cassegrain focal surface, causing the focus of each IFU to degrade away from the center of the on-telescope unit's focal plane. To correct this, small corrections are applied to the telescope's secondary mirror (termed the focus offset) using the telescope control system, ensuring optimal image quality. These techniques allow the GHOST Cassegrain Unit to maintain optimal centering and focus, enabling precise astronomical observations despite challenges such as flexure and uncertainties in probe mapping. Due to this offset, single targets observed at the center of the focal plane, or dual targets observed equidistant from each other will have set the optimal focus.

Overall, the accuracy on point-like sources has been empirically determined to be better than 0.1$\arcsec$ during nighttime operations, when acquiring directly (before centering), sufficient for providing an array of guiding options (such as blind offsets, companion guiding) using the Gemini observing software and PWFS2. 

GHOST also allows for a blind spiral search for faint targets with poorly known coordinates. Commonly, faint objects prove challenging to acquire and guide during nighttime observations. This is simply because there is insufficient flux to visibly distinguish the object from the background even to the trained eye, leading to acquisition and guiding issues. For many targets of opportunities (ToO), this issue is compounded, as they are not only faint but also have poorly known coordinates. To address these issues, GHOST is equipped with a feature to automatically center and acquire faint targets, that are outside it’s entrance aperture, but within a circular area of $\sim10\arcsec$ around a given coordinate. This translates approximately to positional errors of 3$\arcsec$. This feature, dubbed the spiral search can be enabled in any mode.

\begin{figure}
\plotone{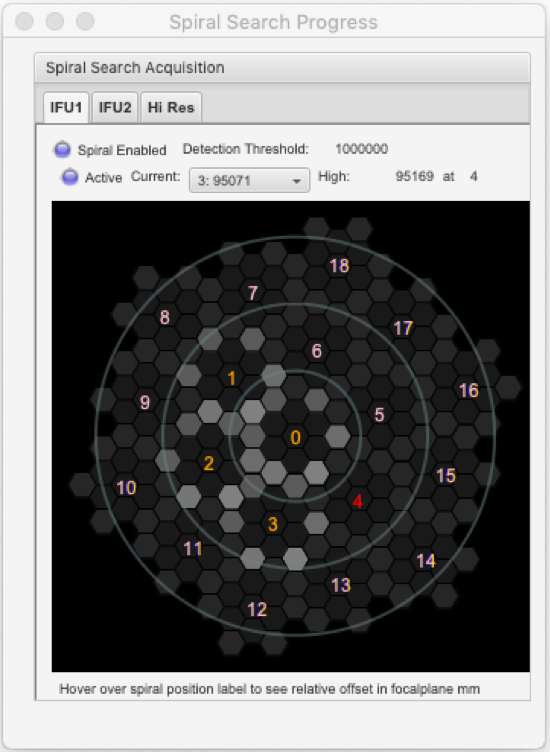}
\caption{Spiral search acquisition sequence dialog. When flux levels are below a pre-determined detection threshold on the presumed target coordinates, then an automated sequence of offsets is initiated, around the initial location. Each concentric circle indicates positional errors of 1$\sigma$, 2$\sigma$, and 3$\sigma$ from the center respectively, where $\sigma$ is 1$\arcsec$. In this example, the search has been initiated. The highest flux level was found at position 4 in the previous iteration as shown in the dialog at the top. In the current iteration, positions 0--3 have been completed (in orange), and 4 is next to be taken (in red) in the process. The IFU will now repeat till the center of the IFU has the highest flux level. \label{spiralsearch}}
\end{figure}

In spiral search mode, initially, acquisition is attempted in the normal fashion by calculating the maximum flux across each microlens, and recenter at the highest amount of flux. However, if insufficient flux is detected, the spiral search feature is activated. This process involves taking an automated sequence of exposures around the given location. This iterative process is repeated either till none of the microlenses have a flux above a user-defined threshold (usually given by the sky background), or the center of the object IFU achieves the maximum flux. An example of the spiral search acquisition is shown in Fig.\,\ref{spiralsearch}, depicting the presumed target coordinates, and the spiral search This ensures an object is centered in the object IFU, and allows for maximum SNR to be achieved.

\subsection{Spectral format}

The spectrograph reformats the delivered microlens image to increase spectral resolution over a wide wavelength range given its large input aperture. The image is also split by a dichroic (with an overlap between 520--543\,nm between the red, and blue detectors), with redward wavelengths falling onto a red and blue detector (salient detector properties are given in Table\,3). In the dual target configuration for SR mode, which acquires science spectra from two IFUs simultaneously, the same wavelength range is covered with target fibers sitting adjacent vertically on the detector. Figure\,\ref{format} shows the spectral formats of both the blue and red arms in standard resolution, with the zoom-in in the SR mode demonstrating the quality of the image in separating the fibers. The minimum separation between orders is around 50 pixels, allowing for inter-order background evaluation. The spectral ranges on the detector and blaze wavelength for each order agree well with the expected format described in \cite{pazder16}.

\begin{figure}
    \centering
    \includegraphics[width=1\columnwidth]{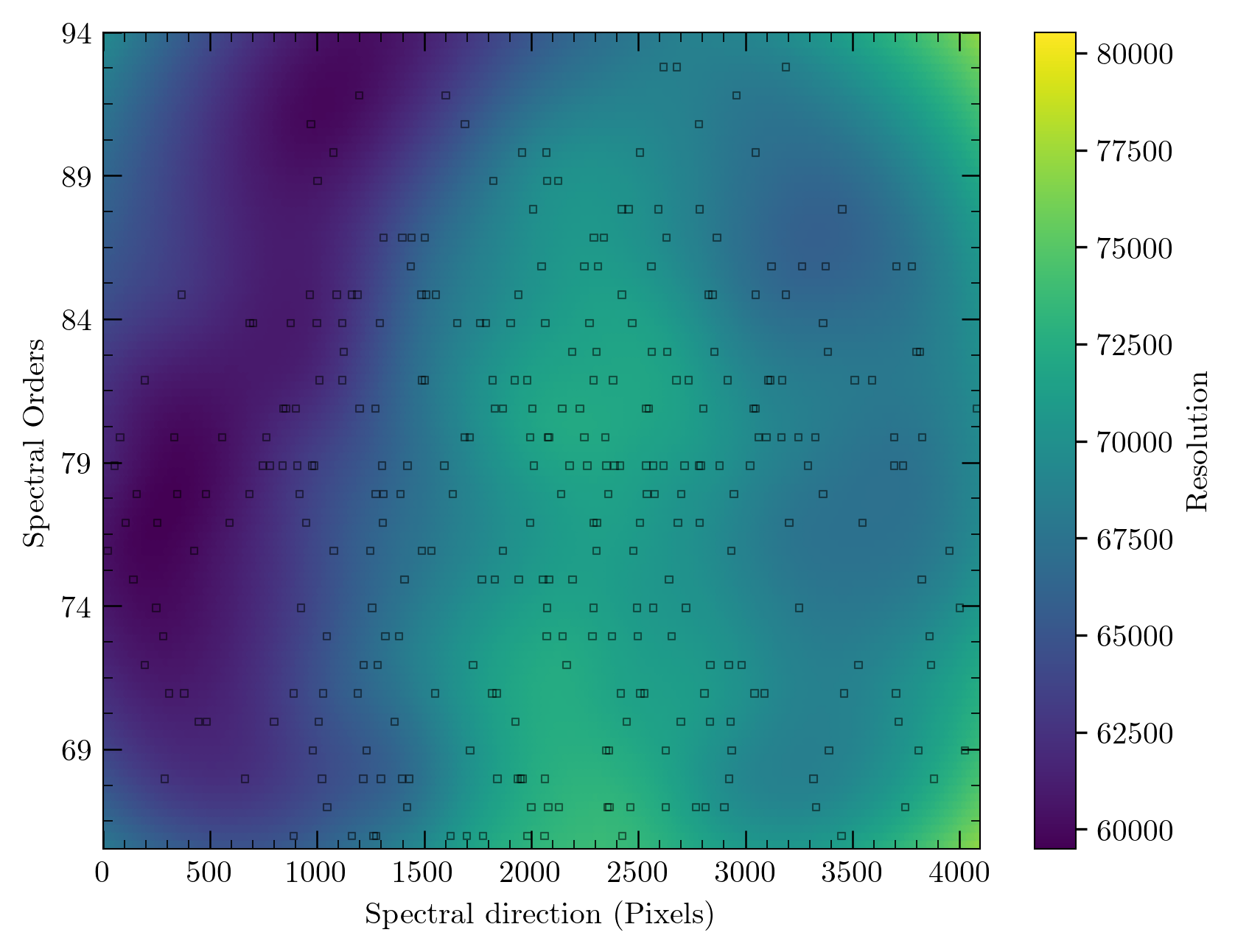}
        \includegraphics[width=1\columnwidth]{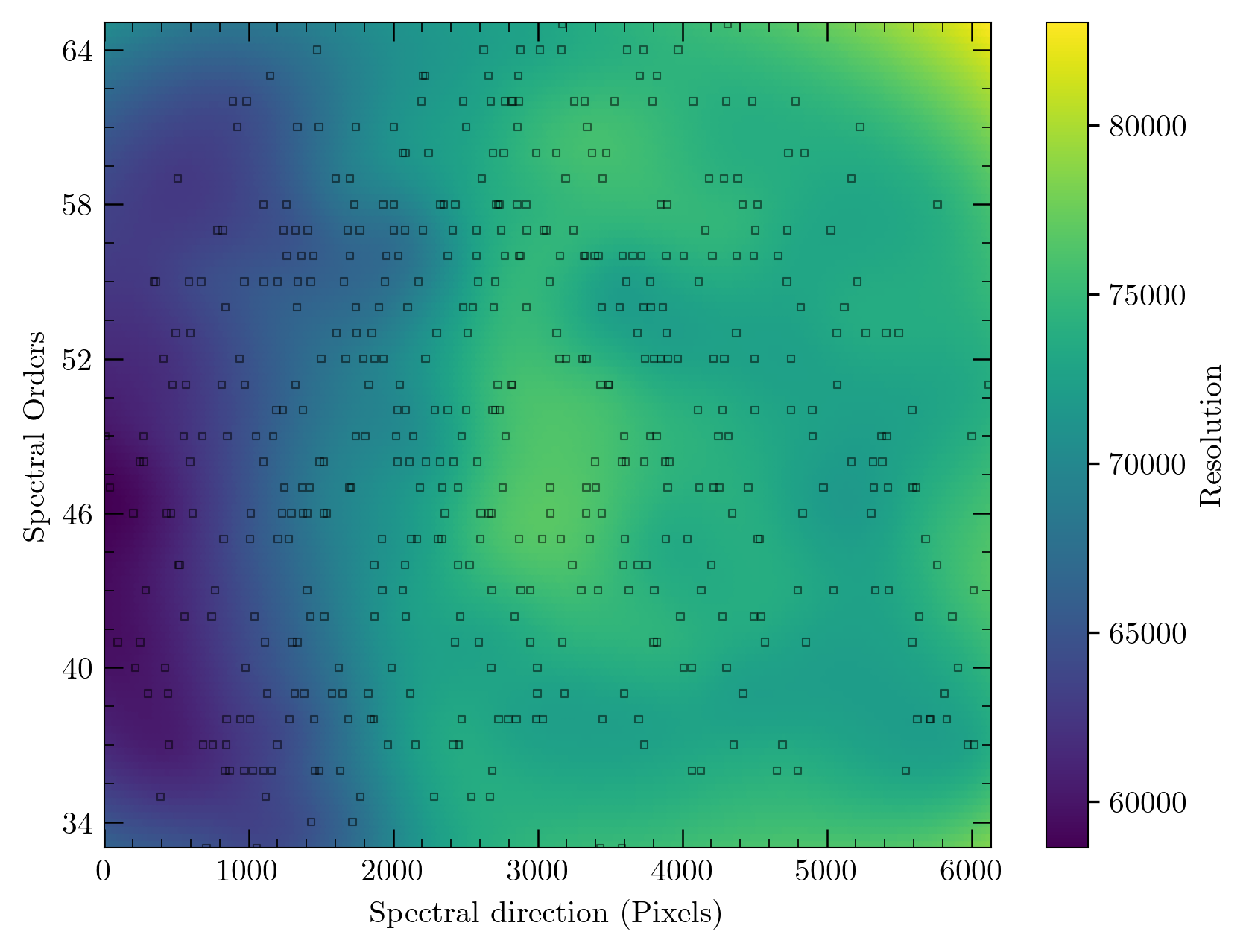}
    \caption{Measured resolution in the HR mode displayed in the echelle format for the blue (top) and red (bottom) detectors. The ordinate refers to the order number, and abscissa the pixel value. Standard resolution mode follows roughly similar resolution patterns on the detector. Measured line locations are marked. Corners of the map are extrapolated, and not necessarily representative of the achieved resolution. The colour bar refers to the achieved resolution, smoothed and mapped using a multi-quadratic function. }
    \label{resolution}
\end{figure}

\subsection{Spectral resolution}

To measure the achieved spectral resolving power, we took multiple arc lamp exposures in the two resolution modes across the full spectrum. A Thorium-Argon hollow cathode lamp was used for this purpose, and we assume that in both modes, individual emission lines are not, or only partially resolved. For lines that were not saturated, and had sufficiently high SNR, we measured the FWHM (full width at half maximum) for well-fit lines using a Gaussian. The FWHM was calculated by using a conversion factor for the boxy instrument line profile, so as to not underestimate the FWHM (the reader is referred to \citealt{robertson} for further details). The resolutions are given in Table\,4. Fig.\,\ref{resolution} displays the results for both detectors in high resolution mode as an interpolated map in the echelle format. The variation of the FWHM across the detector, and orders is presented, and shows that maximum resolving power is reached towards the center, but decreases towards either side of the detector, but more so to the left. This pattern appears similar to both detectors. These values agree with the measurements of the order center presented in \cite{mcconn24} from the commissioning phase, suggesting a stable instrument and sampling. Binning in the spectral direction in the high-resolution mode is not recommended as it under samples the line spread function. The resulting values are slightly higher than the requirements of the spectrograph in the design phase (of around 50\,000 in standard, and 75\,000 in high-resolution mode).

\subsection{Instrumental throughput}

\begin{figure}
\includegraphics[width=1\columnwidth]{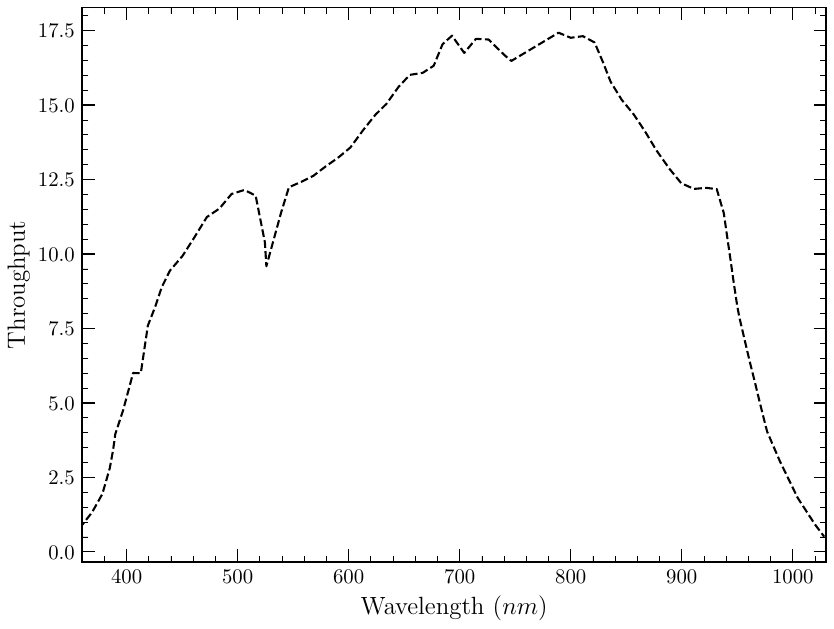}
\caption{Measured total throughput of GHOST at Gemini South from selected spectrophotometric observations, from top of the atmosphere to the detector. For completeness, note that the values on the vertical axis are in per\,cent. The instrument during these observations was located at the bottom port. When mounted on the side-port, the throughput is slightly lower owing to the addition of the science fold mirror reflection. The drop around 530\,nm at the dichroic is also visible. 
\label{thrusr}}
\end{figure}

The total instrumental throughput is defined for GHOST as the ratio between the photons measured on the spectrograph detectors at the blaze wavelength of each order at the top of the atmosphere given the Gemini Telescope collection area, assuming photometric conditions and corrected to an airmass of 1, to the expected number of photons from the source at the top of the atmosphere. The measured efficiency at the blaze wavelengths of each order is shown in Fig.\,\ref{thrusr}, for the standard, and high-resolution modes respectively. These measurements were made with on-target seeing conditions of 0.8$\arcsec$ or better (IQ70 following the observing conditions constraints of the telescope\footnote{https://www.gemini.edu/observing/telescopes-and-sites/sites}), in photometric skies. 

The peak throughput of GHOST is around 12\% in the blue (at 450\,nm), and 17\% in the red (at 650\,nm) in the SR mode. Note that this is similar for both IFU1, and IFU2 since the only difference is the physical fiber bundles, and the differences in throughput between the two IFU's are at the level of less than a few percent. For the HR mode, the throughput is slightly lower in similar conditions, given the smaller microlens aperture, and larger filling factor losses. It is around 7\% in the blue, and 9\% in the red. At the bluest wavelengths, GHOST performs slightly lower than theoretical expectations (given in Sheinis et al. 2016), but exceeds the predicted performance in the red channel. The throughput of GHOST is regularly monitored during nighttime operations to ensure that the efficiency of the instrument is maintained.

\begin{figure}
\includegraphics[width=1\columnwidth]{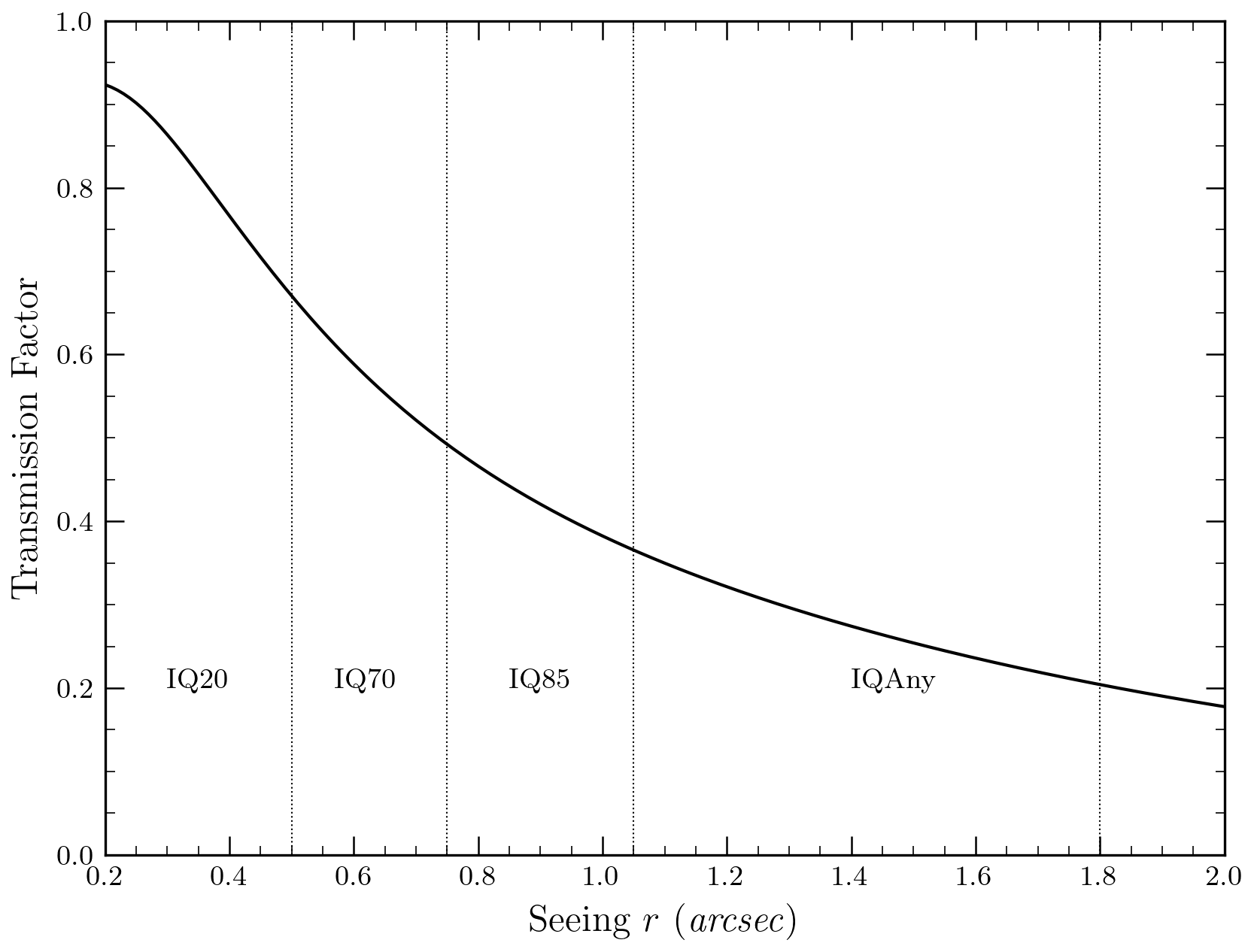}
\caption{The aperture transmission function as a function of seeing, accounting also for filling factor losses in the IFU configuration. Shown also are the seeing bins adopted at Gemini South during queue mode. 
 \label{slitloss}}
\end{figure}

Given the fixed aperture of GHOST, the achieved pseudo-slit transmission function is solely a function of seeing, independent of slit size. The final throughput therefore varies given the nighttime seeing. To estimate the aperture losses, we calculated the approximate transmission function by modeling the input light as a point-spread function using a Moffat profile. This estimation includes the microlens metrology and their filling factors and is given in Fig.\,\ref{slitloss}. The estimates agree well with the throughput achieved of selected spectrophotometric standards at the center of the red and blue detector, compared and scaled to an idealized seeing of 0.4$\arcsec$ predicted using the instrument integration time calculator (see Section 4.4.1 for a discussion of this); and also the design expectations \citep{sheinis16}. We note that while the trends indicate significant aperture losses at higher seeing values, GHOST is still able to produce good SNR spectra in even poor weather conditions (see Section 5.4 for an example) of bright targets, suitable for an array of science cases. Observations in queue mode over the first few years of operations will help paint a fuller picture of the throughput stability and slit transmission over the range of conditions seen at Gemini South. 

\subsubsection{Comparison with integration time calculator}

A simulated integration time calculator (ITC) is available for the whole spectrum of instrument options, with the details tabulated in the relevant webpages\footnote{https://www.gemini.edu/instrumentation/ghost/exposure-time-estimation\#ITC}. The integration time calculator accounts for the initial light entrance by modeling the input aperture and the associated metrology of the microlens, and the tracing of light into the pseudo-slit and through the spectrograph. The final spectrum should be representative of an extracted object spectrum given the input object, and instrument parameters.

To estimate the accuracy of the performance of the ITC with nighttime observations, we compared the ITC predictions with standard star observations across a variety of observing conditions. Spectrophotometric standard stars from \cite{hamuy1994}, or the {\it Hubble Space Telescope} primary white dwarf standards were observed for these purposes. These targets were observed in various settings and combinations, in both the standard resolution and high-resolution mode, including single target, dual target, and target with sky mode. These observational results can be compared with the predicted results of the ITC. 

\begin{figure}
\includegraphics[width=1\columnwidth]{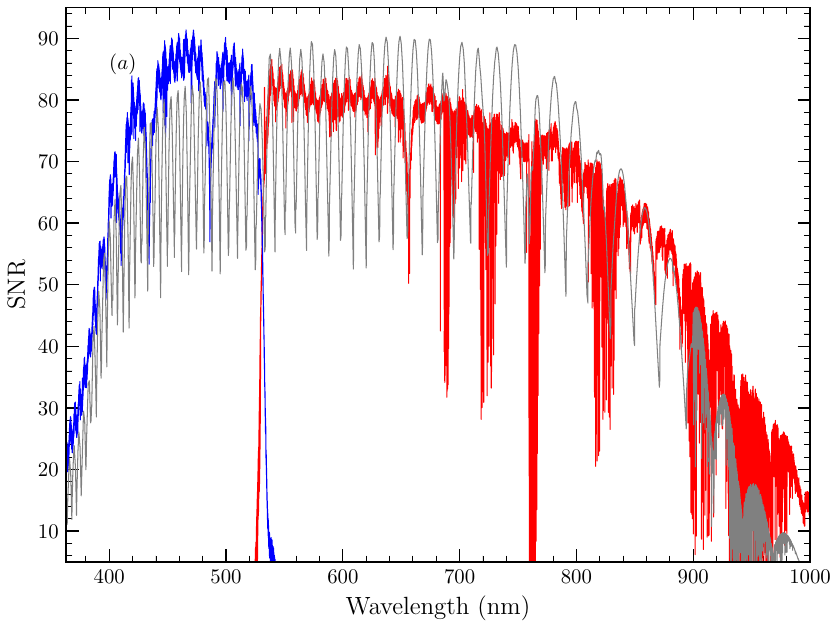}
\includegraphics[width=1\columnwidth]{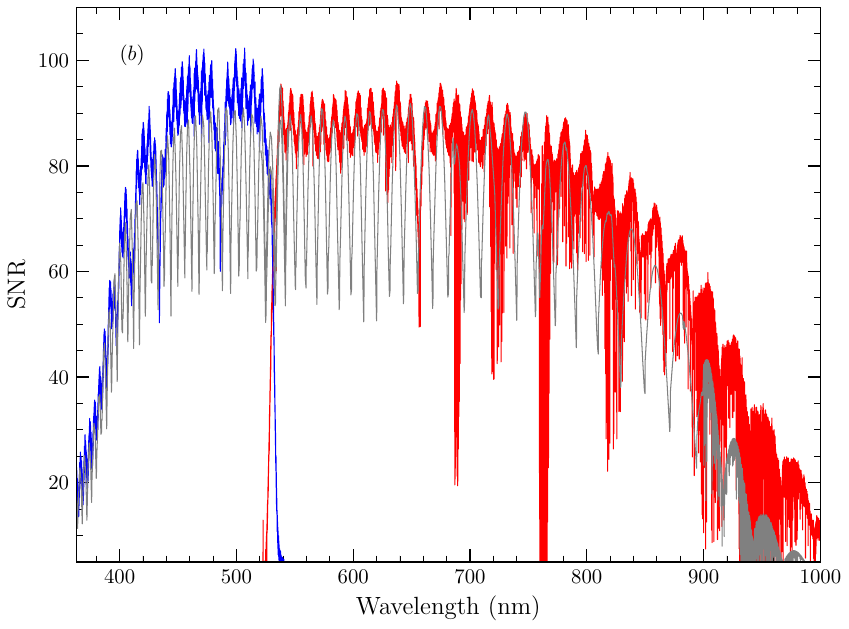}

\caption{A comparison of the performance of the Gemini integration time calculator to measured SNR per resolution element. Panel (a) shows the comparison of the observed spectra (blue and red) taken in standard resolution, target and sky mode with a binning of two in the spectral direction. The ITC simulated spectrum in approximately similar seeing, cloud cover, and sky brightness conditions at the same instrument settings in also shown (gray solid line). Panel (b) shows the same comparison, but for the high resolution mode unbinned. Observations are not corrected for telluric contamination. \label{itc}}
\end{figure}

All data were reduced using the reduction pipeline, and seeing conditions at the time of observation were estimated using the GHOST reconstructed slit viewing images. An example of our results is given in Fig.\,\ref{itc}. Here, we show the comparison between the observed results for the flux standard EG\,274, and the predicted SNR per resolution element from the ITC under identical instrument settings during two different nights, under different seeing, cloud, and sky conditions. The estimated final S/N when compared to observations agrees within an error of 5\% in clear conditions at the central wavelengths of the detectors. In conditions with higher levels of cirrus, the estimated SNR are more varied, possibly since the absolute level of cloud cover must be estimated by the observer by eye using all-sky cameras during the night, while other parameters such as seeing are measured throughout the night either by the reconstructed slit-viewing images, or using the dedicated seeing monitor. Overall, results from spectrophotometric standards correspond well with the predictions, suggesting the ITC has demonstrated its effectiveness in determining the exposure times required to achieve the desired SNR under most observational settings and weather conditions.

\subsection{Radial velocity performance}

The standard and high-resolution modes allow for radial velocity precision around 50--100s of m\,s$^{-1}$, owing to the highly stable environment in which the instrument is located, and the lack of any moving parts besides a slit mask in the spectrograph itself. Here, we report on the instrument stability for such science and the performance from pipeline-reduced data. Note that the stability of performance during initial operations may be compromised due to continuous adjustment of the instrument software system for better operations performance, and the hardware components (for e.g., the slit-viewing camera was replaced due to excessive noise; or the guider configurations have been continuously updated to improve performance).  

Precise monitoring of the spectrograph's environment, and how the stability is affected by it is essential. To test this, we continuously monitor when feasible the performance of the instrument over a variation of environmental conditions, using continuous sequences of calibration frames taken with the internal lamp at periodic intervals.

\begin{figure}
\includegraphics[width=1\columnwidth]{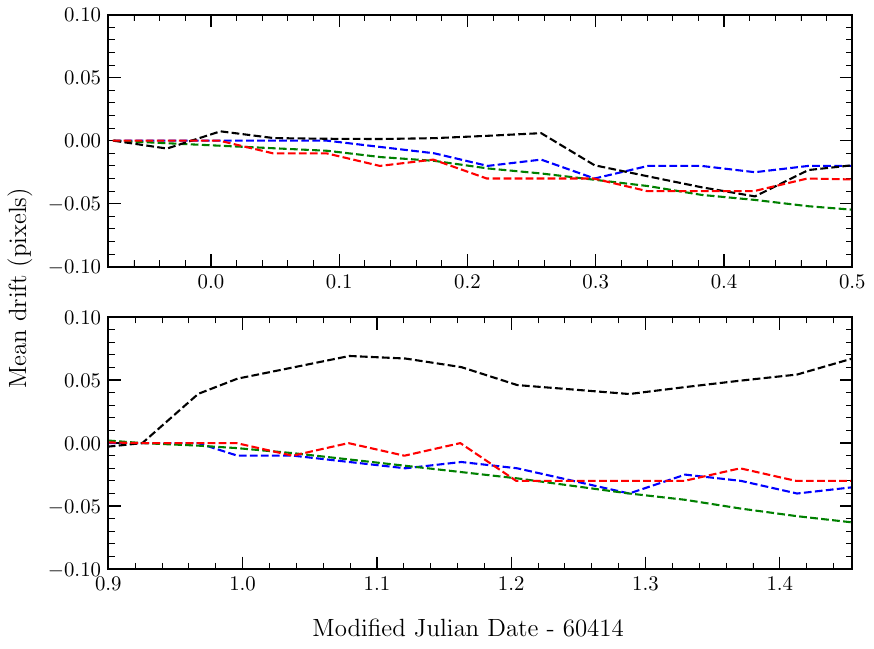}
\caption{Mean pixel drift compared to the grating temperature and enclosure pressure over two nights in April 2024. Dotted lines mark the change in pixel value in the blue detector (blue), red detector (red). The grating temperature (green) is given in units of Kelvin, and enclosure pressure (black) in hPa. All values are normalized to the initial value, with pressure divided by a factor of 20 for display.  \label{drift}}
\end{figure}

The precise pixel location of an arc line varies with pressure and temperature changes, and can be seen changing over two days in April, 2024 in Fig.\,\ref{drift} along with the corresponding change in temperature and pressure. Here, we measured the change between arc calibration frames using cross-correlation. The results suggest a drift of around 0.5\,pixel for changes of 1$^\circ$C in temperature, and 0.25\,pixel for changes of 10hPa in temperature. The dependence of the spectrograph sampling with ambient pressure and temperature is monitored using arc frames every few months. Preliminary results indicate (in addition to the dependence mentioned above) a variation with the cyclic seasonal changes in temperature (which also affects the temperature of the detectors).    

We also tested the precision achieved by the system on-sky by monitoring the RV standard star HD\,21693. We performed between one to three observations per night for a total of six consecutive nights, where the exposure time was changed according to the conditions to achieve an average signal to noise of 200 per resolution element (unbinned spectral data), and observed between airmasses of 1 and 1.4. The resulting relative radial velocities measured as a function of time are shown in Fig.\,\ref{rvtest}. For high-resolution measurements, arcs were taken either before or after each observation. For standard resolution measurements, arcs were taken during the daytime calibrations, which were usually around 2--4\,hours before the science observations in these cases. Note that the during these measurements, the instrument had been in operation for nearly two weeks, and the instrument inner enclosure temperature was relatively stable. The radial velocities were measured using the IRAF{\footnote{NOIRLab IRAF is distributed by the Community Science and Data Center at NSF's NOIRLab, which is managed by the Association of Universities for Research in Astronomy (AURA) under a cooperative agreement with the National Science Foundation.} cross-correlation task {\it fxcor}. Our initial results suggest that the RV precision over this time period is around 150\,m\,s$^{-1}$ in the standard resolution mode, and around 50\,m\,s$^{-1}$ in the high resolution mode. An instrument monitoring campaign has begun along with queue operations to monitor the radial velocity stability over longer time frames in all modes. 


\begin{figure}
\includegraphics[width=1\columnwidth]{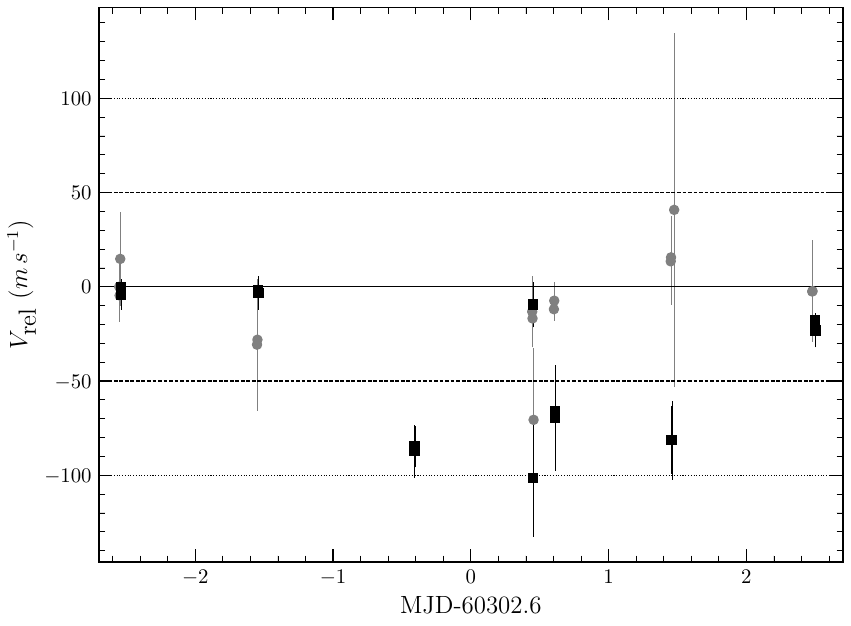}
\caption{Relative radial velocity shift measured over a duration of 6 nights on the standard star HD\,21693 in both standard (solid black squares), and high resolution mode (light gray circles). The dashed and dotted lines indicate a variation of 50 and 100\,m\,s$^{-1}$, respectively.  \label{rvtest}}
\end{figure}

\subsection{Performance at faint-end}

A key component of the instrument design of GHOST is the high efficiency of the instrument. The throughput on sky of both the standard resolution and high resolution are given in Section\,4.4, and demonstrate the ability to achieve a clear detection (around a SNR per resolution element of 10) for a $\sim$19\,mag object in an hour in median conditions. Observations beyond this are more limited by cosmic rays in the final reduction. Alongside with the spiral search mode, this suggests GHOST would be an excellent instrument to observe faint targets, including faint targets of opportunity. 

The performance of GHOST in the faint regime (in a highly binned mode 2$\times$8, representing spectral and spatial directions respectively; see also \citealt{ireland18}) is good given its low read noise and dark current of the spectrograph detectors. In this significantly binned mode, a long exposure (1 hour) translates to roughly around 4.5${e^-}$ in read noise, and a dark current of 8${e^-}$, both sufficiently smaller than the sky background that chiefly dominates these use cases, allowing observations of faint targets (up to\,$V\lesssim$21\,mag) in photometric, dark sky conditions with good seeing.

\begin{figure}
\includegraphics[width=1\columnwidth]{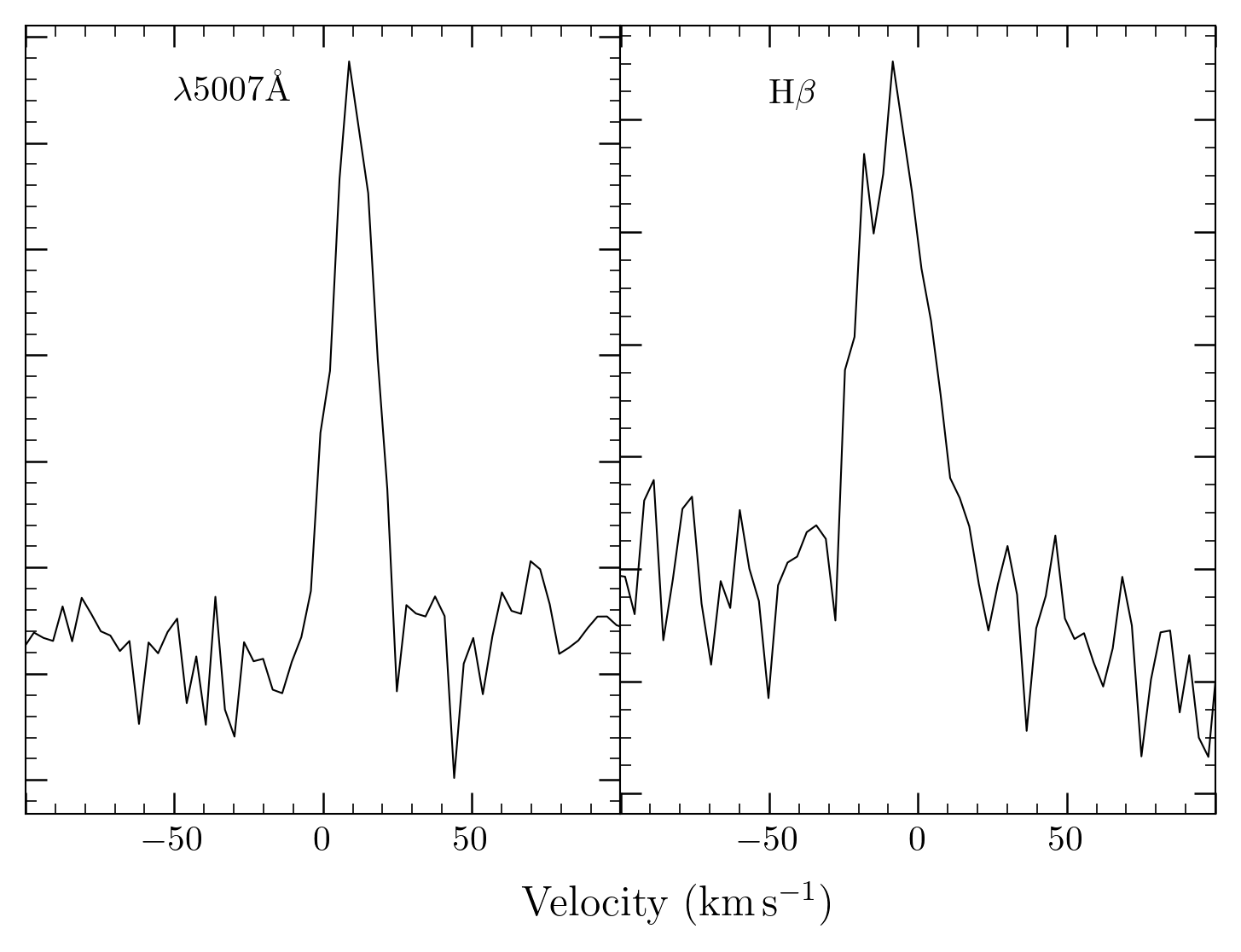}
\caption{Forbidden [\ion{O}{I}] nebular emission at $\lambda$5007\AA\ and H$\beta$ emission line profiles seen towards the star SextansA\,s2 ($V=20.8$\,mag). The final SNR per resolution element in the red was around the continuum in the blue (around H$\beta$) was 8, and the in red (around H$\alpha$ line) was 11 in three hours of total exposure time. The SNR in the continuum varied across the wavelength. \label{sexA}}
\end{figure}

An example of pushing GHOST to its faint limit is shown in Fig.\,\ref{sexA}. Here, we show the H$\beta$ and H$\alpha$ emission line profiles of the extra-galactic O3-O5\,V star, SextansA\,s2. This star represents one of the earliest known spectral type at a metallicity of around a tenth solar \citep{garcia19}. SextansA\,s2 has $V$=20.8\,mag, and is located in the Sextans\,A dwarf galaxy. From low-resolution spectra, it is clear that the star shows bright emission lines at those wavelengths, having a single peaked profile in H$\beta$. From the low-resolution spectra, it remained unclear if the Balmer emission originated from nebular contamination, or the target. We took as a demonstration of the faint limit during commissioning, three 1-hour exposures in clear dark skies, with a seeing of around 0.7$\arcsec$ during the course of the observations. We are limited not only by the faint limit of the spectrograph but the sky brightness as well whose limit is around 20.8\,mag\,arcsec$^{-2}$ in these conditions\footnote{https://www.gemini.edu/observing/telescopes-and-sites/sites\#SkyBackground}. The data were binned on-chip in the spectral direction by a factor of 2, and in the spatial direction by a factor of 8. From the final spectra, we obtained an SNR (per resolution element) per one-hour exposure of around 7, with a total combined SNR $\sim$10 at wavelengths longer than 450\,nm. This was sufficient only to detect the strong emission lines in the blue confidently but not absroption lines in the continuum, and those are displayed in Fig.\,\ref{sexA}. 

\begin{figure}
\includegraphics[width=1\columnwidth]{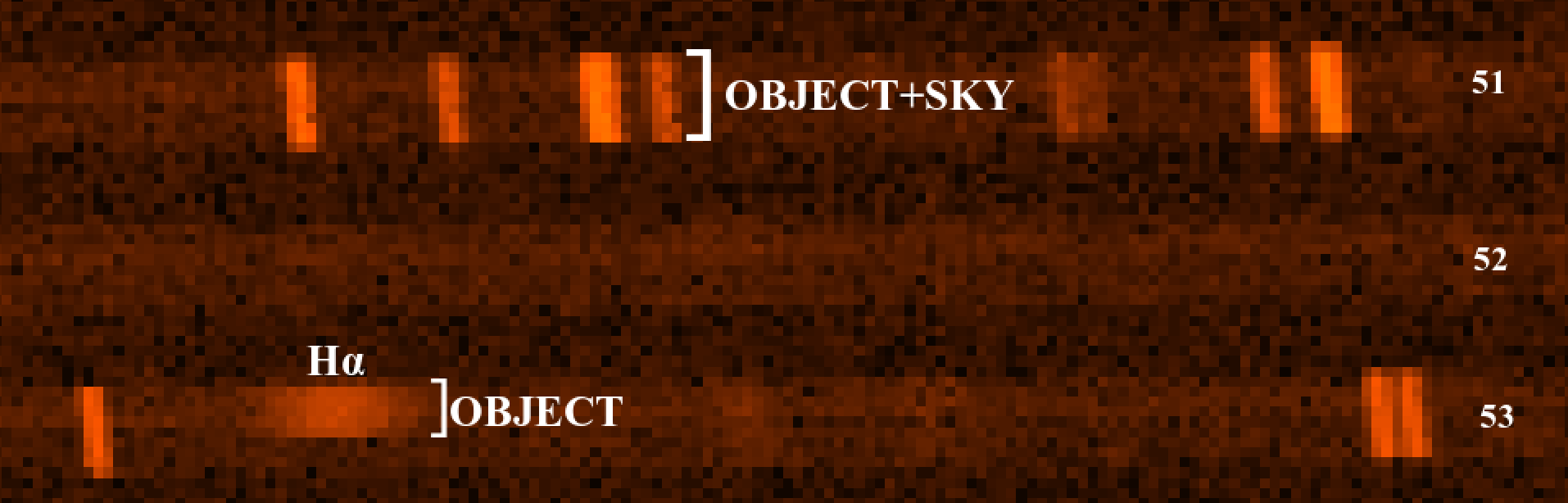}
\caption{Redcued combined red echelleogram observed towards Sextans\,A\,s2. The orders shown are 53--51, covering H$\alpha$ and the telluric lines in order 51. The width of the H$\alpha$ emission is well-defined, as well as the absence of the contamination in the sky, and secondary SRIFU fibers suggesting the emission originates from the object itself. \label{echs2}}
\end{figure}

While GHOST can feasibly (over longer exposures) observe such faint targets, its main benefit lies in exploiting a final high-resolution spectrum (in this case around 47\,000) in the optical, for such a faint target. In this case, similar long exposures at lower resolution ($R\sim$1000) have revealed the spectral nature of this object. But the line profiles in the blue (for e.g. H$\beta$) shown here are more clearly defined than in \cite{garcia19}. The width of the emission lines in H$\alpha$ seen in the echelleogram in Fig.\,\ref{echs2} also suggests that the Balmer emission comes from the source itself, which could not be conclusively determined from low-resolution spectra of \cite{garcia19}. This demonstrates how GHOST allows to explore the nature of such faint objects. Our experiment demonstrates that GHOST can feasibly acquire and obtain enough signal on faint object spectra to enable detection.

\section{First science results }

After the installation and integration of the instrument on-site at Cerro Pach{\'o}n in July 2022, multiple commissioning runs were conducted to verify the system performance, and obtain first science results \citep{hayes, sestito}. In May 2023, a system verification run was conducted by the Gemini management with various National Gemini Office participation, which was then followed by a shared risk call to the wider community with interesting results \citep{sv1,sv2}. These early observations focused on observing a wide variety of targets, spread in brightness, morphology, and scientific interest; and also tested system integration and operations across the full spectrum of conditions found at Pach{\'o}n. Here, we describe some of the interesting scientific results from the first observations in view of the instrument science drivers and performance regimes, offering a snapshot of the instrumental capabilities and performance. In Section 5.1, we describe use cases of PI-driven science at Gemini using examples in both the standard and high-resolution modes including analysis of stellar elemental abundances, and velocity profiles. Section 5.2 is an example of quasar high-resolution spectroscopy demonstrating the use of GHOST for extra-galactic and fundamental science. Section 5.3 contains an example of the radial velocity stability of GHOST for the identification of planetary transits and radial velocity studies. Finally, Section 5.4 includes examples of using GHOST in poor weather conditions, especially for monitoring campaigns. Overall, our results detailed here are intended to demonstrate the applicability of the instrument for the key science cases described in Section 1.

\subsection{Examples of PI driven science}

 \begin{figure}
\includegraphics[width=1\columnwidth]{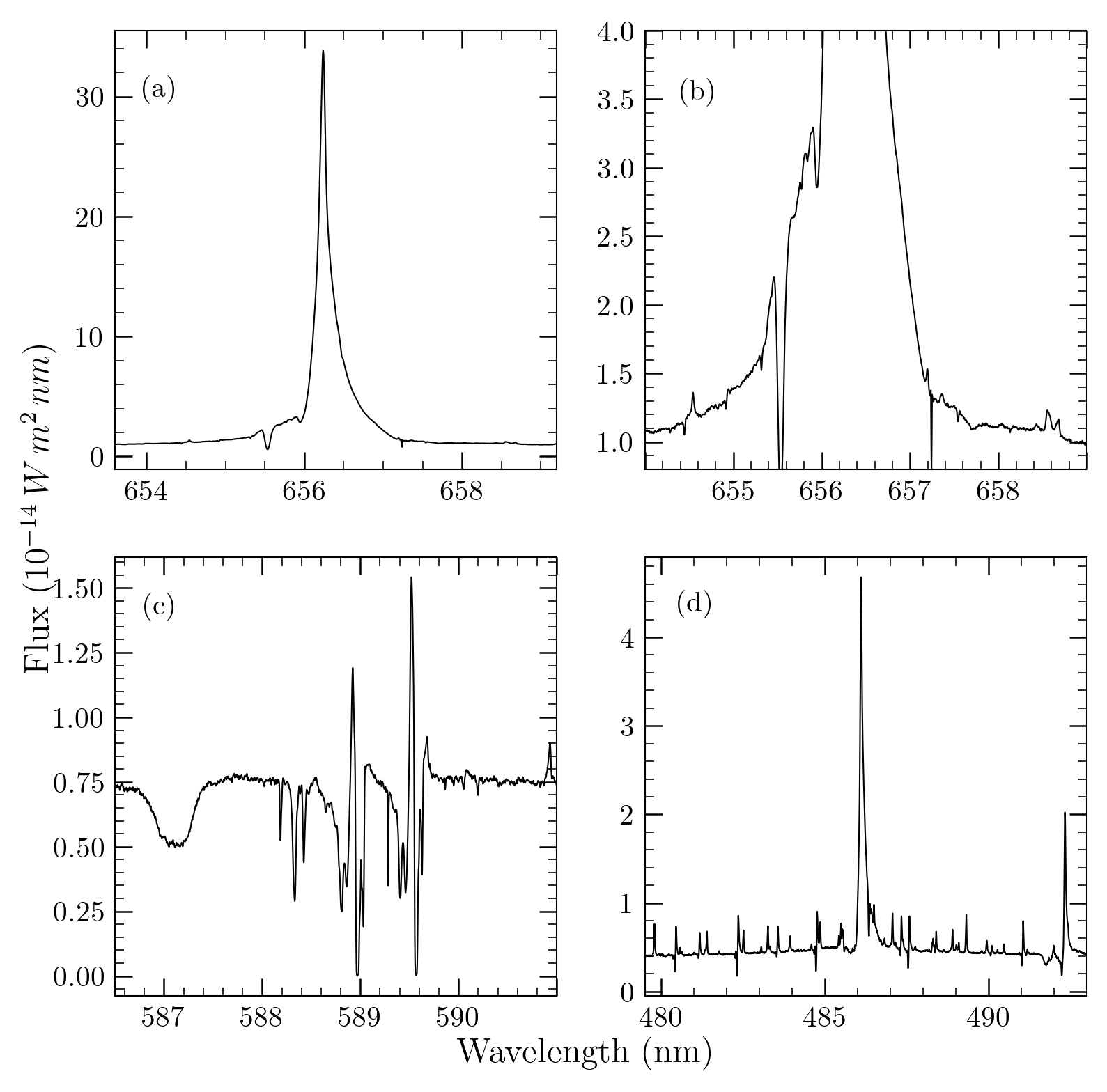}
\caption{Standard resolution spectra of the ``Iron star" XX\,Oph. Panel (a) shows the region around the H$\alpha$ line, displaying the strength of the emission, while (b) shows the finer detail around the line. The Na I doublet lines are displayed in (c). (d) displays the H$\beta$ line, and nearby metal ion emissions, which also display P\,Cygni absorptions. \label{xxophlines}}
\end{figure}

A number of early science cases for using the instrument will be PI-driven studies of well-known objects that require follow-up utilizing the exquisite resolution and wavelength coverage offered by the instrument. Here, we present details of two observations, the ``Iron star" XX\,Oph taken in the standard resolution mode, and TW\,Hydrae taken in the high resolution mode. 

\subsubsection{Revisiting XX\,Oph: Standard resolution observations}
First discovered by \cite{merrill51,merrill24}, XX\,Oph ($V=$8.6\,mag) is one of two known ``Iron stars". The name arises from the array of Fe emission and absorption lines seen in its spectrum, indicative of the simultaneous existence of several different environmental conditions within the star system. Multiple observational campaigns in recent years have cataloged the multitude of lines visible \citep{cool05}, attempting to gain insight into the nature of the star and its environment, and suggest that XX\,Oph is likely a very long period eclipsing binary, having an orbital period greater than 30 years. While two studies using optical spectra at $R$ of 20\,000 \citep{tomasella10}, and 60\,000 \citep{goswami01} have been performed, no in-depth study on the nature of XX\,Oph has been previously conducted using high-resolution spectroscopy. These high-resolution observations (as well as previous lower-resolution observations by \citealt{howell09} and references therein) note the variable line shapes, widths, and velocities visible in the spectrum, consistent with the binary star model proposed by \cite{howell09} and \cite{cool05}.

\begin{figure*}
\includegraphics[width=1\textwidth]{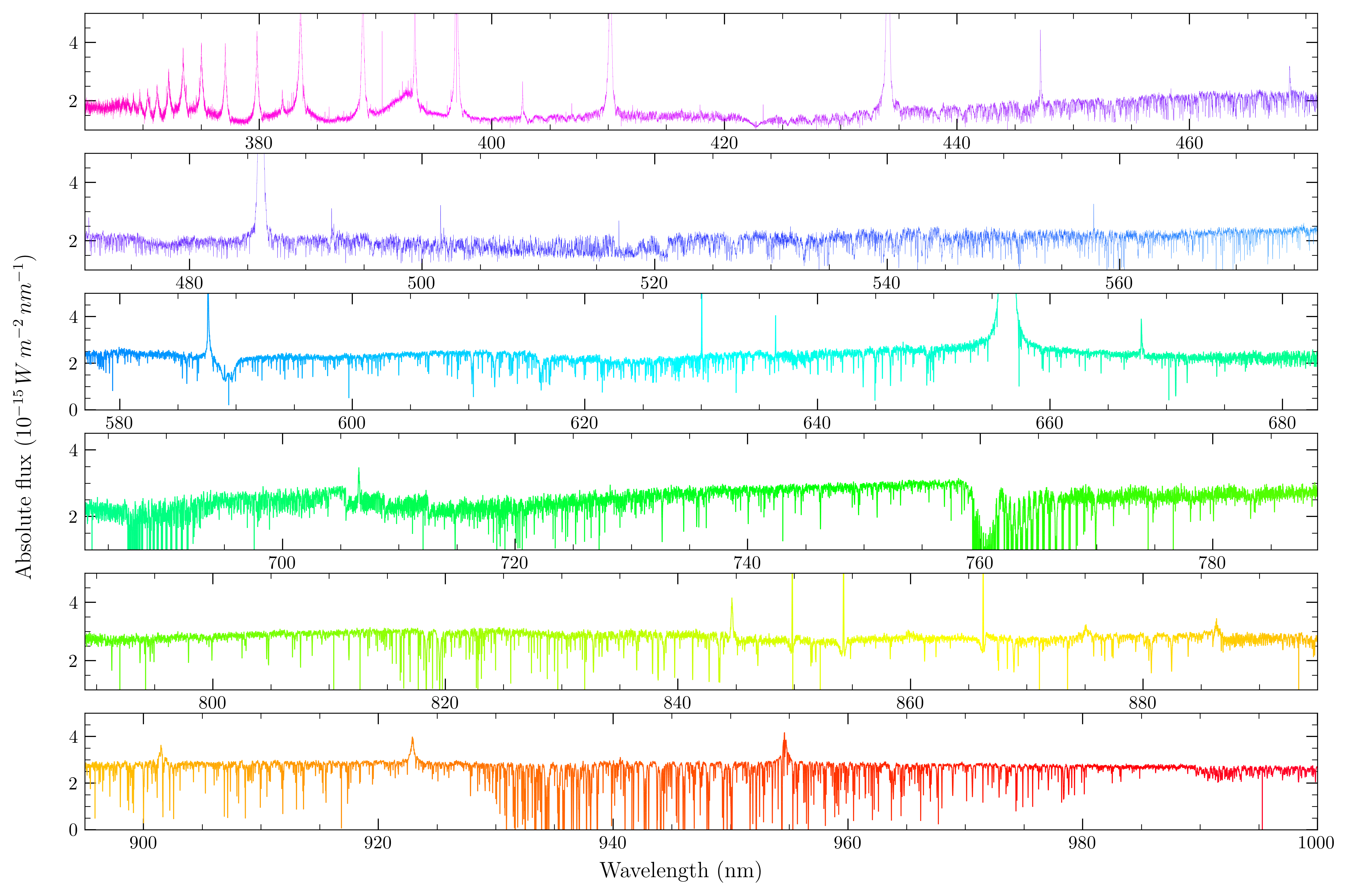}
\caption{GHOST high-resolution spectra of TW\,Hya, covering nearly the entire wavelength range of the instrument. Spectra are in absolute flux units, but not corrected for telluric lines. Strong emission lines are truncated for visual purposes. The colors are mapped based on a shifting multi-color scale from violet to red, aimed to be representative of the region of the spectrum plotted. \label{twhya}}
\end{figure*}

During instrument commissioning, multiple GHOST spectra of XX\,Oph were obtained on the 16 and 17 of April 2023, in the standard resolution mode. The observing conditions were photometric, with a seeing of 1$\arcsec$. The median delivered SNR is about 110 in the blue and about 270 in the red (outside of the telluric absorption features). To demonstrate the quality of the final spectra, in Fig.\,\ref{xxophlines} we plot selected emission and absorption features. A comprehensive study of the remarkable nature of XX\,Oph from the GHOST spectra will be presented in Howell et al. (in prep.). Instead, we restrict the analysis here to a few emission and absorption lines which can demonstrate to the interested user the quality of the spectra, and the analysis possible using GHOST. 

In Fig.\,\ref{xxophlines}a, we display the full extent of the H$\alpha$ line, exhibiting two P Cygni absorption dips on the blueward wing, the bluest located at $-$345\,km\,s$^{-1}$ (from H$\alpha$ at 656.3\,nm), approximately the same velocity as the single broad dip noted by \cite{tomasella10}. This indicates that past medium spectral resolution studies are limited in revealing such detail, and even the previous higher resolution studies did not have the SNR in the line wings/continuum for sensitive detections. The FWHM of the H$\alpha$ profile is 2\,\AA\, with the full extent of the line emission reaching $\pm$2800\,km\,s$^{-1}$. Figure\,\ref{xxophlines}b shows the zoomed-in profile of the region around H$\alpha$. We show in Fig.\,\ref{xxophlines}c, a view of the complex spectral region near the \ion{Na}{I} doublet lines including the broad \ion{He}{I} $\lambda$5871\,\AA\ absorption, with a displacement of $-$232\,km\,s$^{-1}$. In Fig.\,\ref{xxophlines}d, we display the similar far-reaching H$\beta$ emission line wings as well as illustrating the very complex array of emissions mainly due to \ion{Fe}{II}, \ion{Cr}{II}, \ion{Ti}{II}, and \ion{He}{I}. P Cygni absorptions are present in many of the lines, attributed to expanding wind structures within the binary (see \citealt{cool05}). These remarkable detections help to place the extraordinary GHOST spectrum into context amid previous works as well as within the binary model for XX Oph. 

\subsubsection{TW\,Hya: High-resolution observations of Lithium}

TW\,Hya ($V=$10.5\,mag) is an evolutionary enigma and presents a problem for the widely accepted model in which young stars disperse their natal disks within a few million years. This is because the estimated age of TW\,Hya is around $\sim$8-10\,Myr \citep{herc23, herbig78}, far longer than the age by which pre-main sequence stars are thought to disperse their discs (generally most late-type stars disperse their discs within a 3\,Myr; see \citealt{fedeledisc, haischdisc}). However, there remains some uncertainty on the spectral type of TW\,Hya, depending on the wavelength of the observations used to infer the type, and the results affect the derived age estimate (see the discussions in \citealt{herc23, vacca11}). The proper determination of the spectral type in pre-main sequence stars such as TW\,Hya is difficult, as the emission from these objects includes not only a stellar component, but an accreting component producing emission lines and excess continuum emission (shortwards of the Balmer jump), and a component from a circumstellar disc consisting of hot dust whose emission results in the veiling of the stellar absorption lines. In addition, some stellar features may be enhanced by the strength of the stellar magnetic field \citep{sokal18, yang05}. Differences in the effective temperatures (and therefore spectral types) of pre-main sequence stars derived from near-infrared spectra relative to those from optical spectra are discussed by \cite{flores22,flores20}.

Significant insight into the properties of such a well-studied object as TW Hya can be gleaned from the high spectral resolution and the long wavelength coverage offered by GHOST. An example spectrum of TW\,Hya, obtained with the high-resolution mode of GHOST, is displayed in Fig.\,\ref{twhya}. This spectrum was taken on the night of April 17, 2023, with an exposure time of 3$\times$240\,s in the red camera, and 900\,s in the blue camera, and achieved a SNR of around 100 in each exposure (at 450\,nm, and 650\,nm). 

\begin{figure}
\plotone{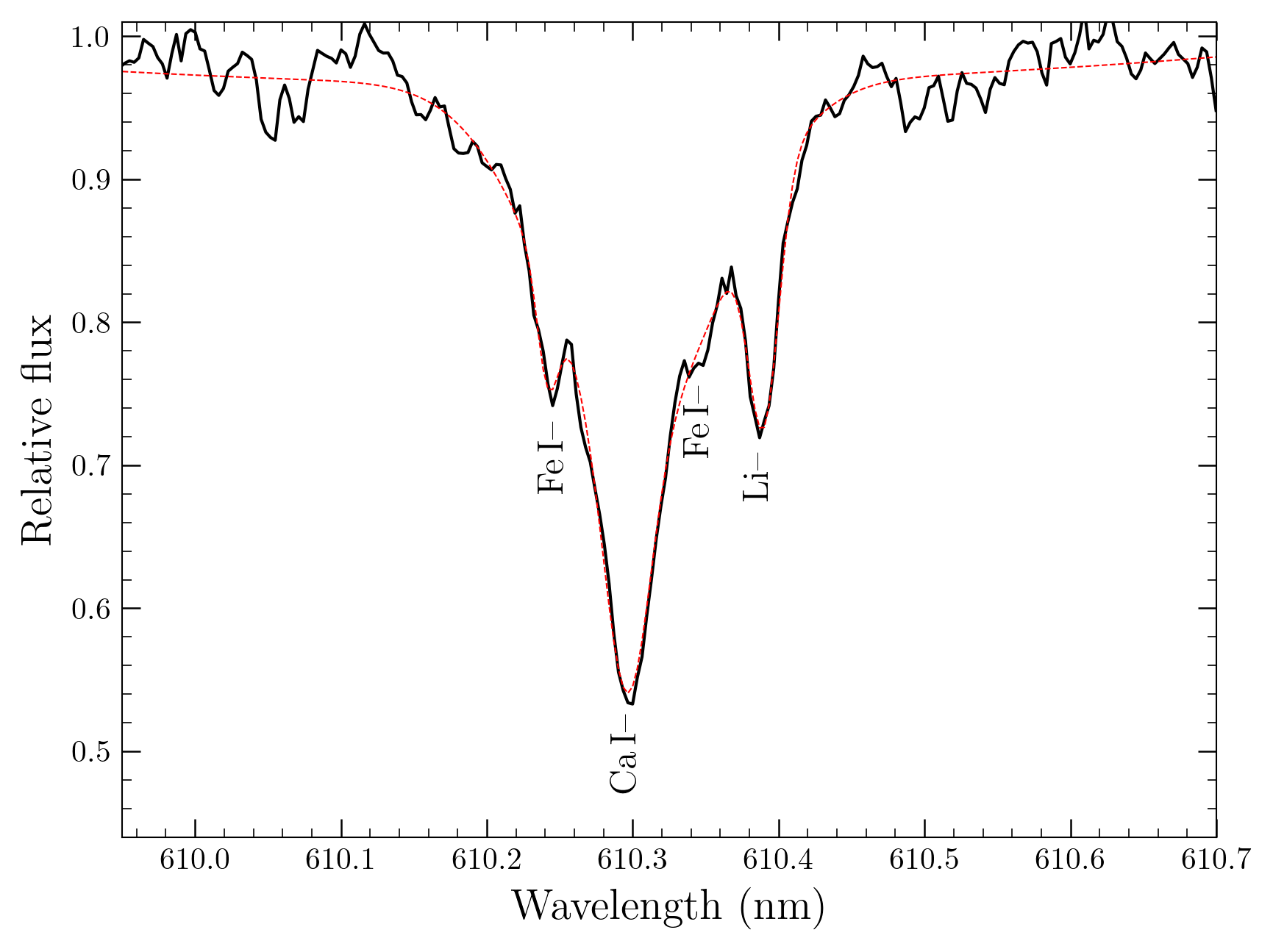}
\caption{Observed spectrum (black solid), and fit (red dashed) around the Li\,$\lambda$6103\AA\ region for TW\,Hya. The best multi-component (four) Gaussian fit is poorer at the redder \ion{Fe}{I} line, but fits well the remaining lines, including the Li line.  \label{li}}
\end{figure}

From the TW\,Hya spectrum shown in Fig.\,\ref{twhya}, one can estimate the accretion luminosity from the line luminosity of various accretion indicators, as well as characterize the line profiles. In particular, we draw attention to the H$\epsilon$ line in emission, which is blended with the \ion{Ca}{II} line at $\lambda$3969\AA\, both in emission. The Balmer, and He lines are visible in emission, although H$\alpha$ is saturated. Also visible in emission is the forbidden [\ion{O}{I}] $\lambda$6300\AA\ line emanating from the inner disc and disc wind. Measurements of the line flux from \ion{He}{I}, and [\ion{O}{I}] lines, and adopting the relationships given in \cite{herc23} give us accretion rates similar to the lower limit measured in that study. To demonstrate the utility of GHOST on such well-studied objects, and also exemplify its use for abundance determinations we highlight the detection of the Li$\lambda$6708\AA\ resonance line, and the subordinate transition at $\lambda$6103\AA. The bluer Li transition is usually not well resolved from nearby metal lines to estimate the line width, but can in principle be used along with the resonance line to estimate Li abundances \citep{duncan}. In our GHOST spectra, the $\lambda$6103.8\AA\ is sufficiently resolved to be deblended from the nearby \ion{Fe}{I}, and \ion{Ca}{II} lines. In Fig.\,\ref{li} we display the line, along with a multi-component Gaussian fit deblending the Li line. The measured line width from the $\lambda$6103\AA\ line is 0.1$\pm$0.02\,\AA\, and the resonance line is 0.45\AA. The latter value falls within the range of values listed from multiple spectra in \cite{barrado}. One can in principle estimate the Li abundance using curves of growth, if the stellar effective temperature is known (e.g.,  see \citealt{zapatero}). Here, we assume an effective temperature of 4000\,K, and adopt the data from \cite{duncan} to estimate a Li abundance of 2.6$\pm$0.1, and 2.7$\pm$0.2 (on the usual scale of 12+log[$N$(Li)/$N($H)]) from the $\lambda$6103 and $\lambda$6708\AA\ lines respectively. The errors originate from the errors on the equivalent width. 


\subsection{Observing faint targets}

As described in Section\,4.6, in good conditions GHOST can be used to observe such targets. Here we present an example of using GHOST to explore more distant extra-galactic objects.

\subsubsection{Observations of quasar J1215-0034}
 
\begin{figure}
\includegraphics[width=1\columnwidth]{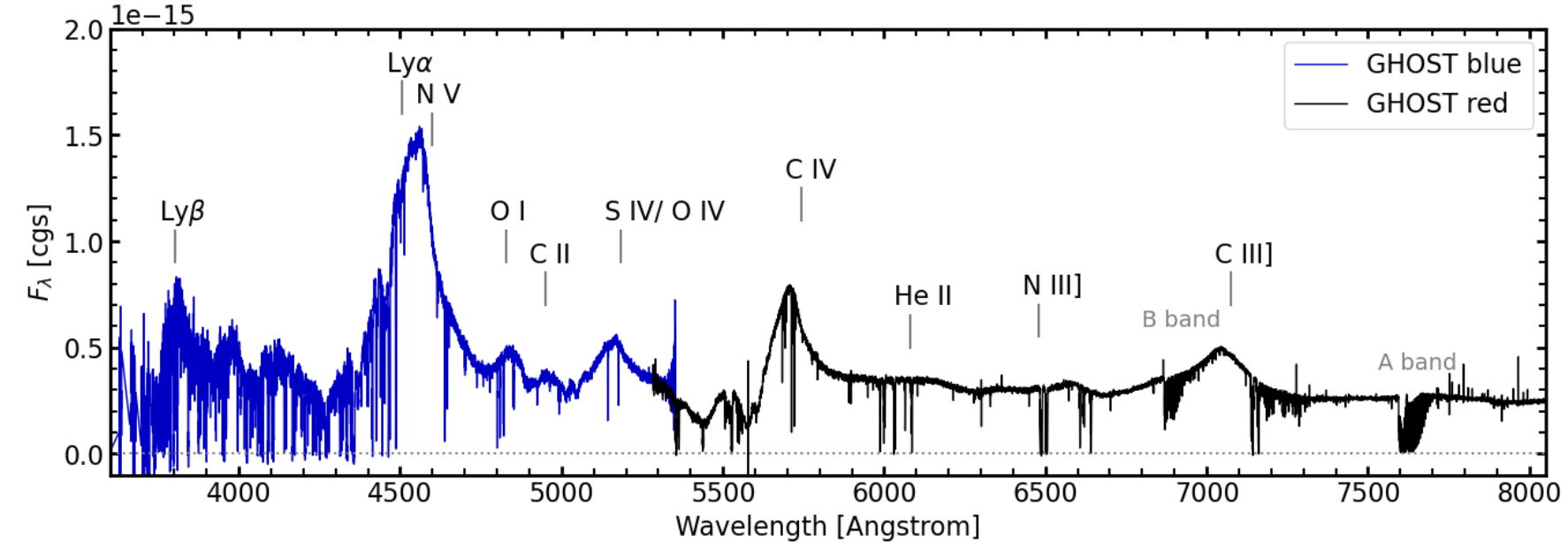}
\caption{Flux-calibrated blue spectrum of QSO\,J1215-0034 taken using GHOST plotted over a selected wavelength range, including the Ly\,$\alpha$ line at 450.4\,nm. \label{quasar}}
\end{figure}

%

To demonstrate the use of GHOST for such science, the quasi-stellar object (QSO) SDSS\,J121549.80$-$003432.1 (hereafter J1215$-$0034) was observed in GHOST's standard resolution mode. With 3$\times$1200\,s exposures in each arm, we achieved a median SNR at the peak of each detector of $\sim$30. Observations of the spectrophotometric standard, EG\,21 were obtained for flux calibration. QSO\,J1215$-$0034 is a faint ($V$=17.5\,mag) object at a redshift of 2.7, that has been observed as part of both the Sloan Digital Sky Survey (SDSS; \citealt{sdss}), and the Dark Energy Spectroscopic Instrument (DESI) survey \citep{desi} around $R\sim$3000, providing a sanity check of GHOST high-resolution spectra obtained against absolute flux calibrated low-resolution spectra, while adding enhanced information of the QSO thanks to the increased spectral resolution.
 
\begin{figure}
\includegraphics[width=1\columnwidth]{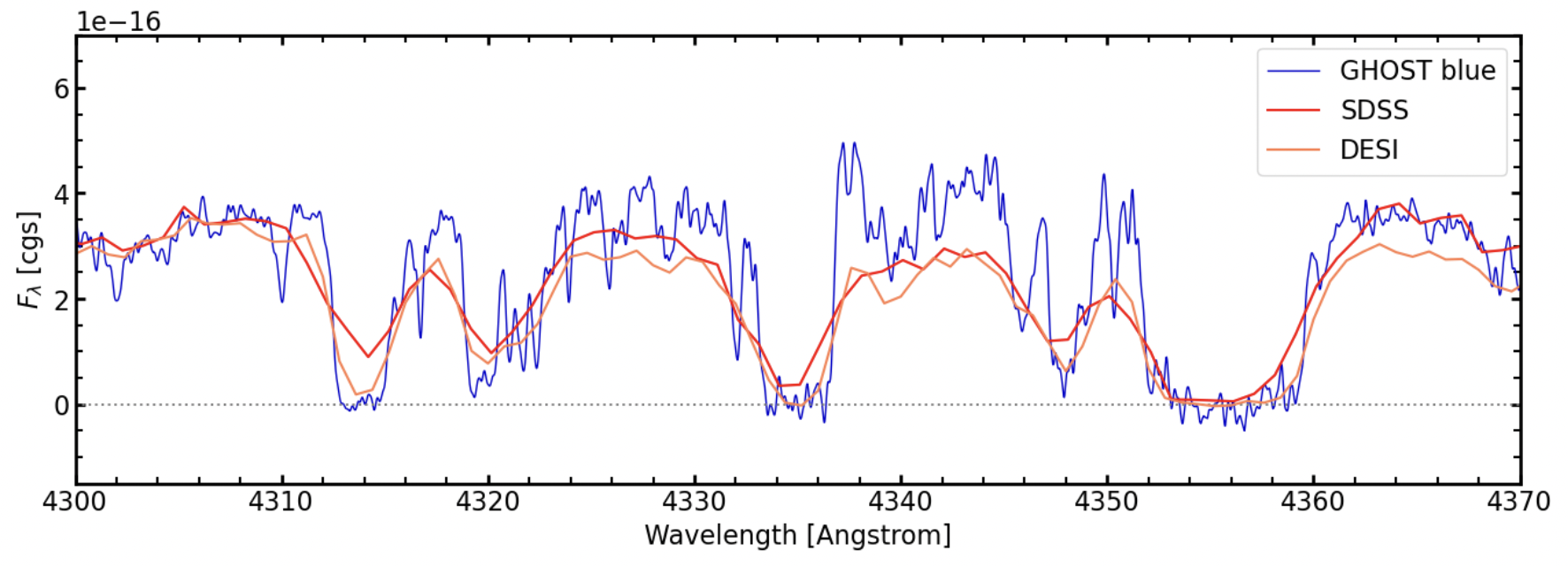}
\caption{Ly$\alpha$ forest of J1215$-$0034 is shown, with the coarser resolution spectra from SDSS, and DESI overplotted. The displayed spectra are in the observer's frame of reference.   \label{lineprof}}
\end{figure}

In Fig.\,\ref{quasar}, we display a section of the GHOST spectrum, overplotted with the lower-resolution DESI early relase data \citep{desiedr}. Immediately visible is the general agreement of the flux calibration, but also the deeper, narrower lines and the noisier spectrum with GHOST. To compare the former, we convolved the GHOST spectrum with the appropriate SDSS filter response curves, and found that the GHOST fluxes were within 0.5\% of SDSS magnitudes from \cite{sdss} in $gri$. For $uz$, an extrapolation for wavelengths not covered by GHOST was used, and the differences were 2\% in absolute magnitudes. This suggests that despite GHOST having a fixed aperture of around 1$\arcsec$, absolute flux calibration even for faint objects is still possible in suitable conditions, and the instrument response is stable. 

J1215$-$0034 is a previously known narrow absorption lines quasar (NAL), having line widths $\leq$500\,$km\,s^{-1}$ making it well suited for high-resolution observations \citep{chenqso, napolitano}. The benefits of higher resolution in resolving key ionic species and Ly$\alpha$ forest are revealed in Fig.\,\ref{lineprof}, allowing for the characterization of narrow quasar absorption profiles. \cite{itohqso} utilize the ground and excited metal line ratios to estimate gas density, and radial distance from the central source using SDSS spectra. We showcase how high-resolution spectra in Fig.\,\ref{lineprof}, such as from GHOST can be useful by comparing these line ratios to low-resolution DESI spectra. Both the depth and width of these key lines are better characterized for this NAL system using GHOST. Similarly, GHOST observed a bright ($G$=17\,mag) quasar during commissioning, where the absorption lines associated with a metal-poor damped Lyman $\alpha$ system were characterized chemically and kinematically \citep{berg}


\subsection{Time-series observations}
\label{sec:ccfstability_ghostpaper}

A principal use envisaged for GHOST is for high-precision measurements of celestial radial velocities, at levels between tens to hunderds of m\,s$^{-1}$ over short and long timescales. This is motivated by the need of exoplanetary and stellar binary science cases. Therefore a key early goal was to measure the stability and precision of on-sky radial velocity measurements. 

\subsubsection{GHOST's short-term line profile and RV stability based on dual target observations}
The dual object mode of GHOST offers additional valuable information for such measurements, by allowing for simultaneous observations fed to the same detector, of a target star and a calibrator. This method was employed for observations of the transit of the hot Jupiter WASP-108~b around the F9 host, WASP-108 \citep{Anderson2014}. For calibration, the reference star TYC8254-1616-1 (A9V, $V=$11.56\,mag) was observed simultaneously during the transit in the dual object standard resolution mode. It is not expected to change its mean line profile or its velocity during the observations and therefore is better suited to assess the instrument's stability during the transit of WASP-108~b. Spectra were obtained with 600s and 300s of integration time for the blue and red cameras respectively, binned by a factor of 2 in the spectral direction over a five-hour time series covering the known transit of WASP-108~b. To assess the radial velocity stability, and precision we employ the Cross-Correlation Function (CCF) technique as described by \cite{pepe2002}. The CCF is a powerful technique for extracting mean line profiles, which enables measurements of important stellar quantities like rotation velocity and activity indicators, and it is particularly valuable for precise Doppler shift measurements due to radial velocity variations, often employed in detecting and characterizing exoplanets \citep[e.g.,][]{Mayor1995}. The CCFs are calculated using the reduction procedures described in \cite{Martioli2022}, where we used a CCF mask obtained from lines detected in a synthetic spectrum generated by the stellar atmosphere code PHOENIX \citep{Hauschildt1999} using the parameters that match those of the observed stars. 

Here, we only present the results for TYC8254-1616-1. The results for WASP-108 are presented in a dedicated paper by Martioli et al. (in prep.), where the CCF is expected to change due to the Rossiter-McLaughlin effect \citep{Triaud2018} caused by the transit of its WASP-108~b. In Fig.\,\ref{fig:ghostccf}, we present the CCF measurements for TYC8254-1616-1, derived from GHOST spectra, highlighting significant instrument stability. To quantitatively assess CCF stability, we calculate the rms dispersion of residual CCFs. The results indicate values of 0.014\% and 0.021\% for the blue and red channels, respectively. This stability in the CCF profiles is reflected in consistent measurements of RVs, full width at half maximum (FWHM), and bisector span (BIS), as shown in Fig.\,\ref{fig:ghostrvstability}. The final $rms$ values are 13 m\,s$^{-1}$ for the blue channel, and 12 m\,s$^{-1}$ for the red channel, and are in good agreement. Notably, there appears to be a systematic correlation between the RVs and airmass, which does not correlate with any of the CCF morphological parameters (FWHM and BIS). This suggests that these shifts may be attributed to instrumental drifts resulting from temperature changes or variations in guiding performance, particularly as airmasses change during the observations. Moreover, the stability of morphological parameters at the level of a few m\,s$^{-1}$ provides strong evidence that the instrument profile remains stable throughout several hours of observation. 

  \begin{figure}
   \centering
       \includegraphics[width=1\columnwidth]{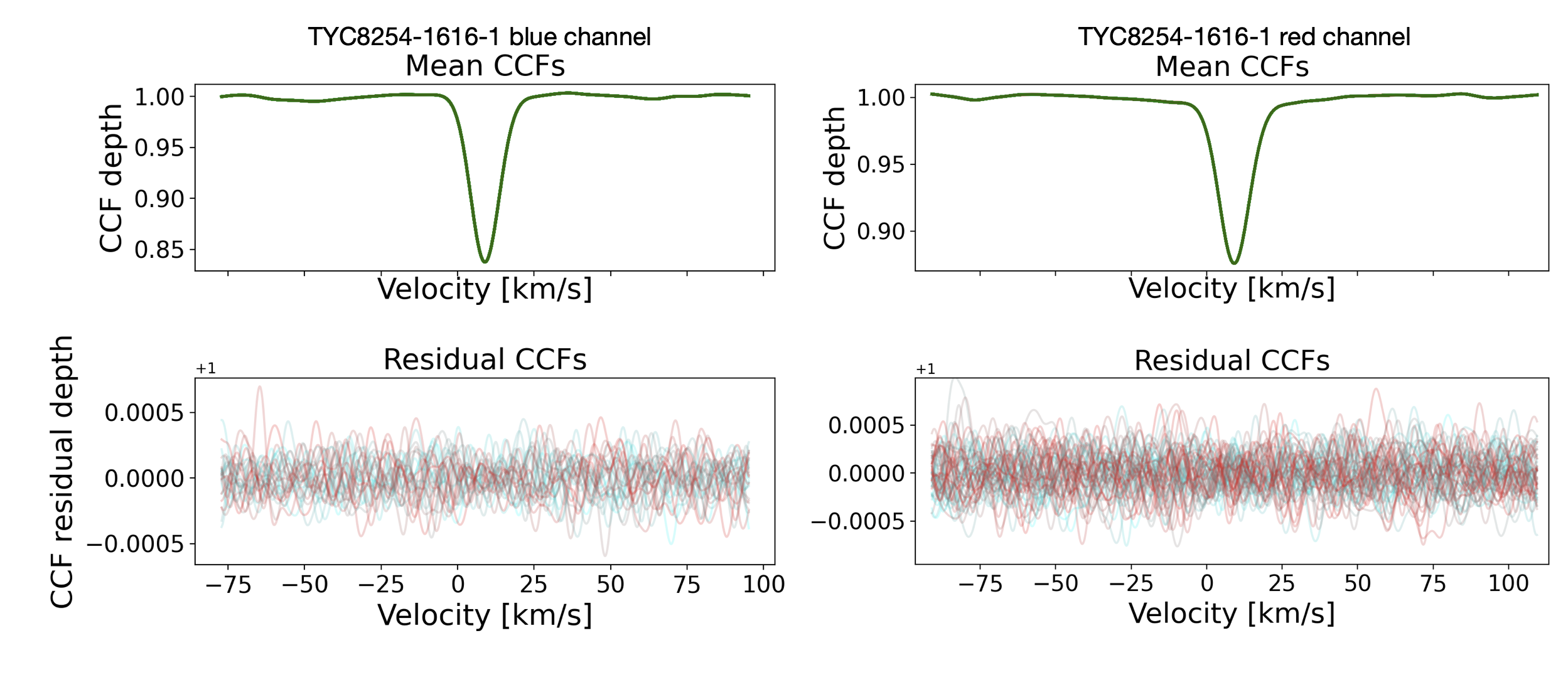}
      \caption{Cross-correlation function (CCF) of TYC8254-1616-1 derived from GHOST spectra in both the blue (left panels) and red (right panels) channels. In the left panels, the top shows the mean CCFs represented by green lines, while the bottom displays the CCF data with the mean CCF subtracted. The color code indicates the first epoch in dark blue and the last epoch in dark red.
      }
        \label{fig:ghostccf}
  \end{figure}

  \begin{figure}
   \centering
       \includegraphics[width=1.1\columnwidth]{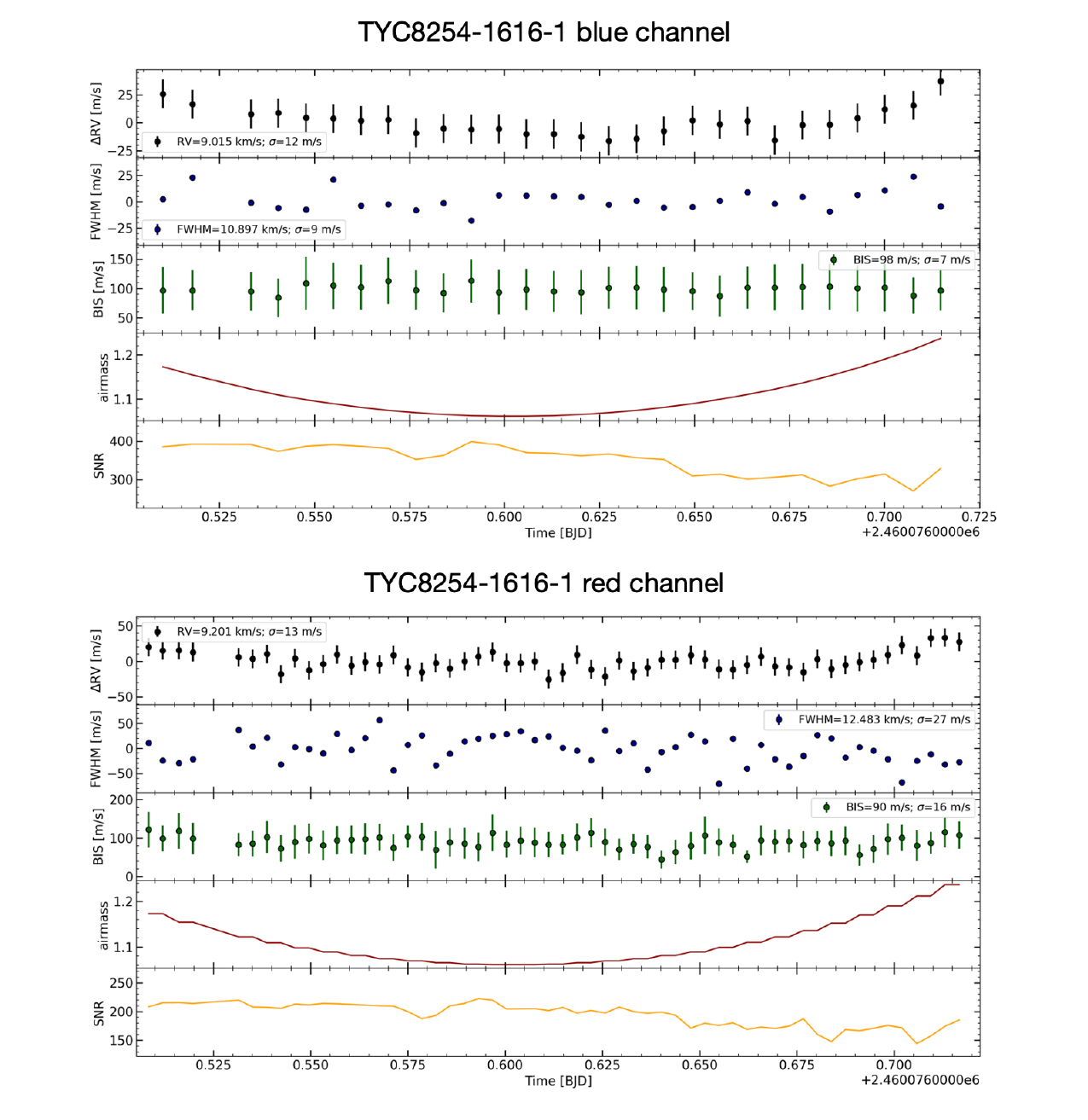}
      \caption{
      Quantities derived from the CCFs of TYC8254-1616-1's GHOST spectra. These measurements were obtained from both the blue (top panels) and red (bottom panels) channels. For each channel, the panels display the time series of RVs, FWHMs, bisector span, airmass, and maximum SNR from top to bottom.
      }
        \label{fig:ghostrvstability}
  \end{figure}

\subsection{Examples of poor weather observations}

GHOST's ability to capture high SNR over the entire optical wavelength in a single exposure, even in poor conditions for relatively bright objects makes the instrument extremely useful for spectroscopic monitoring observations spread out over a period of time thanks to Gemini's queue flexibility. 
As a demonstration of such a class of observations, we discuss in the following section single epoch standard resolution spectra of IK Hya taken in moderate observing conditions. Here, the average seeing was $\sim$1$''$ (corrected for airmass, falling into the IQAny percentile bin{\footnote {http://www.gemini.edu/observing/telescopes-and-sites/sites}}) in moderate cirrus, during dark time. 

\subsubsection{The RR Lyrae star IK Hya}

IK Hya ($V=$10.2\,mag), belongs to the well-known class of pulsating variables, RR Lyrae. RR Lyrae are Population II solar-type stars found in the horizontal branch, with periods typically ranging between a few hours to one day, magnitudes varying in the visible by as much as 2\,mag \citep[e.g.][]{catelan15}, and period-luminosity relations similar to Cepheid variables \citep[e.g.][]{catelan04}. They have been used to measure accurate distances to external galaxies, globular clusters and the Galactic center \citep[see e.g.][for a review]{beaton18}. A subset of RR Lyrae are considered to exhibit the Blazhko effect \citep{blazhko07}, a secular variation in their amplitude. IK Hya is thought to be similar to the prototype of its class, RR Lyrae, as it exhibits a $\sim$4 year long-term cycle, where the Blazhko and pulsation characteristics change like RR Lyrae \citep{skarka14}. From the spectrum (in Fig.\,\ref{ikhya}), the Balmer lines H$\epsilon$ and H$\alpha$ are double-lined, but not any of the other Balmer lines, suggesting the spectrum was taken at the maximum of the Blazhko phase. The origin of these lines are thought to be due to a shock wave in the atmosphere. It is curious to note the absence of doubling in H$\beta$, compared to the other Balmer lines; doubling of H$\beta$ was expected \citep{chadid13}. 

With high-resolution optical spectroscopy of instruments such as GHOST, one can capture the entire optical spectrum (and thereby all the visible Balmer lines) simultaneously. Such monitoring observations, for bright targets in poor conditions can be a significant use of the new instrument. With these spectra, we demonstrate the signal achieved in moderate to poor observing conditions with short-duration exposures, necessary to avoid phase smearing in stars with rapidly changing atmospheres and identify the doubling of some Balmer lines in IK Hya as a scientific outcome.

\begin{figure}
\includegraphics[width=1\columnwidth]{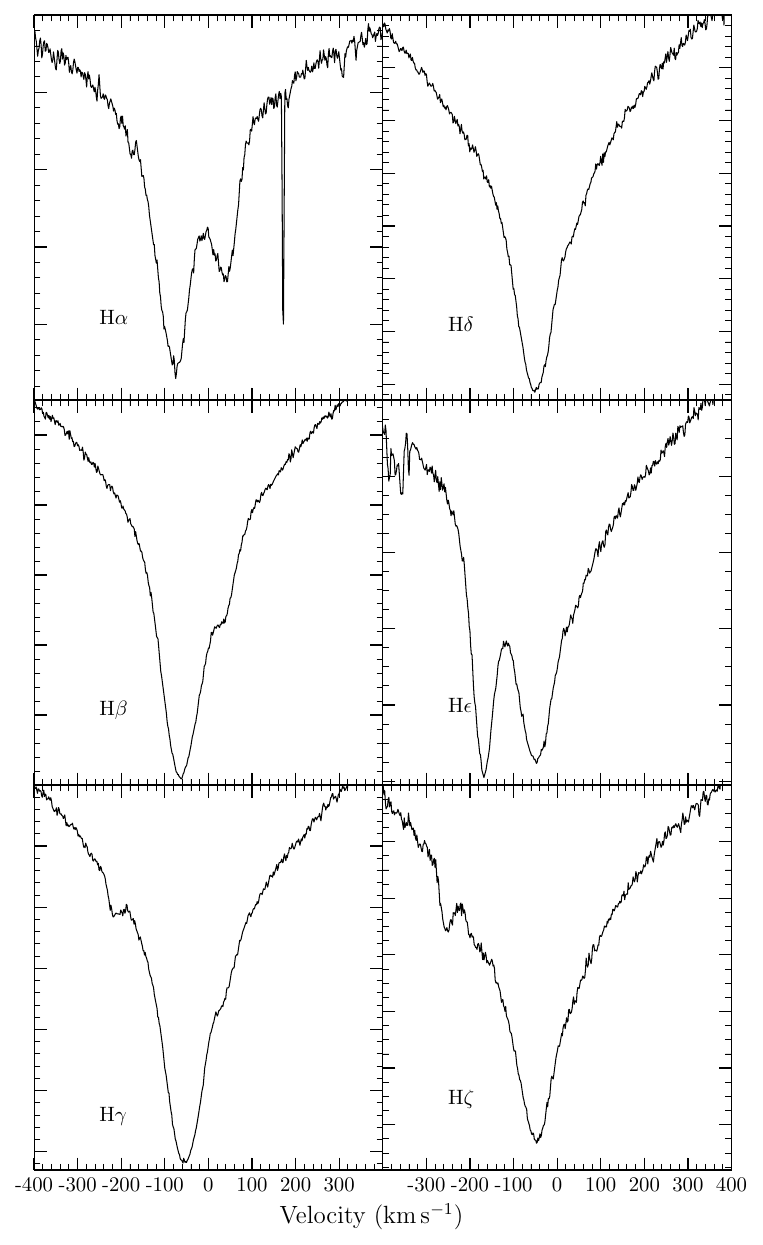}
\caption{Doubling of H$\alpha$ and H$\epsilon$ Balmer lines clearly visible in the GHOST spectra of IK Hya, with the detail on other Balmer lines also shown. The entire spectrum was captured simultaneously. The spectral resolution and signal achieved in these weather conditions demonstrate the usefulness of GHOST at Gemini as a monitoring instrument for relatively bright targets. \label{ikhya}}
\end{figure}

\section{Next Steps} \label{future }

\subsection{Precision radial velocity mode}

GHOST is equipped with an internal calibration source (a Thorium Xenon lamp) that can be operated only in high-resolution mode, along with a fiber agitator to reduce modal noise in extremely high SNR regimes. A brief overview of the mode is found in \cite{ireland16}. These form the hardware component of the precision radial velocity (PRV) mode and are currently installed, and have been tested on the telescope. The internal lamp is read directly into the pseudo-slit, forming an output next to the microlens number 61 (separated by an empty microlens from the science and sky microlenses). The spectrum is eventually read out onto the detector along each echelle order. The intention of the internal calibration source is to allow tracking of any wavelength drifts (see also Section\,4.5 for a description of the same), and allow simultaneous wavelength calibration. The lamp is equipped with a variety of neutral density filters in a dedicated filter wheel, to allow for enough counts and avoid saturation for a wide variety of science exposure times. This allows the user to configure the instrument as necessary for science in the high-resolution mode, and achieve a simultaneous arc calibration frame. Although GHOST is not located in a pressure-controlled environment, monitoring of pressure changes indicates that radial velocity precision at the sub 10\,m\,s$^{-1}$ scale is possible. 

However, there are currently a few challenges to achieve such precision and long-term stability. The instrument profile (IP) is largely shaped by the amount of focal ratio degradation (FRD) which may vary due to stresses on the fiber train, or small angular mis-alignments. 

In addition to variations in the IP, small offsets between fibers in the slit lead to net systematic wavelength offsets (of less than few 10\,m\,s$^{-1}$). In principle the weighting function derived from the spatially resolved fiber image compensates, but the velocity offsets between fibers are not accurately known at each point. Finally, the ThXe lines from the simultaneous calibration source saturate in the red arm in long exposure times even when using a neutral density filter. These lines if bled into neighboring orders can affect the overall calibration. 

Gemini is planning to undergo commissioning of the PRV mode beginning in late 2024 \citep{kalari24} with the aim of estimating the precision of radial velocities achievable with the current hardware and adapted data reduction techniques.  
The main items undergoing development and testing will be (i) the data reduction pipeline to work with the PRV observations (ii) conducting nighttime observations to determine the range of exposure times, object magnitudes, and SNR possible in the PRV mode, and (iii) estimate the radial velocity precision of the PRV mode over a sufficient time baseline. While some observations using the PRV mode have already been conducted, currently Gemini is in the process of completing many of the remaining items and estimating the radial velocity precision achieved in the PRV mode. Depending on the initial assessment of precision achieved, the mode is planned to be either released to the community, or assesed if further hardware to improve simultaneous calibration (for e.g. an etalon) may be required.

\subsection{Spectropolarimetry}

  \begin{figure}
   \centering
       \includegraphics[width=1\columnwidth]{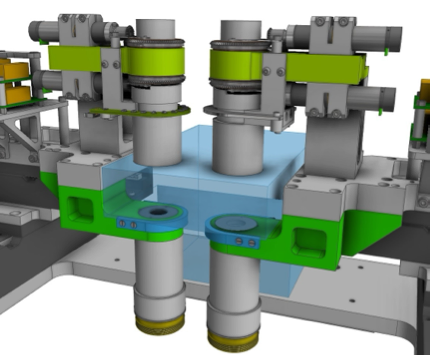}
      \caption{ Computer generated image displaying the location of possible future optics (blue envelope). The space is restricted by the IFU and ADC assembly, but has been measured to be sufficient for wave plates to capture circular polarization. }
        \label{spaceenv}
  \end{figure}

In addition to a future upgrade bringing GHOST into the extreme radial velocity precision regime, GHOST may be equipped with the capability to perform spectropolarimetry. The on-sky Cassegrain unit of GHOST contains a space envelope with enough size, located in between the ADC's and the IFU (see Fig.\,\ref{spaceenv}), allowing for measurement of circularly polarized light (i.e. the Stokes $V$ parameter) before the light reaches the pseudo-slit entrance train, and used on the bottom port of Gemini South's Cassegrain focus. Although one can in principle obtain all the Stokes parameters accounting for the intrinsic polarization of the telescope optics, doing so requires optical components larger than the remaining space envelope. Instead, the space envelope inside GHOST's Cassegrain unit when mounted on the up-looking port of the Cassegrain focus is only sufficient to install wave plates necessary for measurement of circular polarization. This development would be undertaken depending on community demand.


The addition of a spectropolarimeter to an echelle instrument on an 8-meter class of telescope will add significant depth to what is photon-hungry science, and fill a void in current astronomical instrumentation, especially in the Southern Hemisphere. Such capabilities would allow for high-precision measurements of magnetic fields lines, relying now on Zeeman splitting. This would be extremely useful in the measurement of magnetic fields with high precision for objects fainter than can be reached by the current generation of spectropolarimeters \citep{donati}.

\section{Conclusions}

We presented the motivation, inception, design and build phases for a high-resolution optical spectrograph available to the Gemini astronomical community in the Southern hemisphere. GHOST is a highly efficient fiber-fed spectrograph residing in the telescope pier lab at Gemini South. It allows users to capture either a single target (and additional sky) spectra of $R\sim$76\,000 in high-resolution mode; or two targets (and an additional sky spectrum) of $R\sim$56\,000 in standard mode. The limiting magnitude to achieve SNR of 10 is $V\simeq$19 and 20\,mag in the two different modes in 1 hour, respectively. 

In this paper we report and discuss the on-sky performance and some of the first science results of GHOST following successful commissioning and systems integration into the Gemini South telescope. The results demonstrate that the instrument meets the efficiency, resolution, and radial velocity precision requirements; and meets exacting stability requirements. We also showcase the exquisite spectra produced and unique science highlights. 

Following the successful integration of GHOST into Gemini regular science operations, we discuss the next steps planned including the commissioning of the precision radial velocity mode, and the future upgrades possible for the instrument. Finally, given the unique nature of GHOST, and Gemini's multi-messenger objectives, we highlighted the suitability of GHOST to follow-up a variety of transient objects. With the advent of GHOST, the community's need for a high-resolution efficient spectrograph in the southern sky is met and also opens a new avenue of exploration using high-resolution spectroscopy. GHOST has been offered to the community in late 2023, and is part of the Gemini facility instrument suite from now on wards.

\begin{acknowledgments}

This work was supported by and based on observations obtained at the international Gemini Observatory, a program of NSF’s NOIRLab, which is managed by the Association of Universities for Research in Astronomy (AURA) under a cooperative agreement with the National Science Foundation on behalf of the Gemini Observatory partnership: the National Science Foundation (United States), National Research Council (Canada), Agencia Nacional de Investigaci\'{o}n y Desarrollo (Chile), Ministerio de Ciencia, Tecnolog\'{i}a e Innovaci\'{o}n (Argentina), Minist\'{e}rio da Ci\^{e}ncia, Tecnologia, Inova\c{c}\~{o}es e Comunica\c{c}\~{o}es (Brazil), and Korea Astronomy and Space Science Institute (Republic of Korea). GHOST was built by a collaboration between Australian Astronomical Optics at Macquarie University, National Research Council Herzberg of Canada, the Australian National University and Gemini Observatory. The authors would like to acknowledge the contributions of the GHOST instrument build team, the Gemini GHOST instrument team, the full systems verification team, and the Gemini operations team. We thank the Gemini directorate for their support. V.M.K. thanks J.S.Vink for useful comments. E.M. acknowledges funding from FAPEMIG under project number APQ-02493-22 and a research productivity grant number 309829/2022-4 awarded by the CNPq, Brazil

\end{acknowledgments}

%

\vspace{5mm}
\facilities{Gemini}





\bibliography{sample631}{}

\begin{thebibliography}{}
\expandafter\ifx\csname natexlab\endcsname\relax\def\natexlab#1{#1}\fi
\providecommand{\url}[1]{\href{#1}{#1}}
\providecommand{\dodoi}[1]{doi:~\href{http://doi.org/#1}{\nolinkurl{#1}}}
\providecommand{\doeprint}[1]{\href{http://ascl.net/#1}{\nolinkurl{http://ascl.net/#1}}}
\providecommand{\doarXiv}[1]{\href{https://arxiv.org/abs/#1}{\nolinkurl{https://arxiv.org/abs/#1}}}

\bibitem[{{Abbott} {et~al.}(2017){Abbott}, {Abbott}, {Abbott}, {Acernese},
  {Ackley}, {Adams}, {Adams}, {Addesso}, {Adhikari}, {Adya}, {Affeldt},
  {Afrough}, {Agarwal}, {Agathos}, {Agatsuma}, {Aggarwal}, {Aguiar}, {Aiello},
  {Ain}, {Ajith}, {Allen}, {Allen}, {Allocca}, {Altin}, {Amato}, {Ananyeva},
  {Anderson}, {Anderson}, {Angelova}, {Antier}, {Appert}, {Arai}, {Araya},
  {Areeda}, {Arnaud}, {Arun}, {Ascenzi}, {Ashton}, {Ast}, {Aston}, {Astone},
  {Atallah}, {Aufmuth}, {Aulbert}, {AultONeal}, {Austin}, {Avila-Alvarez},
  {Babak}, {Bacon}, {Bader}, {Bae}, {Baker}, {Baldaccini}, {Ballardin},
  {Ballmer}, {Banagiri}, {Barayoga}, {Barclay}, {Barish}, {Barker}, {Barkett},
  {Barone}, {Barr}, {Barsotti}, {Barsuglia}, {Barta}, {Barthelmy}, {Bartlett},
  {Bartos}, {Bassiri}, {Basti}, {Batch}, {Bawaj}, {Bayley}, {Bazzan},
  {B{\'e}csy}, {Beer}, {Bejger}, {Belahcene}, {Bell}, {Berger}, {Bergmann},
  {Bero}, {Berry}, {Bersanetti}, {Bertolini}, {Betzwieser}, {Bhagwat},
  {Bhandare}, {Bilenko}, {Billingsley}, {Billman}, {Birch}, {Birney},
  {Birnholtz}, {Biscans}, {Biscoveanu}, {Bisht}, {Bitossi}, {Biwer},
  {Bizouard}, {Blackburn}, {Blackman}, {Blair}, {Blair}, {Blair}, {Bloemen},
  {Bock}, {Bode}, {Boer}, {Bogaert}, {Bohe}, {Bondu}, {Bonilla}, {Bonnand},
  {Boom}, {Bork}, {Boschi}, {Bose}, {Bossie}, {Bouffanais}, {Bozzi},
  {Bradaschia}, {Brady}, {Branchesi}, {Brau}, {Briant}, {Brillet}, {Brinkmann},
  {Brisson}, {Brockill}, {Broida}, {Brooks}, {Brown}, {Brown}, {Brunett},
  {Buchanan}, {Buikema}, {Bulik}, {Bulten}, {Buonanno}, {Buskulic}, {Buy},
  {Byer}, {Cabero}, {Cadonati}, {Cagnoli}, {Cahillane}, {Calder{\'o}n
  Bustillo}, {Callister}, {Calloni}, {Camp}, {Canepa}, {Canizares}, {Cannon},
  {Cao}, {Cao}, {Capano}, {Capocasa}, {Carbognani}, {Caride}, {Carney},
  {Casanueva Diaz}, {Casentini}, {Caudill}, {Cavagli{\`a}}, {Cavalier},
  {Cavalieri}, {Cella}, {Cepeda}, {Cerd{\'a}-Dur{\'a}n}, {Cerretani},
  {Cesarini}, {Chamberlin}, {Chan}, {Chao}, {Charlton}, {Chase},
  {Chassande-Mottin}, {Chatterjee}, {Chatziioannou}, {Cheeseboro}, {Chen},
  {Chen}, {Chen}, {Cheng}, {Chia}, {Chincarini}, {Chiummo}, {Chmiel}, {Cho},
  {Cho}, {Chow}, {Christensen}, {Chu}, {Chua}, {Chua}, {Chung}, {Chung},
  {Ciani}, {Ciolfi}, {Cirelli}, {Cirone}, {Clara}, {Clark}, {Clearwater},
  {Cleva}, {Cocchieri}, {Coccia}, {Cohadon}, {Cohen}, {Colla}, {Collette},
  {Cominsky}, {Constancio}, {Conti}, {Cooper}, {Corban}, {Corbitt},
  {Cordero-Carri{\'o}n}, {Corley}, {Cornish}, {Corsi}, {Cortese}, {Costa},
  {Coughlin}, {Coughlin}, {Coulon}, {Countryman}, {Couvares}, {Covas}, {Cowan},
  {Coward}, {Cowart}, {Coyne}, {Coyne}, {Creighton}, {Creighton}, {Cripe},
  {Crowder}, {Cullen}, {Cumming}, {Cunningham}, {Cuoco}, {Dal Canton},
  {D{\'a}lya}, {Danilishin}, {D'Antonio}, {Danzmann}, {Dasgupta}, {Da Silva
  Costa}, {Dattilo}, {Dave}, {Davier}, {Davis}, {Daw}, {Day}, {De}, {DeBra},
  {Degallaix}, {De Laurentis}, {Del{\'e}glise}, {Del Pozzo}, {Demos}, {Denker},
  {Dent}, {De Pietri}, {Dergachev}, {De Rosa}, {DeRosa}, {De Rossi}, {DeSalvo},
  {de Varona}, {Devenson}, {Dhurandhar}, {D{\'\i}az}, {Di Fiore}, {Di
  Giovanni}, {Di Girolamo}, {Di Lieto}, {Di Pace}, {Di Palma}, {Di Renzo},
  {Doctor}, {Dolique}, {Donovan}, {Dooley}, {Doravari}, {Dorrington},
  {Douglas}, {Dovale {\'A}lvarez}, {Downes}, {Drago}, {Dreissigacker},
  {Driggers}, {Du}, {Ducrot}, {Dupej}, {Dwyer}, {Edo}, {Edwards}, {Effler},
  {Ehrens}, {Eichholz}, {Eikenberry}, {Eisenstein}, {Essick}, {Estevez},
  {Etienne}, {Etzel}, {Evans}, {Evans}, {Factourovich}, {Fafone}, {Fair},
  {Fairhurst}, {Fan}, {Farinon}, {Farr}, {Farr}, {Fauchon-Jones}, {Favata},
  {Fays}, {Fee}, {Fehrmann}, {Feicht}, {Fejer}, {Fernandez-Galiana},
  {Ferrante}, {Ferreira}, {Ferrini}, {Fidecaro}, {Finstad}, {Fiori},
  {Fiorucci}, {Fishbach}, {Fisher}, {Fitz-Axen}, {Flaminio}, {Fletcher},
  {Fong}, {Font}, {Forsyth}, {Forsyth}, {Fournier}, {Frasca}, {Frasconi},
  {Frei}, {Freise}, {Frey}, {Frey}, {Fries}, {Fritschel}, {Frolov}, {Fulda},
  {Fyffe}, {Gabbard}, {Gadre}, {Gaebel}, {Gair}, {Gammaitoni}, {Ganija},
  {Gaonkar}, {Garcia-Quiros}, {Garufi}, {Gateley}, {Gaudio}, {Gaur},
  {Gayathri}, {Gehrels}, {Gemme}, {Genin}, {Gennai}, {George}, {George},
  {Gergely}, {Germain}, {Ghonge}, {Ghosh}, {Ghosh}, {Ghosh}, {Giaime},
  {Giardina}, {Giazotto}, {Gill}, {Glover}, {Goetz}, {Goetz}, {Gomes},
  {Goncharov}, {Gonz{\'a}lez}, {Gonzalez Castro}, {Gopakumar}, {Gorodetsky},
  {Gossan}, {Gosselin}, {Gouaty}, {Grado}, {Graef}, {Granata}, {Grant}, {Gras},
  {Gray}, {Greco}, {Green}, {Gretarsson}, {Griswold}, {Groot}, {Grote},
  {Grunewald}, {Gruning}, {Guidi}, {Guo}, {Gupta}, {Gupta}, {Gushwa},
  {Gustafson}, {Gustafson}, {Halim}, {Hall}, {Hall}, {Hamilton}, {Hammond},
  {Haney}, {Hanke}, {Hanks}, {Hanna}, {Hannam}, {Hannuksela}, {Hanson},
  {Hardwick}, {Harms}, {Harry}, {Harry}, {Hart}, {Haster}, {Haughian}, {Healy},
  {Heidmann}, {Heintze}, {Heitmann}, {Hello}, {Hemming}, {Hendry}, {Heng},
  {Hennig}, {Heptonstall}, {Heurs}, {Hild}, {Hinderer}, {Hoak}, {Hofman},
  {Holt}, {Holz}, {Hopkins}, {Horst}, {Hough}, {Houston}, {Howell}, {Hreibi},
  {Hu}, {Huerta}, {Huet}, {Hughey}, {Husa}, {Huttner}, {Huynh-Dinh}, {Indik},
  {Inta}, {Intini}, {Isa}, {Isac}, {Isi}, {Iyer}, {Izumi}, {Jacqmin}, {Jani},
  {Jaranowski}, {Jawahar}, {Jim{\'e}nez-Forteza}, {Johnson}, {Jones}, {Jones},
  {Jonker}, {Ju}, {Junker}, {Kalaghatgi}, {Kalogera}, {Kamai}, {Kandhasamy},
  {Kang}, {Kanner}, {Kapadia}, {Karki}, {Karvinen}, {Kasprzack}, {Katolik},
  {Katsavounidis}, {Katzman}, {Kaufer}, {Kawabe}, {K{\'e}f{\'e}lian}, {Keitel},
  {Kemball}, {Kennedy}, {Kent}, {Key}, {Khalili}, {Khan}, {Khan}, {Khan},
  {Khazanov}, {Kijbunchoo}, {Kim}, {Kim}, {Kim}, {Kim}, {Kim}, {Kim},
  {Kimbrell}, {King}, {King}, {Kinley-Hanlon}, {Kirchhoff}, {Kissel},
  {Kleybolte}, {Klimenko}, {Knowles}, {Koch}, {Koehlenbeck}, {Koley},
  {Kondrashov}, {Kontos}, {Korobko}, {Korth}, {Kowalska}, {Kozak},
  {Kr{\"a}mer}, {Kringel}, {Krishnan}, {Kr{\'o}lak}, {Kuehn}, {Kumar}, {Kumar},
  {Kumar}, {Kuo}, {Kutynia}, {Kwang}, {Lackey}, {Lai}, {Landry}, {Lang},
  {Lange}, {Lantz}, {Lanza}, {Larson}, {Lartaux-Vollard}, {Lasky}, {Laxen},
  {Lazzarini}, {Lazzaro}, {Leaci}, {Leavey}, {Lee}, {Lee}, {Lee}, {Lee}, {Lee},
  {Lehmann}, {Lenon}, {Leonardi}, {Leroy}, {Letendre}, {Levin}, {Li}, {Linker},
  {Littenberg}, {Liu}, {Lo}, {Lockerbie}, {London}, {Lord}, {Lorenzini},
  {Loriette}, {Lormand}, {Losurdo}, {Lough}, {Lousto}, {Lovelace}, {L{\"u}ck},
  {Lumaca}, {Lundgren}, {Lynch}, {Ma}, {Macas}, {Macfoy}, {Machenschalk},
  {MacInnis}, {Macleod}, {Maga{\~n}a Hernandez}, {Maga{\~n}a-Sandoval},
  {Maga{\~n}a Zertuche}, {Magee}, {Majorana}, {Maksimovic}, {Man}, {Mandic},
  {Mangano}, {Mansell}, {Manske}, {Mantovani}, {Marchesoni}, {Marion},
  {M{\'a}rka}, {M{\'a}rka}, {Markakis}, {Markosyan}, {Markowitz}, {Maros},
  {Marquina}, {Marsh}, {Martelli}, {Martellini}, {Martin}, {Martin},
  {Martynov}, {Mason}, {Massera}, {Masserot}, {Massinger}, {Masso-Reid},
  {Mastrogiovanni}, {Matas}, {Matichard}, {Matone}, {Mavalvala}, {Mazumder},
  {McCarthy}, {McClelland}, {McCormick}, {McCuller}, {McGuire}, {McIntyre},
  {McIver}, {McManus}, {McNeill}, {McRae}, {McWilliams}, {Meacher}, {Meadors},
  {Mehmet}, {Meidam}, {Mejuto-Villa}, {Melatos}, {Mendell}, {Mercer}, {Merilh},
  {Merzougui}, {Meshkov}, {Messenger}, {Messick}, {Metzdorff}, {Meyers},
  {Miao}, {Michel}, {Middleton}, {Mikhailov}, {Milano}, {Miller}, {Miller},
  {Miller}, {Millhouse}, {Milovich-Goff}, {Minazzoli}, {Minenkov}, {Ming},
  {Mishra}, {Mitra}, {Mitrofanov}, {Mitselmakher}, {Mittleman}, {Moffa},
  {Moggi}, {Mogushi}, {Mohan}, {Mohapatra}, {Montani}, {Moore}, {Moraru},
  {Moreno}, {Morriss}, {Mours}, {Mow-Lowry}, {Mueller}, {Muir}, {Mukherjee},
  {Mukherjee}, {Mukherjee}, {Mukund}, {Mullavey}, {Munch}, {Mu{\~n}iz},
  {Muratore}, {Murray}, {Napier}, {Nardecchia}, {Naticchioni}, {Nayak},
  {Neilson}, {Nelemans}, {Nelson}, {Nery}, {Neunzert}, {Nevin}, {Newport},
  {Newton}, {Ng}, {Nguyen}, {Nguyen}, {Nichols}, {Nielsen}, {Nissanke}, {Nitz},
  {Noack}, {Nocera}, {Nolting}, {North}, {Nuttall}, {Oberling}, {O'Dea},
  {Ogin}, {Oh}, {Oh}, {Ohme}, {Okada}, {Oliver}, {Oppermann}, {Oram},
  {O'Reilly}, {Ormiston}, {Ortega}, {O'Shaughnessy}, {Ossokine}, {Ottaway},
  {Overmier}, {Owen}, {Pace}, {Page}, {Page}, {Pai}, {Pai}, {Palamos},
  {Palashov}, {Palomba}, {Pal-Singh}, {Pan}, {Pan}, {Pang}, {Pang}, {Pankow},
  {Pannarale}, {Pant}, {Paoletti}, {Paoli}, {Papa}, {Parida}, {Parker},
  {Pascucci}, {Pasqualetti}, {Passaquieti}, {Passuello}, {Patil}, {Patricelli},
  {Pearlstone}, {Pedraza}, {Pedurand}, {Pekowsky}, {Pele}, {Penn}, {Perez},
  {Perreca}, {Perri}, {Pfeiffer}, {Phelps}, {Piccinni}, {Pichot},
  {Piergiovanni}, {Pierro}, {Pillant}, {Pinard}, {Pinto}, {Pirello}, {Pitkin},
  {Poe}, {Poggiani}, {Popolizio}, {Porter}, {Post}, {Powell}, {Prasad},
  {Pratt}, {Pratten}, {Predoi}, {Prestegard}, {Price}, {Prijatelj}, {Principe},
  {Privitera}, {Prodi}, {Prokhorov}, {Puncken}, {Punturo}, {Puppo},
  {P{\"u}rrer}, {Qi}, {Quetschke}, {Quintero}, {Quitzow-James}, {Raab},
  {Rabeling}, {Radkins}, {Raffai}, {Raja}, {Rajan}, {Rajbhandari}, {Rakhmanov},
  {Ramirez}, {Ramos-Buades}, {Rapagnani}, {Raymond}, {Razzano}, {Read},
  {Regimbau}, {Rei}, {Reid}, {Reitze}, {Ren}, {Reyes}, {Ricci}, {Ricker},
  {Rieger}, {Riles}, {Rizzo}, {Robertson}, {Robie}, {Robinet}, {Rocchi},
  {Rolland}, {Rollins}, {Roma}, {Romano}, {Romel}, {Romie}, {Rosi{\'n}ska},
  {Ross}, {Rowan}, {R{\"u}diger}, {Ruggi}, {Rutins}, {Ryan}, {Sachdev},
  {Sadecki}, {Sadeghian}, {Sakellariadou}, {Salconi}, {Saleem}, {Salemi},
  {Samajdar}, {Sammut}, {Sampson}, {Sanchez}, {Sanchez}, {Sanchis-Gual},
  {Sandberg}, {Sanders}, {Sassolas}, {Sathyaprakash}, {Saulson}, {Sauter},
  {Savage}, {Sawadsky}, {Schale}, {Scheel}, {Scheuer}, {Schmidt}, {Schmidt},
  {Schnabel}, {Schofield}, {Sch{\"o}nbeck}, {Schreiber}, {Schuette}, {Schulte},
  {Schutz}, {Schwalbe}, {Scott}, {Scott}, {Seidel}, {Sellers}, {Sengupta},
  {Sentenac}, {Sequino}, {Sergeev}, {Shaddock}, {Shaffer}, {Shah}, {Shahriar},
  {Shaner}, {Shao}, {Shapiro}, {Shawhan}, {Sheperd}, {Shoemaker}, {Shoemaker},
  {Siellez}, {Siemens}, {Sieniawska}, {Sigg}, {Silva}, {Singer}, {Singh},
  {Singhal}, {Sintes}, {Slagmolen}, {Smith}, {Smith}, {Smith}, {Somala}, {Son},
  {Sonnenberg}, {Sorazu}, {Sorrentino}, {Souradeep}, {Spencer}, {Srivastava},
  {Staats}, {Staley}, {Steinke}, {Steinlechner}, {Steinlechner}, {Steinmeyer},
  {Stevenson}, {Stone}, {Stops}, {Strain}, {Stratta}, {Strigin}, {Strunk},
  {Sturani}, {Stuver}, {Summerscales}, {Sun}, {Sunil}, {Suresh}, {Sutton},
  {Swinkels}, {Szczepa{\'n}czyk}, {Tacca}, {Tait}, {Talbot}, {Talukder},
  {Tanner}, {T{\'a}pai}, {Taracchini}, {Tasson}, {Taylor}, {Taylor}, {Tewari},
  {Theeg}, {Thies}, {Thomas}, {Thomas}, {Thomas}, {Thorne}, {Thorne}, {Thrane},
  {Tiwari}, {Tiwari}, {Tokmakov}, {Toland}, {Tonelli}, {Tornasi},
  {Torres-Forn{\'e}}, {Torrie}, {T{\"o}yr{\"a}}, {Travasso}, {Traylor},
  {Trinastic}, {Tringali}, {Trozzo}, {Tsang}, {Tse}, {Tso}, {Tsukada}, {Tsuna},
  {Tuyenbayev}, {Ueno}, {Ugolini}, {Unnikrishnan}, {Urban}, {Usman},
  {Vahlbruch}, {Vajente}, {Valdes}, {van Bakel}, {van Beuzekom}, {van den
  Brand}, {Van Den Broeck}, {Vander-Hyde}, {van der Schaaf}, {van Heijningen},
  {van Veggel}, {Vardaro}, {Varma}, {Vass}, {Vas{\'u}th}, {Vecchio},
  {Vedovato}, {Veitch}, {Veitch}, {Venkateswara}, {Venugopalan}, {Verkindt},
  {Vetrano}, {Vicer{\'e}}, {Viets}, {Vinciguerra}, {Vine}, {Vinet}, {Vitale},
  {Vo}, {Vocca}, {Vorvick}, {Vyatchanin}, {Wade}, {Wade}, {Wade}, {Walet},
  {Walker}, {Wallace}, {Walsh}, {Wang}, {Wang}, {Wang}, {Wang}, {Wang}, {Ward},
  {Warner}, {Was}, {Watchi}, {Weaver}, {Wei}, {Weinert}, {Weinstein}, {Weiss},
  {Wen}, {Wessel}, {Wessels}, {Westerweck}, {Westphal}, {Wette}, {Whelan},
  {Whitcomb}, {Whiting}, {Whittle}, {Wilken}, {Williams}, {Williams},
  {Williamson}, {Willis}, {Willke}, {Wimmer}, {Winkler}, {Wipf}, {Wittel},
  {Woan}, {Woehler}, {Wofford}, {Wong}, {Worden}, {Wright}, {Wu}, {Wysocki},
  {Xiao}, {Yamamoto}, {Yancey}, {Yang}, {Yap}, {Yazback}, {Yu}, {Yu}, {Yvert},
  {Zadro{\.z}ny}, {Zanolin}, {Zelenova}, {Zendri}, {Zevin}, {Zhang}, {Zhang},
  {Zhang}, {Zhang}, {Zhao}, {Zhou}, {Zhou}, {Zhu}, {Zhu}, {Zimmerman},
  {Zucker}, {Zweizig}, {LIGO Scientific Collaboration}, {Virgo Collaboration},
  {Wilson-Hodge}, {Bissaldi}, {Blackburn}, {Briggs}, {Burns}, {Cleveland},
  {Connaughton}, {Gibby}, {Giles}, {Goldstein}, {Hamburg}, {Jenke}, {Hui},
  {Kippen}, {Kocevski}, {McBreen}, {Meegan}, {Paciesas}, {Poolakkil}, {Preece},
  {Racusin}, {Roberts}, {Stanbro}, {Veres}, {von Kienlin}, {GBM}, {Savchenko},
  {Ferrigno}, {Kuulkers}, {Bazzano}, {Bozzo}, {Brandt}, {Chenevez},
  {Courvoisier}, {Diehl}, {Domingo}, {Hanlon}, {Jourdain}, {Laurent}, {Lebrun},
  {Lutovinov}, {Martin-Carrillo}, {Mereghetti}, {Natalucci}, {Rodi}, {Roques},
  {Sunyaev}, {Ubertini}, {INTEGRAL}, {Aartsen}, {Ackermann}, {Adams},
  {Aguilar}, {Ahlers}, {Ahrens}, {Samarai}, {Altmann}, {Andeen}, {Anderson},
  {Ansseau}, {Anton}, {Arg{\"u}elles}, {Auffenberg}, {Axani}, {Bagherpour},
  {Bai}, {Barron}, {Barwick}, {Baum}, {Bay}, {Beatty}, {Becker Tjus},
  {Bernardini}, {Besson}, {Binder}, {Bindig}, {Blaufuss}, {Blot}, {Bohm},
  {B{\"o}rner}, {Bos}, {Bose}, {B{\"o}ser}, {Botner}, {Bourbeau}, {Bourbeau},
  {Bradascio}, {Braun}, {Brayeur}, {Brenzke}, {Bretz}, {Bron},
  {Brostean-Kaiser}, {Burgman}, {Carver}, {Casey}, {Casier}, {Cheung},
  {Chirkin}, {Christov}, {Clark}, {Classen}, {Coenders}, {Collin}, {Conrad},
  {Cowen}, {Cross}, {Day}, {de Andr{\'e}}, {De Clercq}, {DeLaunay},
  {Dembinski}, {De Ridder}, {Desiati}, {de Vries}, {de Wasseige}, {de With},
  {DeYoung}, {D{\'\i}az-V{\'e}lez}, {di Lorenzo}, {Dujmovic}, {Dumm},
  {Dunkman}, {Dvorak}, {Eberhardt}, {Ehrhardt}, {Eichmann}, {Eller}, {Evenson},
  {Fahey}, {Fazely}, {Felde}, {Filimonov}, {Finley}, {Flis}, {Franckowiak},
  {Friedman}, {Fuchs}, {Gaisser}, {Gallagher}, {Gerhardt}, {Ghorbani}, {Giang},
  {Glauch}, {Gl{\"u}senkamp}, {Goldschmidt}, {Gonzalez}, {Grant}, {Griffith},
  {Haack}, {Hallgren}, {Halzen}, {Hanson}, {Hebecker}, {Heereman}, {Helbing},
  {Hellauer}, {Hickford}, {Hignight}, {Hill}, {Hoffman}, {Hoffmann},
  {Hokanson-Fasig}, {Hoshina}, {Huang}, {Huber}, {Hultqvist}, {H{\"u}nnefeld},
  {In}, {Ishihara}, {Jacobi}, {Japaridze}, {Jeong}, {Jero}, {Jones},
  {Kalaczynski}, {Kang}, {Kappes}, {Karg}, {Karle}, {Kauer}, {Keivani},
  {Kelley}, {Kheirandish}, {Kim}, {Kim}, {Kintscher}, {Kiryluk}, {Kittler},
  {Klein}, {Kohnen}, {Koirala}, {Kolanoski}, {K{\"o}pke}, {Kopper}, {Kopper},
  {Koschinsky}, {Koskinen}, {Kowalski}, {Krings}, {Kroll}, {Kr{\"u}ckl},
  {Kunnen}, {Kunwar}, {Kurahashi}, {Kuwabara}, {Kyriacou}, {Labare},
  {Lanfranchi}, {Larson}, {Lauber}, {Lesiak-Bzdak}, {Leuermann}, {Liu}, {Lu},
  {L{\"u}nemann}, {Luszczak}, {Madsen}, {Maggi}, {Mahn}, {Mancina}, {Maruyama},
  {Mase}, {Maunu}, {McNally}, {Meagher}, {Medici}, {Meier}, {Menne}, {Merino},
  {Meures}, {Miarecki}, {Micallef}, {Moment{\'e}}, {Montaruli}, {Moore},
  {Moulai}, {Nahnhauer}, {Nakarmi}, {Naumann}, {Neer}, {Niederhausen},
  {Nowicki}, {Nygren}, {Obertacke Pollmann}, {Olivas}, {O'Murchadha},
  {Palczewski}, {Pandya}, {Pankova}, {Peiffer}, {Pepper}, {P{\'e}rez de los
  Heros}, {Pieloth}, {Pinat}, {Price}, {Przybylski}, {Raab}, {R{\"a}del},
  {Rameez}, {Rawlins}, {Rea}, {Reimann}, {Relethford}, {Relich}, {Resconi},
  {Rhode}, {Richman}, {Robertson}, {Rongen}, {Rott}, {Ruhe}, {Ryckbosch},
  {Rysewyk}, {S{\"a}lzer}, {Sanchez Herrera}, {Sandrock}, {Sandroos},
  {Santander}, {Sarkar}, {Sarkar}, {Satalecka}, {Schlunder}, {Schmidt},
  {Schneider}, {Schoenen}, {Sch{\"o}neberg}, {Schumacher}, {Seckel},
  {Seunarine}, {Soedingrekso}, {Soldin}, {Song}, {Spiczak}, {Spiering},
  {Stachurska}, {Stamatikos}, {Stanev}, {Stasik}, {Stettner}, {Steuer},
  {Stezelberger}, {Stokstad}, {St{\"o}ssl}, {Strotjohann}, {Stuttard},
  {Sullivan}, {Sutherland}, {Taboada}, {Tatar}, {Tenholt}, {Ter-Antonyan},
  {Terliuk}, {Te{\v{s}}i{\'c}}, {Tilav}, {Toale}, {Tobin}, {Toscano}, {Tosi},
  {Tselengidou}, {Tung}, {Turcati}, {Turley}, {Ty}, {Unger}, {Usner},
  {Vandenbroucke}, {Van Driessche}, {van Eijndhoven}, {Vanheule}, {van Santen},
  {Vehring}, {Vogel}, {Vraeghe}, {Walck}, {Wallace}, {Wallraff}, {Wandler},
  {Wandkowsky}, {Waza}, {Weaver}, {Weiss}, {Wendt}, {Werthebach}, {Whelan},
  {Wiebe}, {Wiebusch}, {Wille}, {Williams}, {Wills}, {Wolf}, {Wood}, {Woolsey},
  {Woschnagg}, {Xu}, {Xu}, {Xu}, {Yanez}, {Yodh}, {Yoshida}, {Yuan}, {Zoll},
  {IceCube Collaboration}, {Balasubramanian}, {Mate}, {Bhalerao},
  {Bhattacharya}, {Vibhute}, {Dewangan}, {Rao}, {Vadawale}, {AstroSat Cadmium
  Zinc Telluride Imager Team}, {Svinkin}, {Hurley}, {Aptekar}, {Frederiks},
  {Golenetskii}, {Kozlova}, {Lysenko}, {Oleynik}, {Tsvetkova}, {Ulanov},
  {Cline}, {IPN Collaboration}, {Li}, {Xiong}, {Zhang}, {Lu}, {Song}, {Cao},
  {Chang}, {Chen}, {Chen}, {Chen}, {Chen}, {Chen}, {Chen}, {Cui}, {Cui},
  {Deng}, {Dong}, {Du}, {Fu}, {Gao}, {Gao}, {Gao}, {Ge}, {Gu}, {Guan}, {Guo},
  {Han}, {Hu}, {Huang}, {Huo}, {Jia}, {Jiang}, {Jiang}, {Jin}, {Jin}, {Li},
  {Li}, {Li}, {Li}, {Li}, {Li}, {Li}, {Li}, {Li}, {Li}, {Li}, {Liang}, {Liao},
  {Liu}, {Liu}, {Liu}, {Liu}, {Liu}, {Liu}, {Liu}, {Lu}, {Lu}, {Luo}, {Ma},
  {Meng}, {Nang}, {Nie}, {Ou}, {Qu}, {Sai}, {Sun}, {Tan}, {Tao}, {Tao}, {Tuo},
  {Wang}, {Wang}, {Wang}, {Wang}, {Wang}, {Wen}, {Wu}, {Wu}, {Xiao}, {Xu},
  {Xu}, {Yan}, {Yang}, {Yang}, {Yang}, {Zhang}, {Zhang}, {Zhang}, {Zhang},
  {Zhang}, {Zhang}, {Zhang}, {Zhang}, {Zhang}, {Zhang}, {Zhang}, {Zhang},
  {Zhang}, {Zhang}, {Zhang}, {Zhang}, {Zhang}, {Zhang}, {Zhao}, {Zhao}, {Zhao},
  {Zheng}, {Zhu}, {Zhu}, {Zou}, {Insight-HXMT Collaboration}, {Albert},
  {Andr{\'e}}, {Anghinolfi}, {Ardid}, {Aubert}, {Aublin}, {Avgitas}, {Baret},
  {Barrios-Mart{\'\i}}, {Basa}, {Belhorma}, {Bertin}, {Biagi}, {Bormuth},
  {Bourret}, {Bouwhuis}, {Br{\^a}nza{\c{s}}}, {Bruijn}, {Brunner}, {Busto},
  {Capone}, {Caramete}, {Carr}, {Celli}, {Cherkaoui El Moursli}, {Chiarusi},
  {Circella}, {Coelho}, {Coleiro}, {Coniglione}, {Costantini}, {Coyle},
  {Creusot}, {D{\'\i}az}, {Deschamps}, {De Bonis}, {Distefano}, {Di Palma},
  {Domi}, {Donzaud}, {Dornic}, {Drouhin}, {Eberl}, {El Bojaddaini}, {El
  Khayati}, {Els{\"a}sser}, {Enzenh{\"o}fer}, {Ettahiri}, {Fassi}, {Felis},
  {Fusco}, {Gay}, {Giordano}, {Glotin}, {Gr{\'e}goire}, {Ruiz}, {Graf},
  {Hallmann}, {van Haren}, {Heijboer}, {Hello}, {Hern{\'a}ndez-Rey},
  {H{\"o}ssl}, {Hofest{\"a}dt}, {Hugon}, {Illuminati}, {James}, {de Jong},
  {Jongen}, {Kadler}, {Kalekin}, {Katz}, {Kiessling}, {Kouchner}, {Kreter},
  {Kreykenbohm}, {Kulikovskiy}, {Lachaud}, {Lahmann}, {Lef{\`e}vre}, {Leonora},
  {Lotze}, {Loucatos}, {Marcelin}, {Margiotta}, {Marinelli},
  {Mart{\'\i}nez-Mora}, {Mele}, {Melis}, {Michael}, {Migliozzi}, {Moussa},
  {Navas}, {Nezri}, {Organokov}, {P{\u{a}}v{\u{a}}la{\c{s}}}, {Pellegrino},
  {Perrina}, {Piattelli}, {Popa}, {Pradier}, {Quinn}, {Racca}, {Riccobene},
  {S{\'a}nchez-Losa}, {Salda{\~n}a}, {Salvadori}, {Samtleben}, {Sanguineti},
  {Sapienza}, {Sieger}, {Spurio}, {Stolarczyk}, {Taiuti}, {Tayalati},
  {Trovato}, {Turpin}, {T{\"o}nnis}, {Vallage}, {Van Elewyck}, {Versari},
  {Vivolo}, {Vizzoca}, {Wilms}, {Zornoza}, {Z{\'u}{\~n}iga}, {ANTARES
  Collaboration}, {Beardmore}, {Breeveld}, {Burrows}, {Cenko}, {Cusumano},
  {D'A{\`\i}}, {de Pasquale}, {Emery}, {Evans}, {Giommi}, {Gronwall}, {Kennea},
  {Krimm}, {Kuin}, {Lien}, {Marshall}, {Melandri}, {Nousek}, {Oates},
  {Osborne}, {Pagani}, {Page}, {Palmer}, {Perri}, {Siegel}, {Sbarufatti},
  {Tagliaferri}, {Tohuvavohu}, {Swift Collaboration}, {Tavani}, {Verrecchia},
  {Bulgarelli}, {Evangelista}, {Pacciani}, {Feroci}, {Pittori}, {Giuliani},
  {Del Monte}, {Donnarumma}, {Argan}, {Trois}, {Ursi}, {Cardillo}, {Piano},
  {Longo}, {Lucarelli}, {Munar-Adrover}, {Fuschino}, {Labanti}, {Marisaldi},
  {Minervini}, {Fioretti}, {Parmiggiani}, {Gianotti}, {Trifoglio}, {Di Persio},
  {Antonelli}, {Barbiellini}, {Caraveo}, {Cattaneo}, {Costa}, {Colafrancesco},
  {D'Amico}, {Ferrari}, {Morselli}, {Paoletti}, {Picozza}, {Pilia}, {Rappoldi},
  {Soffitta}, {Vercellone}, {AGILE Team}, {Foley}, {Coulter}, {Kilpatrick},
  {Drout}, {Piro}, {Shappee}, {Siebert}, {Simon}, {Ulloa}, {Kasen}, {Madore},
  {Murguia-Berthier}, {Pan}, {Prochaska}, {Ramirez-Ruiz}, {Rest},
  {Rojas-Bravo}, {1M2H Team}, {Berger}, {Soares-Santos}, {Annis}, {Alexander},
  {Allam}, {Balbinot}, {Blanchard}, {Brout}, {Butler}, {Chornock}, {Cook},
  {Cowperthwaite}, {Diehl}, {Drlica-Wagner}, {Drout}, {Durret}, {Eftekhari},
  {Finley}, {Fong}, {Frieman}, {Fryer}, {Garc{\'\i}a-Bellido}, {Gruendl},
  {Hartley}, {Herner}, {Kessler}, {Lin}, {Lopes}, {Louren{\c{c}}o}, {Margutti},
  {Marshall}, {Matheson}, {Medina}, {Metzger}, {Mu{\~n}oz}, {Muir}, {Nicholl},
  {Nugent}, {Palmese}, {Paz-Chinch{\'o}n}, {Quataert}, {Sako}, {Sauseda},
  {Schlegel}, {Scolnic}, {Secco}, {Smith}, {Sobreira}, {Villar}, {Vivas},
  {Wester}, {Williams}, {Yanny}, {Zenteno}, {Zhang}, {Abbott}, {Banerji},
  {Bechtol}, {Benoit-L{\'e}vy}, {Bertin}, {Brooks}, {Buckley-Geer}, {Burke},
  {Capozzi}, {Carnero Rosell}, {Carrasco Kind}, {Castander}, {Crocce}, {Cunha},
  {D'Andrea}, {da Costa}, {Davis}, {DePoy}, {Desai}, {Dietrich}, {Eifler},
  {Fernandez}, {Flaugher}, {Fosalba}, {Gaztanaga}, {Gerdes}, {Giannantonio},
  {Goldstein}, {Gruen}, {Gschwend}, {Gutierrez}, {Honscheid}, {James},
  {Jeltema}, {Johnson}, {Johnson}, {Kent}, {Krause}, {Kron}, {Kuehn}, {Lahav},
  {Lima}, {Maia}, {March}, {Martini}, {McMahon}, {Menanteau}, {Miller},
  {Miquel}, {Mohr}, {Nichol}, {Ogando}, {Plazas}, {Romer}, {Roodman}, {Rykoff},
  {Sanchez}, {Scarpine}, {Schindler}, {Schubnell}, {Sevilla-Noarbe}, {Sheldon},
  {Smith}, {Smith}, {Stebbins}, {Suchyta}, {Swanson}, {Tarle}, {Thomas},
  {Troxel}, {Tucker}, {Vikram}, {Walker}, {Wechsler}, {Weller}, {Carlin},
  {Gill}, {Li}, {Marriner}, {Neilsen}, {Dark Energy Camera GW-EM
  Collaboration}, {DES Collaboration}, {Haislip}, {Kouprianov}, {Reichart},
  {Sand}, {Tartaglia}, {Valenti}, {Yang}, {DLT40 Collaboration}, {Benetti},
  {Brocato}, {Campana}, {Cappellaro}, {Covino}, {D'Avanzo}, {D'Elia}, {Getman},
  {Ghirlanda}, {Ghisellini}, {Limatola}, {Nicastro}, {Palazzi}, {Pian},
  {Piranomonte}, {Possenti}, {Rossi}, {Salafia}, {Tomasella}, {Amati},
  {Antonelli}, {Bernardini}, {Bufano}, {Capaccioli}, {Casella}, {Dadina}, {De
  Cesare}, {Di Paola}, {Giuffrida}, {Giunta}, {Israel}, {Lisi}, {Maiorano},
  {Mapelli}, {Masetti}, {Pescalli}, {Pulone}, {Salvaterra}, {Schipani},
  {Spera}, {Stamerra}, {Stella}, {Testa}, {Turatto}, {Vergani}, {Aresu},
  {Bachetti}, {Buffa}, {Burgay}, {Buttu}, {Caria}, {Carretti}, {Casasola},
  {Castangia}, {Carboni}, {Casu}, {Concu}, {Corongiu}, {Deiana}, {Egron},
  {Fara}, {Gaudiomonte}, {Gusai}, {Ladu}, {Loru}, {Leurini}, {Marongiu},
  {Melis}, {Melis}, {Migoni}, {Milia}, {Navarrini}, {Orlati}, {Ortu}, {Palmas},
  {Pellizzoni}, {Perrodin}, {Pisanu}, {Poppi}, {Righini}, {Saba}, {Serra},
  {Serrau}, {Stagni}, {Surcis}, {Vacca}, {Vargiu}, {Hunt}, {Jin}, {Klose},
  {Kouveliotou}, {Mazzali}, {M{\o}ller}, {Nava}, {Piran}, {Selsing}, {Vergani},
  {Wiersema}, {Toma}, {Higgins}, {Mundell}, {di Serego Alighieri}, {G{\'o}tz},
  {Gao}, {Gomboc}, {Kaper}, {Kobayashi}, {Kopac}, {Mao}, {Starling}, {Steele},
  {van der Horst}, {GRAWITA: GRAvitational Wave Inaf TeAm}, {Acero}, {Atwood},
  {Baldini}, {Barbiellini}, {Bastieri}, {Berenji}, {Bellazzini}, {Bissaldi},
  {Blandford}, {Bloom}, {Bonino}, {Bottacini}, {Bregeon}, {Buehler}, {Buson},
  {Cameron}, {Caputo}, {Caraveo}, {Cavazzuti}, {Chekhtman}, {Cheung}, {Chiang},
  {Ciprini}, {Cohen-Tanugi}, {Cominsky}, {Costantin}, {Cuoco}, {D'Ammando}, {de
  Palma}, {Digel}, {Di Lalla}, {Di Mauro}, {Di Venere}, {Dubois}, {Fegan},
  {Focke}, {Franckowiak}, {Fukazawa}, {Funk}, {Fusco}, {Gargano}, {Gasparrini},
  {Giglietto}, {Giordano}, {Giroletti}, {Glanzman}, {Green}, {Grondin},
  {Guillemot}, {Guiriec}, {Harding}, {Horan}, {J{\'o}hannesson}, {Kamae},
  {Kensei}, {Kuss}, {La Mura}, {Latronico}, {Lemoine-Goumard}, {Longo},
  {Loparco}, {Lovellette}, {Lubrano}, {Magill}, {Maldera}, {Manfreda},
  {Mazziotta}, {McEnery}, {Meyer}, {Michelson}, {Mirabal}, {Monzani},
  {Moretti}, {Morselli}, {Moskalenko}, {Negro}, {Nuss}, {Ojha}, {Omodei},
  {Orienti}, {Orlando}, {Palatiello}, {Paliya}, {Paneque}, {Pesce-Rollins},
  {Piron}, {Porter}, {Principe}, {Rain{\`o}}, {Rando}, {Razzano}, {Razzaque},
  {Reimer}, {Reimer}, {Reposeur}, {Rochester}, {Saz Parkinson}, {Sgr{\`o}},
  {Siskind}, {Spada}, {Spandre}, {Suson}, {Takahashi}, {Tanaka}, {Thayer},
  {Thayer}, {Thompson}, {Tibaldo}, {Torres}, {Torresi}, {Troja}, {Venters},
  {Vianello}, {Zaharijas}, {Fermi Large Area Telescope Collaboration},
  {Allison}, {Bannister}, {Dobie}, {Kaplan}, {Lenc}, {Lynch}, {Murphy},
  {Sadler}, {Australia Telescope Compact Array}, {Hotan}, {James}, {Oslowski},
  {Raja}, {Shannon}, {Whiting}, {Australian SKA Pathfinder}, {Arcavi},
  {Howell}, {McCully}, {Hosseinzadeh}, {Hiramatsu}, {Poznanski}, {Barnes},
  {Zaltzman}, {Vasylyev}, {Maoz}, {Las Cumbres Observatory Group}, {Cooke},
  {Bailes}, {Wolf}, {Deller}, {Lidman}, {Wang}, {Gendre}, {Andreoni}, {Ackley},
  {Pritchard}, {Bessell}, {Chang}, {M{\"o}ller}, {Onken}, {Scalzo},
  {Ridden-Harper}, {Sharp}, {Tucker}, {Farrell}, {Elmer}, {Johnston},
  {Venkatraman Krishnan}, {Keane}, {Green}, {Jameson}, {Hu}, {Ma}, {Sun}, {Wu},
  {Wang}, {Shang}, {Hu}, {Ashley}, {Yuan}, {Li}, {Tao}, {Zhu}, {Zhang},
  {Suntzeff}, {Zhou}, {Yang}, {Orange}, {Morris}, {Cucchiara}, {Giblin},
  {Klotz}, {Staff}, {Thierry}, {Schmidt}, {OzGrav}, {(Deeper}, {Wider},
  {program}, {AST3}, {CAASTRO Collaborations}, {Tanvir}, {Levan}, {Cano}, {de
  Ugarte-Postigo}, {Gonz{\'a}lez-Fern{\'a}ndez}, {Greiner}, {Hjorth}, {Irwin},
  {Kr{\"u}hler}, {Mandel}, {Milvang-Jensen}, {O'Brien}, {Rol}, {Rosetti},
  {Rosswog}, {Rowlinson}, {Steeghs}, {Th{\"o}ne}, {Ulaczyk}, {Watson}, {Bruun},
  {Cutter}, {Figuera Jaimes}, {Fujii}, {Fruchter}, {Gompertz}, {Jakobsson},
  {Hodosan}, {J{\`e}rgensen}, {Kangas}, {Kann}, {Rabus}, {Schr{\o}der},
  {Stanway}, {Wijers}, {VINROUGE Collaboration}, {Lipunov}, {Gorbovskoy},
  {Kornilov}, {Tyurina}, {Balanutsa}, {Kuznetsov}, {Vlasenko}, {Podesta},
  {Lopez}, {Podesta}, {Levato}, {Saffe}, {Mallamaci}, {Budnev}, {Gress},
  {Kuvshinov}, {Gorbunov}, {Vladimirov}, {Zimnukhov}, {Gabovich}, {Yurkov},
  {Sergienko}, {Rebolo}, {Serra-Ricart}, {Tlatov}, {Ishmuhametova}, {MASTER
  Collaboration}, {Abe}, {Aoki}, {Aoki}, {Asakura}, {Baar}, {Barway}, {Bond},
  {Doi}, {Finet}, {Fujiyoshi}, {Furusawa}, {Honda}, {Itoh}, {Kanda},
  {Kawabata}, {Kawabata}, {Kim}, {Koshida}, {Kuroda}, {Lee}, {Liu},
  {Matsubayashi}, {Miyazaki}, {Morihana}, {Morokuma}, {Motohara}, {Murata},
  {Nagai}, {Nagashima}, {Nagayama}, {Nakaoka}, {Nakata}, {Ohsawa}, {Ohshima},
  {Ohta}, {Okita}, {Saito}, {Saito}, {Sako}, {Sekiguchi}, {Sumi}, {Tajitsu},
  {Takahashi}, {Takayama}, {Tamura}, {Tanaka}, {Tanaka}, {Terai}, {Tominaga},
  {Tristram}, {Uemura}, {Utsumi}, {Yamaguchi}, {Yasuda}, {Yoshida}, {Zenko},
  {J-GEM}, {Adams}, {Anupama}, {Bally}, {Barway}, {Bellm}, {Blagorodnova},
  {Cannella}, {Chandra}, {Chatterjee}, {Clarke}, {Cobb}, {Cook}, {Copperwheat},
  {De}, {Emery}, {Feindt}, {Foster}, {Fox}, {Frail}, {Fremling}, {Frohmaier},
  {Garcia}, {Ghosh}, {Giacintucci}, {Goobar}, {Gottlieb}, {Grefenstette},
  {Hallinan}, {Harrison}, {Heida}, {Helou}, {Ho}, {Horesh}, {Hotokezaka}, {Ip},
  {Itoh}, {Jacobs}, {Jencson}, {Kasen}, {Kasliwal}, {Kassim}, {Kim}, {Kiran},
  {Kuin}, {Kulkarni}, {Kupfer}, {Lau}, {Madsen}, {Mazzali}, {Miller},
  {Miyasaka}, {Mooley}, {Myers}, {Nakar}, {Ngeow}, {Nugent}, {Ofek},
  {Palliyaguru}, {Pavana}, {Perley}, {Peters}, {Pike}, {Piran}, {Qi}, {Quimby},
  {Rana}, {Rosswog}, {Rusu}, {Sadler}, {Van Sistine}, {Sollerman}, {Xu}, {Yan},
  {Yatsu}, {Yu}, {Zhang}, {Zhao}, {GROWTH}, {JAGWAR}, {Caltech-NRAO},
  {TTU-NRAO}, {NuSTAR Collaborations}, {Chambers}, {Huber}, {Schultz},
  {Bulger}, {Flewelling}, {Magnier}, {Lowe}, {Wainscoat}, {Waters}, {Willman},
  {Pan-STARRS}, {Ebisawa}, {Hanyu}, {Harita}, {Hashimoto}, {Hidaka}, {Hori},
  {Ishikawa}, {Isobe}, {Iwakiri}, {Kawai}, {Kawai}, {Kawamuro}, {Kawase},
  {Kitaoka}, {Makishima}, {Matsuoka}, {Mihara}, {Morita}, {Morita}, {Nakahira},
  {Nakajima}, {Nakamura}, {Negoro}, {Oda}, {Sakamaki}, {Sasaki}, {Serino},
  {Shidatsu}, {Shimomukai}, {Sugawara}, {Sugita}, {Sugizaki}, {Tachibana},
  {Takao}, {Tanimoto}, {Tomida}, {Tsuboi}, {Tsunemi}, {Ueda}, {Ueno}, {Yamada},
  {Yamaoka}, {Yamauchi}, {Yatabe}, {Yoneyama}, {Yoshii}, {MAXI Team}, {Coward},
  {Crisp}, {Macpherson}, {Andreoni}, {Laugier}, {Noysena}, {Klotz}, {Gendre},
  {Thierry}, {Turpin}, {Consortium}, {Im}, {Choi}, {Kim}, {Yoon}, {Lim}, {Lee},
  {Lee}, {Kim}, {Ko}, {Joe}, {Kwon}, {Kim}, {Lim}, {Choi}, {KU Collaboration},
  {Fynbo}, {Malesani}, {Xu}, {Optical Telescope}, {Smartt}, {Jerkstrand},
  {Kankare}, {Sim}, {Fraser}, {Inserra}, {Maguire}, {Leloudas}, {Magee},
  {Shingles}, {Smith}, {Young}, {Kotak}, {Gal-Yam}, {Lyman}, {Homan},
  {Agliozzo}, {Anderson}, {Angus}, {Ashall}, {Barbarino}, {Bauer}, {Berton},
  {Botticella}, {Bulla}, {Cannizzaro}, {Cartier}, {Cikota}, {Clark}, {De Cia},
  {Della Valle}, {Dennefeld}, {Dessart}, {Dimitriadis}, {Elias-Rosa}, {Firth},
  {Fl{\"o}rs}, {Frohmaier}, {Galbany}, {Gonz{\'a}lez-Gait{\'a}n}, {Gromadzki},
  {Guti{\'e}rrez}, {Hamanowicz}, {Harmanen}, {Heintz}, {Hernandez}, {Hodgkin},
  {Hook}, {Izzo}, {James}, {Jonker}, {Kerzendorf}, {Kostrzewa-Rutkowska},
  {Kromer}, {Kuncarayakti}, {Lawrence}, {Manulis}, {Mattila}, {McBrien},
  {M{\"u}ller}, {Nordin}, {O'Neill}, {Onori}, {Palmerio}, {Pastorello},
  {Patat}, {Pignata}, {Podsiadlowski}, {Razza}, {Reynolds}, {Roy}, {Ruiter},
  {Rybicki}, {Salmon}, {Pumo}, {Prentice}, {Seitenzahl}, {Smith}, {Sollerman},
  {Sullivan}, {Szegedi}, {Taddia}, {Taubenberger}, {Terreran}, {Van Soelen},
  {Vos}, {Walton}, {Wright}, {Wyrzykowski}, {Yaron}, {pre=''(''>ePESSTO},
  {Chen}, {Kr{\"u}hler}, {Schady}, {Wiseman}, {Greiner}, {Rau}, {Schweyer},
  {Klose}, {Nicuesa Guelbenzu}, {GROND}, {Palliyaguru}, {Tech University},
  {Shara}, {Williams}, {Vaisanen}, {Potter}, {Romero Colmenero}, {Crawford},
  {Buckley}, {Mao}, {SALT Group}, {D{\'\i}az}, {Macri}, {Garc{\'\i}a Lambas},
  {Mendes de Oliveira}, {Nilo Castell{\'o}n}, {Ribeiro}, {S{\'a}nchez},
  {Schoenell}, {Abramo}, {Akras}, {Alcaniz}, {Artola}, {Beroiz}, {Bonoli},
  {Cabral}, {Camuccio}, {Chavushyan}, {Coelho}, {Colazo}, {Costa-Duarte},
  {Cuevas Larenas}, {Dom{\'\i}nguez Romero}, {Dultzin}, {Fern{\'a}ndez},
  {Garc{\'\i}a}, {Girardini}, {Gon{\c{c}}alves}, {Gon{\c{c}}alves}, {Gurovich},
  {Jim{\'e}nez-Teja}, {Kanaan}, {Lares}, {Lopes de Oliveira}, {L{\'o}pez-Cruz},
  {Melia}, {Molino}, {Padilla}, {Pe{\~n}uela}, {Placco}, {Qui{\~n}ones},
  {Ram{\'\i}rez Rivera}, {Renzi}, {Riguccini}, {R{\'\i}os-L{\'o}pez},
  {Rodriguez}, {Sampedro}, {Schneiter}, {Sodr{\'e}}, {Starck}, {Torres-Flores},
  {Tornatore}, {Zadro{\.z}ny}, {Castillo}, {TOROS: Transient Robotic
  Observatory of South Collaboration}, {Castro-Tirado}, {Tello}, {Hu}, {Zhang},
  {Cunniffe}, {Castell{\'o}n}, {Hiriart}, {Caballero-Garc{\'\i}a},
  {Jel{\'\i}nek}, {Kub{\'a}nek}, {P{\'e}rez del Pulgar}, {Park}, {Jeong},
  {Castro Cer{\'o}n}, {Pandey}, {Yock}, {Querel}, {Fan}, {Wang}, {BOOTES
  Collaboration}, {Beardsley}, {Brown}, {Crosse}, {Emrich}, {Franzen},
  {Gaensler}, {Horsley}, {Johnston-Hollitt}, {Kenney}, {Morales}, {Pallot},
  {Sokolowski}, {Steele}, {Tingay}, {Trott}, {Walker}, {Wayth}, {Williams},
  {Wu}, {Murchison Widefield Array}, {Yoshida}, {Sakamoto}, {Kawakubo},
  {Yamaoka}, {Takahashi}, {Asaoka}, {Ozawa}, {Torii}, {Shimizu}, {Tamura},
  {Ishizaki}, {Cherry}, {Ricciarini}, {Penacchioni}, {Marrocchesi}, {CALET
  Collaboration}, {Pozanenko}, {Volnova}, {Mazaeva}, {Minaev}, {Krugov},
  {Kusakin}, {Reva}, {Moskvitin}, {Rumyantsev}, {Inasaridze}, {Klunko},
  {Tungalag}, {Schmalz}, {Burhonov}, {IKI-GW Follow-up Collaboration},
  {Abdalla}, {Abramowski}, {Aharonian}, {Ait Benkhali}, {Ang{\"u}ner},
  {Arakawa}, {Arrieta}, {Aubert}, {Backes}, {Balzer}, {Barnard}, {Becherini},
  {Becker Tjus}, {Berge}, {Bernhard}, {Bernl{\"o}hr}, {Blackwell},
  {B{\"o}ttcher}, {Boisson}, {Bolmont}, {Bonnefoy}, {Bordas}, {Bregeon},
  {Brun}, {Brun}, {Bryan}, {B{\"u}chele}, {Bulik}, {Capasso}, {Caroff},
  {Carosi}, {Casanova}, {Cerruti}, {Chakraborty}, {Chaves}, {Chen},
  {Chevalier}, {Colafrancesco}, {Condon}, {Conrad}, {Davids}, {Decock}, {Deil},
  {Devin}, {deWilt}, {Dirson}, {Djannati-Ata{\"\i}}, {Donath}, {O'C. Drury},
  {Dutson}, {Dyks}, {Edwards}, {Egberts}, {Emery}, {Ernenwein}, {Eschbach},
  {Farnier}, {Fegan}, {Fernandes}, {Fiasson}, {Fontaine}, {Funk},
  {F{\"u}ssling}, {Gabici}, {Gallant}, {Garrigoux}, {Gat{\'e}}, {Giavitto},
  {Giebels}, {Glawion}, {Glicenstein}, {Gottschall}, {Grondin}, {Hahn},
  {Haupt}, {Hawkes}, {Heinzelmann}, {Henri}, {Hermann}, {Hinton}, {Hofmann},
  {Hoischen}, {Holch}, {Holler}, {Horns}, {Ivascenko}, {Iwasaki},
  {Jacholkowska}, {Jamrozy}, {Jankowsky}, {Jankowsky}, {Jingo}, {Jouvin},
  {Jung-Richardt}, {Kastendieck}, {Katarzy{\'n}ski}, {Katsuragawa},
  {Kerszberg}, {Khangulyan}, {Kh{\'e}lifi}, {King}, {Klepser}, {Klochkov},
  {Klu{\'z}niak}, {Komin}, {Kosack}, {Krakau}, {Kraus}, {Kr{\"u}ger}, {Laffon},
  {Lamanna}, {Lau}, {Lees}, {Lefaucheur}, {Lemi{\`e}re}, {Lemoine-Goumard},
  {Lenain}, {Leser}, {Lohse}, {Lorentz}, {Liu}, {Lypova}, {Malyshev},
  {Marandon}, {Marcowith}, {Mariaud}, {Marx}, {Maurin}, {Maxted}, {Mayer},
  {Meintjes}, {Meyer}, {Mitchell}, {Moderski}, {Mohamed}, {Mohrmann},
  {Mor{\r{a}}}, {Moulin}, {Murach}, {Nakashima}, {de Naurois}, {Ndiyavala},
  {Niederwanger}, {Niemiec}, {Oakes}, {O'Brien}, {Odaka}, {Ohm}, {Ostrowski},
  {Oya}, {Padovani}, {Panter}, {Parsons}, {Pekeur}, {Pelletier}, {Perennes},
  {Petrucci}, {Peyaud}, {Piel}, {Pita}, {Poireau}, {Poon}, {Prokhorov},
  {Prokoph}, {P{\"u}hlhofer}, {Punch}, {Quirrenbach}, {Raab}, {Rauth},
  {Reimer}, {Reimer}, {Renaud}, {de los Reyes}, {Rieger}, {Rinchiuso},
  {Romoli}, {Rowell}, {Rudak}, {Rulten}, {Sahakian}, {Saito}, {Sanchez},
  {Santangelo}, {Sasaki}, {Schlickeiser}, {Sch{\"u}ssler}, {Schulz},
  {Schwanke}, {Schwemmer}, {Seglar-Arroyo}, {Settimo}, {Seyffert}, {Shafi},
  {Shilon}, {Shiningayamwe}, {Simoni}, {Sol}, {Spanier}, {Spir-Jacob},
  {Stawarz}, {Steenkamp}, {Stegmann}, {Steppa}, {Sushch}, {Takahashi},
  {Tavernet}, {Tavernier}, {Taylor}, {Terrier}, {Tibaldo}, {Tiziani},
  {Tluczykont}, {Trichard}, {Tsirou}, {Tsuji}, {Tuffs}, {Uchiyama}, {van der
  Walt}, {van Eldik}, {van Rensburg}, {van Soelen}, {Vasileiadis}, {Veh},
  {Venter}, {Viana}, {Vincent}, {Vink}, {Voisin}, {V{\"o}lk}, {Vuillaume},
  {Wadiasingh}, {Wagner}, {Wagner}, {Wagner}, {White}, {Wierzcholska},
  {Willmann}, {W{\"o}rnlein}, {Wouters}, {Yang}, {Zaborov}, {Zacharias},
  {Zanin}, {Zdziarski}, {Zech}, {Zefi}, {Ziegler}, {Zorn}, {{\.Z}ywucka},
  {H.~E.~S.~S. Collaboration}, {Fender}, {Broderick}, {Rowlinson}, {Wijers},
  {Stewart}, {ter Veen}, {Shulevski}, {LOFAR Collaboration}, {Kavic},
  {Simonetti}, {League}, {Tsai}, {Obenberger}, {Nathaniel}, {Taylor}, {Dowell},
  {Liebling}, {Estes}, {Lippert}, {Sharma}, {Vincent}, {Farella}, {Wavelength
  Array}, {Abeysekara}, {Albert}, {Alfaro}, {Alvarez}, {Arceo},
  {Arteaga-Vel{\'a}zquez}, {Avila Rojas}, {Ayala Solares}, {Barber}, {Becerra
  Gonzalez}, {Becerril}, {Belmont-Moreno}, {BenZvi}, {Berley}, {Bernal},
  {Braun}, {Brisbois}, {Caballero-Mora}, {Capistr{\'a}n}, {Carrami{\~n}ana},
  {Casanova}, {Castillo}, {Cotti}, {Cotzomi}, {Couti{\~n}o de Le{\'o}n}, {De
  Le{\'o}n}, {De la Fuente}, {Diaz Hernandez}, {Dichiara}, {Dingus},
  {DuVernois}, {D{\'\i}az-V{\'e}lez}, {Ellsworth}, {Engel},
  {Enr{\'\i}quez-Rivera}, {Fiorino}, {Fleischhack}, {Fraija},
  {Garc{\'\i}a-Gonz{\'a}lez}, {Garfias}, {Gerhardt}, {Gonz{\~o}lez Mu{\~n}oz},
  {Gonz{\'a}lez}, {Goodman}, {Hampel-Arias}, {Harding}, {Hernandez},
  {Hernandez-Almada}, {Hona}, {H{\"u}ntemeyer}, {Iriarte}, {Jardin-Blicq},
  {Joshi}, {Kaufmann}, {Kieda}, {Lara}, {Lauer}, {Lennarz}, {Le{\'o}n Vargas},
  {Linnemann}, {Longinotti}, {Raya}, {Luna-Garc{\'\i}a}, {L{\'o}pez-Coto},
  {Malone}, {Marinelli}, {Martinez}, {Martinez-Castellanos},
  {Mart{\'\i}nez-Castro}, {Mart{\'\i}nez-Huerta}, {Matthews},
  {Miranda-Romagnoli}, {Moreno}, {Mostaf{\'a}}, {Nellen}, {Newbold}, {Nisa},
  {Noriega-Papaqui}, {Pelayo}, {Pretz}, {P{\'e}rez-P{\'e}rez}, {Ren}, {Rho},
  {Rivi{\`e}re}, {Rosa-Gonz{\'a}lez}, {Rosenberg}, {Ruiz-Velasco}, {Salazar},
  {Salesa Greus}, {Sandoval}, {Schneider}, {Schoorlemmer}, {Sinnis}, {Smith},
  {Springer}, {Surajbali}, {Tibolla}, {Tollefson}, {Torres}, {Ukwatta},
  {Weisgarber}, {Westerhoff}, {Wisher}, {Wood}, {Yapici}, {Yodh}, {Younk},
  {Zhou}, {{\'A}lvarez}, {HAWC Collaboration}, {Aab}, {Abreu}, {Aglietta},
  {Albuquerque}, {Albury}, {Allekotte}, {Almela}, {Alvarez Castillo},
  {Alvarez-Mu{\~n}iz}, {Anastasi}, {Anchordoqui}, {Andrada}, {Andringa},
  {Aramo}, {Arsene}, {Asorey}, {Assis}, {Avila}, {Badescu}, {Balaceanu},
  {Barbato}, {Barreira Luz}, {Becker}, {Bellido}, {Berat}, {Bertaina},
  {Bertou}, {Biermann}, {Biteau}, {Blaess}, {Blanco}, {Blazek}, {Bleve},
  {Boh{\'a}{\v{c}}ov{\'a}}, {Bonifazi}, {Borodai}, {Botti}, {Brack}, {Brancus},
  {Bretz}, {Bridgeman}, {Briechle}, {Buchholz}, {Bueno}, {Buitink}, {Buscemi},
  {Caballero-Mora}, {Caccianiga}, {Cancio}, {Canfora}, {Caruso}, {Castellina},
  {Catalani}, {Cataldi}, {Cazon}, {Chavez}, {Chinellato}, {Chudoba}, {Clay},
  {Cobos Cerutti}, {Colalillo}, {Coleman}, {Collica}, {Coluccia},
  {Concei{\c{c}}{\~a}o}, {Consolati}, {Contreras}, {Cooper}, {Coutu},
  {Covault}, {Cronin}, {D'Amico}, {Daniel}, {Dasso}, {Daumiller}, {Dawson},
  {Day}, {de Almeida}, {de Jong}, {De Mauro}, {de Mello Neto}, {De Mitri}, {de
  Oliveira}, {de Souza}, {Debatin}, {Deligny}, {D{\'\i}az Castro}, {Diogo},
  {Dobrigkeit}, {D'Olivo}, {Dorosti}, {Dos Anjos}, {Dova}, {Dundovic}, {Ebr},
  {Engel}, {Erdmann}, {Erfani}, {Escobar}, {Espadanal}, {Etchegoyen}, {Falcke},
  {Farmer}, {Farrar}, {Fauth}, {Fazzini}, {Feldbusch}, {Fenu}, {Fick},
  {Figueira}, {Filip{\v{c}}i{\v{c}}}, {Freire}, {Fujii}, {Fuster},
  {Ga{\"\i}or}, {Garc{\'\i}a}, {Gat{\'e}}, {Gemmeke}, {Gherghel-Lascu}, {Ghia},
  {Giaccari}, {Giammarchi}, {Giller}, {G{\l}as}, {Glaser}, {Golup}, {G{\'o}mez
  Berisso}, {G{\'o}mez Vitale}, {Gonz{\'a}lez}, {Gorgi}, {Gottowik}, {Grillo},
  {Grubb}, {Guarino}, {Guedes}, {Halliday}, {Hampel}, {Hansen}, {Harari},
  {Harrison}, {Harvey}, {Haungs}, {Hebbeker}, {Heck}, {Heimann}, {Herve},
  {Hill}, {Hojvat}, {Holt}, {Homola}, {H{\"o}randel}, {Horvath},
  {Hrabovsk{\'y}}, {Huege}, {Hulsman}, {Insolia}, {Isar}, {Jandt}, {Johnsen},
  {Josebachuili}, {Jurysek}, {K{\"a}{\"a}p{\"a}}, {Kampert}, {Keilhauer},
  {Kemmerich}, {Kemp}, {Kieckhafer}, {Klages}, {Kleifges}, {Kleinfeller},
  {Krause}, {Krohm}, {Kuempel}, {Kukec Mezek}, {Kunka}, {Kuotb Awad}, {Lago},
  {LaHurd}, {Lang}, {Lauscher}, {Legumina}, {Leigui de Oliveira},
  {Letessier-Selvon}, {Lhenry-Yvon}, {Link}, {Lo Presti}, {Lopes}, {L{\'o}pez},
  {L{\'o}pez Casado}, {Lorek}, {Luce}, {Lucero}, {Malacari}, {Mallamaci},
  {Mandat}, {Mantsch}, {Mariazzi}, {Maris}, {Marsella}, {Martello}, {Martinez},
  {Mart{\'\i}nez Bravo}, {Mas{\'\i}as Meza}, {Mathes}, {Mathys}, {Matthews},
  {Matthiae}, {Mayotte}, {Mazur}, {Medina}, {Medina-Tanco}, {Melo},
  {Menshikov}, {Merenda}, {Michal}, {Micheletti}, {Middendorf}, {Miramonti},
  {Mitrica}, {Mockler}, {Mollerach}, {Montanet}, {Morello}, {Morlino},
  {M{\"u}ller}, {M{\"u}ller}, {Muller}, {M{\"u}ller}, {Mussa}, {Naranjo},
  {Nguyen}, {Niculescu-Oglinzanu}, {Niechciol}, {Niemietz}, {Niggemann},
  {Nitz}, {Nosek}, {Novotny}, {No{\v{z}}ka}, {N{\'u}{\~n}ez}, {Oikonomou},
  {Olinto}, {Palatka}, {Pallotta}, {Papenbreer}, {Parente}, {Parra}, {Paul},
  {Pech}, {Pedreira}, {P{\c{e}}kala}, {Pe{\~n}a-Rodriguez}, {Pereira},
  {Perlin}, {Perrone}, {Peters}, {Petrera}, {Phuntsok}, {Pierog}, {Pimenta},
  {Pirronello}, {Platino}, {Plum}, {Poh}, {Porowski}, {Prado}, {Privitera},
  {Prouza}, {Quel}, {Querchfeld}, {Quinn}, {Ramos-Pollan}, {Rautenberg},
  {Ravignani}, {Ridky}, {Riehn}, {Risse}, {Ristori}, {Rizi}, {Rodrigues de
  Carvalho}, {Rodriguez Fernandez}, {Rodriguez Rojo}, {Roncoroni}, {Roth},
  {Roulet}, {Rovero}, {Ruehl}, {Saffi}, {Saftoiu}, {Salamida}, {Salazar},
  {Saleh}, {Salina}, {S{\'a}nchez}, {Sanchez-Lucas}, {Santos}, {Santos},
  {Sarazin}, {Sarmento}, {Sarmiento-Cano}, {Sato}, {Schauer}, {Scherini},
  {Schieler}, {Schimp}, {Schmidt}, {Scholten}, {Schov{\'a}nek}, {Schr{\"o}der},
  {Schr{\"o}der}, {Schulz}, {Schumacher}, {Sciutto}, {Segreto}, {Shadkam},
  {Shellard}, {Sigl}, {Silli}, {{\v{S}}m{\'\i}da}, {Snow}, {Sommers},
  {Sonntag}, {Soriano}, {Squartini}, {Stanca}, {Stani{\v{c}}}, {Stasielak},
  {Stassi}, {Stolpovskiy}, {Strafella}, {Streich}, {Suarez},
  {Suarez-Dur{\'a}n}, {Sudholz}, {Suomij{\"a}rvi}, {Supanitsky},
  {{\v{S}}up{\'\i}k}, {Swain}, {Szadkowski}, {Taboada}, {Taborda},
  {Timmermans}, {Todero Peixoto}, {Tomankova}, {Tom{\'e}}, {Torralba Elipe},
  {Travnicek}, {Trini}, {Tueros}, {Ulrich}, {Unger}, {Urban}, {Vald{\'e}s
  Galicia}, {Vali{\~n}o}, {Valore}, {van Aar}, {van Bodegom}, {van den Berg},
  {van Vliet}, {Varela}, {Vargas C{\'a}rdenas}, {V{\'a}zquez}, {Veberi{\v{c}}},
  {Ventura}, {Vergara Quispe}, {Verzi}, {Vicha}, {Villase{\~n}or}, {Vorobiov},
  {Wahlberg}, {Wainberg}, {Walz}, {Watson}, {Weber}, {Weindl}, {Wiede{\'n}ski},
  {Wiencke}, {Wilczy{\'n}ski}, {Wirtz}, {Wittkowski}, {Wundheiler}, {Yang},
  {Yushkov}, {Zas}, {Zavrtanik}, {Zavrtanik}, {Zepeda}, {Zimmermann},
  {Ziolkowski}, {Zong}, {Zuccarello}, {Pierre Auger Collaboration}, {Kim},
  {Schulze}, {Bauer}, {Corral-Santana}, {de Gregorio-Monsalvo},
  {Gonz{\'a}lez-L{\'o}pez}, {Hartmann}, {Ishwara-Chandra}, {Mart{\'\i}n},
  {Mehner}, {Misra}, {Micha{\l}owski}, {Resmi}, {ALMA Collaboration}, {Paragi},
  {Agudo}, {An}, {Beswick}, {Casadio}, {Frey}, {Jonker}, {Kettenis}, {Marcote},
  {Moldon}, {Szomoru}, {van Langevelde}, {Yang}, {Euro VLBI Team}, {Cwiek},
  {Cwiok}, {Czyrkowski}, {Dabrowski}, {Kasprowicz}, {Mankiewicz}, {Nawrocki},
  {Opiela}, {Piotrowski}, {Wrochna}, {Zaremba}, {{\.Z}arnecki}, {Pi of Sky
  Collaboration}, {Haggard}, {Nynka}, {Ruan}, {Chandra Team at McGill
  University}, {Bland}, {Booler}, {Devillepoix}, {de Gois}, {Hancock}, {Howie},
  {Paxman}, {Sansom}, {Towner}, {Desert Fireball Network}, {Tonry}, {Coughlin},
  {Stubbs}, {Denneau}, {Heinze}, {Stalder}, {Weiland}, {ATLAS}, {Eatough},
  {Kramer}, {Kraus}, {Time Resolution Universe Survey}, {Troja}, {Piro},
  {Becerra Gonz{\'a}lez}, {Butler}, {Fox}, {Khandrika}, {Kutyrev}, {Lee},
  {Ricci}, {Ryan}, {S{\'a}nchez-Ram{\'\i}rez}, {Veilleux}, {Watson},
  {Wieringa}, {Burgess}, {van Eerten}, {Fontes}, {Fryer}, {Korobkin},
  {Wollaeger}, {RIMAS}, {RATIR}, {Camilo}, {Foley}, {Goedhart}, {Makhathini},
  {Oozeer}, {Smirnov}, {Fender}, {Woudt}, \& {South Africa/MeerKAT}}]{nsmerger}
{Abbott}, B.~P., {Abbott}, R., {Abbott}, T.~D., {et~al.} 2017, \apjl, 848, L12,
  \dodoi{10.3847/2041-8213/aa91c9}

\bibitem[{{Anderson} {et~al.}(2014){Anderson}, {Brown}, {Collier Cameron},
  {Delrez}, {Fumel}, {Gillon}, {Hellier}, {Jehin}, {Lendl}, {Maxted},
  {Neveu-VanMalle}, {Pepe}, {Pollacco}, {Queloz}, {Rojo}, {Segransan},
  {Serenelli}, {Smalley}, {Smith}, {Southworth}, {Triaud}, {Turner}, {Udry}, \&
  {West}}]{Anderson2014}
{Anderson}, D.~R., {Brown}, D.~J.~A., {Collier Cameron}, A., {et~al.} 2014,
  arXiv e-prints, arXiv:1410.3449, \dodoi{10.48550/arXiv.1410.3449}

\bibitem[{{Barrado Y Navascu{\'e}s}(2006)}]{barrado}
{Barrado Y Navascu{\'e}s}, D. 2006, \aap, 459, 511,
  \dodoi{10.1051/0004-6361:20065717}

\bibitem[{{Beaton} {et~al.}(2018){Beaton}, {Bono}, {Braga}, {Dall'Ora},
  {Fiorentino}, {Jang}, {Mart{\'\i}nez-V{\'a}zquez}, {Matsunaga}, {Monelli},
  {Neeley}, \& {Salaris}}]{beaton18}
{Beaton}, R.~L., {Bono}, G., {Braga}, V.~F., {et~al.} 2018, \ssr, 214, 113,
  \dodoi{10.1007/s11214-018-0542-1}

\bibitem[{{Berg} {et~al.}(2024){Berg}, {Hayes}, {Cristiani}, {McConnachie},
  {Robertson}, {Sestito}, {Simpson}, {Waller}, {Chin}, {Densmore}, {Diaz},
  {Edgar}, {Fuentes Lettura}, {G{\'o}mez-Jim{\'e}nez}, {Kalari}, {Lawrence},
  {Margheim}, {Pazder}, {Ruiz-Carmona}, {Salinas}, {Silva}, {Silversides}, \&
  {Venn}}]{berg}
{Berg}, T. A.~M., {Hayes}, C.~R., {Cristiani}, S., {et~al.} 2024, \mnras,
  \dodoi{10.1093/mnras/stae1033}

\bibitem[{{Bla{\v{z}}ko}(1907)}]{blazhko07}
{Bla{\v{z}}ko}, S. 1907, Astronomische Nachrichten, 175, 325,
  \dodoi{10.1002/asna.19071752002}

\bibitem[{{Boccas} {et~al.}(2012){Boccas}, {Kleinman}, {Goodsell},
  {Tollestrup}, {Adamson}, {Arriagada}, {Christou}, {Gonzalez}, {Hanna},
  {Hartung}, {Lazo}, {Mason}, {Neichel}, {Perez}, {Simons}, {Walls}, \&
  {White}}]{boccas}
{Boccas}, M., {Kleinman}, S.~J., {Goodsell}, S., {et~al.} 2012, in Society of
  Photo-Optical Instrumentation Engineers (SPIE) Conference Series, Vol. 8446,
  Ground-based and Airborne Instrumentation for Astronomy IV, ed. I.~S.
  {McLean}, S.~K. {Ramsay}, \& H.~{Takami}, 844606, \dodoi{10.1117/12.925516}

\bibitem[{{Catelan} {et~al.}(2004){Catelan}, {Pritzl}, \& {Smith}}]{catelan04}
{Catelan}, M., {Pritzl}, B.~J., \& {Smith}, H.~A. 2004, \apjs, 154, 633,
  \dodoi{10.1086/422916}

\bibitem[{{Catelan} \& {Smith}(2015)}]{catelan15}
{Catelan}, M., \& {Smith}, H.~A. 2015, {Pulsating Stars}

\bibitem[{{Chadid} \& {Preston}(2013)}]{chadid13}
{Chadid}, M., \& {Preston}, G.~W. 2013, \mnras, 434, 552,
  \dodoi{10.1093/mnras/stt1040}

\bibitem[{{Chen} {et~al.}(2018){Chen}, {Yao}, {Pang}, {Yi}, {Lu}, {Liu},
  {Nong}, {Liang}, {Liang}, {Ma}, {Wu}, {Gan}, \& {Zou}}]{chenqso}
{Chen}, Z.-F., {Yao}, M., {Pang}, T.-T., {et~al.} 2018, \apjs, 239, 23,
  \dodoi{10.3847/1538-4365/aaeac3}

\bibitem[{{Churilov} {et~al.}(2018){Churilov}, {Zhelem}, {Case}, {Kondrat},
  {Fiegert}, {Waller}, {Lawrence}, {Edgar}, {Baker}, \& {Ireland}}]{churilov18}
{Churilov}, V., {Zhelem}, R., {Case}, S., {et~al.} 2018, in Society of
  Photo-Optical Instrumentation Engineers (SPIE) Conference Series, Vol. 10702,
  Ground-based and Airborne Instrumentation for Astronomy VII, ed. C.~J.
  {Evans}, L.~{Simard}, \& H.~{Takami}, 1070269, \dodoi{10.1117/12.2312401}

\bibitem[{{Cool} {et~al.}(2005){Cool}, {Howell}, {Pe{\~n}a}, {Adamson}, \&
  {Thompson}}]{cool05}
{Cool}, R.~J., {Howell}, S.~B., {Pe{\~n}a}, M., {Adamson}, A.~J., \&
  {Thompson}, R.~R. 2005, \pasp, 117, 462, \dodoi{10.1086/429701}

\bibitem[{{DESI Collaboration} {et~al.}(2023{\natexlab{a}}){DESI
  Collaboration}, {Adame}, {Aguilar}, {Ahlen}, {Alam}, {Aldering}, {Alexander},
  {Alfarsy}, {Allende Prieto}, {Alvarez}, {Alves}, {Anand}, {Andrade-Oliveira},
  {Armengaud}, {Asorey}, {Avila}, {Aviles}, {Bailey},
  {Balaguera-Antol{\'\i}nez}, {Ballester}, {Baltay}, {Bault}, {Bautista},
  {Behera}, {Beltran}, {BenZvi}, {Beraldo e Silva}, {Bermejo-Climent}, {Berti},
  {Besuner}, {Beutler}, {Bianchi}, {Blake}, {Blum}, {Bolton}, {Brieden},
  {Brodzeller}, {Brooks}, {Brown}, {Buckley-Geer}, {Burtin}, {Cabayol-Garcia},
  {Cai}, {Canning}, {Cardiel-Sas}, {Carnero Rosell}, {Castander},
  {Cervantes-Cota}, {Chabanier}, {Chaussidon}, {Chaves-Montero}, {Chen},
  {Chuang}, {Claybaugh}, {Cole}, {Cooper}, {Cuceu}, {Davis}, {Dawson}, {de
  Belsunce}, {de la Cruz}, {de la Macorra}, {de Mattia}, {Demina},
  {Demirbozan}, {DeRose}, {Dey}, {Dey}, {Dhungana}, {Ding}, {Ding}, {Doel},
  {Doshi}, {Douglass}, {Edge}, {Eftekharzadeh}, {Eisenstein}, {Elliott},
  {Escoffier}, {Fagrelius}, {Fan}, {Fanning}, {Fawcett}, {Ferraro}, {Ereza},
  {Flaugher}, {Font-Ribera}, {Forero-S{\'a}nchez}, {Forero-Romero}, {Frenk},
  {G{\"a}nsicke}, {Garc{\'\i}a}, {Garc{\'\i}a-Bellido}, {Garcia-Quintero},
  {Garrison}, {Gil-Mar{\'\i}n}, {Golden-Marx}, {Gontcho}, {Gonzalez-Morales},
  {Gonzalez-Perez}, {Gordon}, {Graur}, {Green}, {Gruen}, {Guy}, {Hadzhiyska},
  {Hahn}, {Han}, {Hanif}, {Herrera-Alcantar}, {Honscheid}, {Hou}, {Howlett},
  {Huterer}, {Ir{\v{s}}i{\v{c}}}, {Ishak}, {Jacques}, {Jana}, {Jiang},
  {Jimenez}, {Jing}, {Joudaki}, {Jullo}, {Juneau}, {Kizhuprakkat},
  {Kara{\c{c}}ayl{\i}}, {Karim}, {Kehoe}, {Kent}, {Khederlarian}, {Kim},
  {Kirkby}, {Kisner}, {Kitaura}, {Kneib}, {Koposov}, {Kov{\'a}cs}, {Kremin},
  {Krolewski}, {L'Huillier}, {Lambert}, {Lamman}, {Lan}, {Landriau}, {Lang},
  {Lange}, {Lasker}, {Le Guillou}, {Leauthaud}, {Levi}, {Li}, {Linder},
  {Lyons}, {Magneville}, {Manera}, {Manser}, {Margala}, {Martini}, {McDonald},
  {Medina}, {Medina-Varela}, {Meisner}, {Mena-Fern{\'a}ndez}, {Meneses-Rizo},
  {Mezcua}, {Miquel}, {Montero-Camacho}, {Moon}, {Moore}, {Moustakas},
  {Mueller}, {Mundet}, {Mu{\~n}oz-Guti{\'e}rrez}, {Myers}, {Nadathur},
  {Napolitano}, {Neveux}, {Newman}, {Nie}, {Nikutta}, {Niz}, {Norberg},
  {Noriega}, {Paillas}, {Palanque-Delabrouille}, {Palmese}, {Zhiwei},
  {Parkinson}, {Penmetsa}, {Percival}, {P{\'e}rez-Fern{\'a}ndez},
  {P{\'e}rez-R{\`a}fols}, {Pieri}, {Poppett}, {Porredon}, {Pothier}, {Prada},
  {Pucha}, {Raichoor}, {Ram{\'\i}rez-P{\'e}rez}, {Ramirez-Solano},
  {Rashkovetskyi}, {Ravoux}, {Rocher}, {Rockosi}, {Ross}, {Rossi}, {Ruggeri},
  {Ruhlmann-Kleider}, {Sabiu}, {Said}, {Saintonge}, {Samushia}, {Sanchez},
  {Saulder}, {Schaan}, {Schlafly}, {Schlegel}, {Scholte}, {Schubnell}, {Seo},
  {Shafieloo}, {Sharples}, {Sheu}, {Silber}, {Sinigaglia}, {Siudek}, {Slepian},
  {Smith}, {Sprayberry}, {Stephey}, {Su{\'a}rez-P{\'e}rez}, {Sun}, {Tan},
  {Tarl{\'e}}, {Tojeiro}, {Ure{\~n}a-L{\'o}pez}, {Vaisakh}, {Valcin}, {Valdes},
  {Valluri}, {Vargas-Maga{\~n}a}, {Variu}, {Verde}, {Walther}, {Wang}, {Wang},
  {Weaver}, {Weaverdyck}, {Wechsler}, {White}, {Xie}, {Yang}, {Y{\`e}che},
  {Yu}, {Yuan}, {Zhang}, {Zhang}, {Zhao}, {Zheng}, {Zhou}, {Zhou}, {Zou},
  {Zou}, \& {Zu}}]{desi}
{DESI Collaboration}, {Adame}, A.~G., {Aguilar}, J., {et~al.}
  2023{\natexlab{a}}, arXiv e-prints, arXiv:2306.06308,
  \dodoi{10.48550/arXiv.2306.06308}

\bibitem[{{DESI Collaboration} {et~al.}(2023{\natexlab{b}}){DESI
  Collaboration}, {Adame}, {Aguilar}, {Ahlen}, {Alam}, {Aldering}, {Alexander},
  {Alfarsy}, {Allende Prieto}, {Alvarez}, {Alves}, {Anand}, {Andrade-Oliveira},
  {Armengaud}, {Asorey}, {Avila}, {Aviles}, {Bailey},
  {Balaguera-Antol{\'\i}nez}, {Ballester}, {Baltay}, {Bault}, {Bautista},
  {Behera}, {Beltran}, {BenZvi}, {Beraldo e Silva}, {Bermejo-Climent}, {Berti},
  {Besuner}, {Beutler}, {Bianchi}, {Blake}, {Blum}, {Bolton}, {Brieden},
  {Brodzeller}, {Brooks}, {Brown}, {Buckley-Geer}, {Burtin}, {Cabayol-Garcia},
  {Cai}, {Canning}, {Cardiel-Sas}, {Carnero Rosell}, {Castander},
  {Cervantes-Cota}, {Chabanier}, {Chaussidon}, {Chaves-Montero}, {Chen},
  {Chuang}, {Claybaugh}, {Cole}, {Cooper}, {Cuceu}, {Davis}, {Dawson}, {de
  Belsunce}, {de la Cruz}, {de la Macorra}, {de Mattia}, {Demina},
  {Demirbozan}, {DeRose}, {Dey}, {Dey}, {Dhungana}, {Ding}, {Ding}, {Doel},
  {Doshi}, {Douglass}, {Edge}, {Eftekharzadeh}, {Eisenstein}, {Elliott},
  {Escoffier}, {Fagrelius}, {Fan}, {Fanning}, {Fawcett}, {Ferraro}, {Ereza},
  {Flaugher}, {Font-Ribera}, {Forero-S{\'a}nchez}, {Forero-Romero}, {Frenk},
  {G{\"a}nsicke}, {Garc{\'\i}a}, {Garc{\'\i}a-Bellido}, {Garcia-Quintero},
  {Garrison}, {Gil-Mar{\'\i}n}, {Golden-Marx}, {Gontcho}, {Gonzalez-Morales},
  {Gonzalez-Perez}, {Gordon}, {Graur}, {Green}, {Gruen}, {Guy}, {Hadzhiyska},
  {Hahn}, {Han}, {Hanif}, {Herrera-Alcantar}, {Honscheid}, {Hou}, {Howlett},
  {Huterer}, {Ir{\v{s}}i{\v{c}}}, {Ishak}, {Jacques}, {Jana}, {Jiang},
  {Jimenez}, {Jing}, {Joudaki}, {Jullo}, {Juneau}, {Kizhuprakkat},
  {Kara{\c{c}}ayl{\i}}, {Karim}, {Kehoe}, {Kent}, {Khederlarian}, {Kim},
  {Kirkby}, {Kisner}, {Kitaura}, {Kneib}, {Koposov}, {Kov{\'a}cs}, {Kremin},
  {Krolewski}, {L'Huillier}, {Lambert}, {Lamman}, {Lan}, {Landriau}, {Lang},
  {Lange}, {Lasker}, {Le Guillou}, {Leauthaud}, {Levi}, {Li}, {Linder},
  {Lyons}, {Magneville}, {Manera}, {Manser}, {Margala}, {Martini}, {McDonald},
  {Medina}, {Medina-Varela}, {Meisner}, {Mena-Fern{\'a}ndez}, {Meneses-Rizo},
  {Mezcua}, {Miquel}, {Montero-Camacho}, {Moon}, {Moore}, {Moustakas},
  {Mueller}, {Mundet}, {Mu{\~n}oz-Guti{\'e}rrez}, {Myers}, {Nadathur},
  {Napolitano}, {Neveux}, {Newman}, {Nie}, {Nikutta}, {Niz}, {Norberg},
  {Noriega}, {Paillas}, {Palanque-Delabrouille}, {Palmese}, {Zhiwei},
  {Parkinson}, {Penmetsa}, {Percival}, {P{\'e}rez-Fern{\'a}ndez},
  {P{\'e}rez-R{\`a}fols}, {Pieri}, {Poppett}, {Porredon}, {Pothier}, {Prada},
  {Pucha}, {Raichoor}, {Ram{\'\i}rez-P{\'e}rez}, {Ramirez-Solano},
  {Rashkovetskyi}, {Ravoux}, {Rocher}, {Rockosi}, {Ross}, {Rossi}, {Ruggeri},
  {Ruhlmann-Kleider}, {Sabiu}, {Said}, {Saintonge}, {Samushia}, {Sanchez},
  {Saulder}, {Schaan}, {Schlafly}, {Schlegel}, {Scholte}, {Schubnell}, {Seo},
  {Shafieloo}, {Sharples}, {Sheu}, {Silber}, {Sinigaglia}, {Siudek}, {Slepian},
  {Smith}, {Sprayberry}, {Stephey}, {Su{\'a}rez-P{\'e}rez}, {Sun}, {Tan},
  {Tarl{\'e}}, {Tojeiro}, {Ure{\~n}a-L{\'o}pez}, {Vaisakh}, {Valcin}, {Valdes},
  {Valluri}, {Vargas-Maga{\~n}a}, {Variu}, {Verde}, {Walther}, {Wang}, {Wang},
  {Weaver}, {Weaverdyck}, {Wechsler}, {White}, {Xie}, {Yang}, {Y{\`e}che},
  {Yu}, {Yuan}, {Zhang}, {Zhang}, {Zhao}, {Zheng}, {Zhou}, {Zhou}, {Zou},
  {Zou}, \& {Zu}}]{desiedr}
---. 2023{\natexlab{b}}, arXiv e-prints, arXiv:2306.06308,
  \dodoi{10.48550/arXiv.2306.06308}

\bibitem[{{Donati} \& {Landstreet}(2009)}]{donati}
{Donati}, J.~F., \& {Landstreet}, J.~D. 2009, \araa, 47, 333,
  \dodoi{10.1146/annurev-astro-082708-101833}

\bibitem[{{Dovgal} {et~al.}(2024){Dovgal}, {Venn}, {Sestito}, {Hayes},
  {McConnachie}, {Navarro}, {Placco}, {Starkenburg}, {Martin}, {Pazder},
  {Chiboucas}, {Deibert}, {Gamen}, {Heo}, {Kalari}, {Martioli}, {Xu}, {Diaz},
  {Gomez-Jimenez}, {Henderson}, {Prado}, {Quiroz}, {Robertson}, {Ruiz-Carmona},
  {Simpson}, {Urrutia}, {Waller}, {Berg}, {Burley}, {Hartman}, {Ireland},
  {Margheim}, {Perez}, \& {Thomas-Osip}}]{sv1}
{Dovgal}, A., {Venn}, K.~A., {Sestito}, F., {et~al.} 2024, \mnras, 527, 7810,
  \dodoi{10.1093/mnras/stad3673}

\bibitem[{{Duncan}(1991)}]{duncan}
{Duncan}, D.~K. 1991, \apj, 373, 250, \dodoi{10.1086/170044}

\bibitem[{{Fan} {et~al.}(2023){Fan}, {Ba{\~n}ados}, \& {Simcoe}}]{quasarigm}
{Fan}, X., {Ba{\~n}ados}, E., \& {Simcoe}, R.~A. 2023, \araa, 61, 373,
  \dodoi{10.1146/annurev-astro-052920-102455}

\bibitem[{{Fedele} {et~al.}(2010){Fedele}, {van den Ancker}, {Henning},
  {Jayawardhana}, \& {Oliveira}}]{fedeledisc}
{Fedele}, D., {van den Ancker}, M.~E., {Henning}, T., {Jayawardhana}, R., \&
  {Oliveira}, J.~M. 2010, \aap, 510, A72, \dodoi{10.1051/0004-6361/200912810}

\bibitem[{{Flores} {et~al.}(2022){Flores}, {Connelley}, {Reipurth}, \&
  {Duch{\^e}ne}}]{flores22}
{Flores}, C., {Connelley}, M.~S., {Reipurth}, B., \& {Duch{\^e}ne}, G. 2022,
  \apj, 925, 21, \dodoi{10.3847/1538-4357/ac37bd}

\bibitem[{{Flores} {et~al.}(2020){Flores}, {Reipurth}, \&
  {Connelley}}]{flores20}
{Flores}, C., {Reipurth}, B., \& {Connelley}, M.~S. 2020, \apj, 898, 109,
  \dodoi{10.3847/1538-4357/ab9e67}

\bibitem[{{Frebel} \& {Norris}(2015{\natexlab{a}})}]{frebel}
{Frebel}, A., \& {Norris}, J.~E. 2015{\natexlab{a}}, \araa, 53, 631,
  \dodoi{10.1146/annurev-astro-082214-122423}

\bibitem[{{Frebel} \& {Norris}(2015{\natexlab{b}})}]{nearcosmology}
---. 2015{\natexlab{b}}, \araa, 53, 631,
  \dodoi{10.1146/annurev-astro-082214-122423}

\bibitem[{{Gaia Collaboration} {et~al.}(2016){Gaia Collaboration}, {Prusti},
  {de Bruijne}, {Brown}, {Vallenari}, {Babusiaux}, {Bailer-Jones}, {Bastian},
  {Biermann}, {Evans}, {Eyer}, {Jansen}, {Jordi}, {Klioner}, {Lammers},
  {Lindegren}, {Luri}, {Mignard}, {Milligan}, {Panem}, {Poinsignon},
  {Pourbaix}, {Randich}, {Sarri}, {Sartoretti}, {Siddiqui}, {Soubiran},
  {Valette}, {van Leeuwen}, {Walton}, {Aerts}, {Arenou}, {Cropper}, {Drimmel},
  {H{\o}g}, {Katz}, {Lattanzi}, {O'Mullane}, {Grebel}, {Holland}, {Huc},
  {Passot}, {Bramante}, {Cacciari}, {Casta{\~n}eda}, {Chaoul}, {Cheek}, {De
  Angeli}, {Fabricius}, {Guerra}, {Hern{\'a}ndez}, {Jean-Antoine-Piccolo},
  {Masana}, {Messineo}, {Mowlavi}, {Nienartowicz}, {Ord{\'o}{\~n}ez-Blanco},
  {Panuzzo}, {Portell}, {Richards}, {Riello}, {Seabroke}, {Tanga},
  {Th{\'e}venin}, {Torra}, {Els}, {Gracia-Abril}, {Comoretto},
  {Garcia-Reinaldos}, {Lock}, {Mercier}, {Altmann}, {Andrae}, {Astraatmadja},
  {Bellas-Velidis}, {Benson}, {Berthier}, {Blomme}, {Busso}, {Carry},
  {Cellino}, {Clementini}, {Cowell}, {Creevey}, {Cuypers}, {Davidson}, {De
  Ridder}, {de Torres}, {Delchambre}, {Dell'Oro}, {Ducourant}, {Fr{\'e}mat},
  {Garc{\'\i}a-Torres}, {Gosset}, {Halbwachs}, {Hambly}, {Harrison}, {Hauser},
  {Hestroffer}, {Hodgkin}, {Huckle}, {Hutton}, {Jasniewicz}, {Jordan},
  {Kontizas}, {Korn}, {Lanzafame}, {Manteiga}, {Moitinho}, {Muinonen},
  {Osinde}, {Pancino}, {Pauwels}, {Petit}, {Recio-Blanco}, {Robin}, {Sarro},
  {Siopis}, {Smith}, {Smith}, {Sozzetti}, {Thuillot}, {van Reeven}, {Viala},
  {Abbas}, {Abreu Aramburu}, {Accart}, {Aguado}, {Allan}, {Allasia},
  {Altavilla}, {{\'A}lvarez}, {Alves}, {Anderson}, {Andrei}, {Anglada Varela},
  {Antiche}, {Antoja}, {Ant{\'o}n}, {Arcay}, {Atzei}, {Ayache}, {Bach},
  {Baker}, {Balaguer-N{\'u}{\~n}ez}, {Barache}, {Barata}, {Barbier}, {Barblan},
  {Baroni}, {Barrado y Navascu{\'e}s}, {Barros}, {Barstow}, {Becciani},
  {Bellazzini}, {Bellei}, {Bello Garc{\'\i}a}, {Belokurov}, {Bendjoya},
  {Berihuete}, {Bianchi}, {Bienaym{\'e}}, {Billebaud}, {Blagorodnova},
  {Blanco-Cuaresma}, {Boch}, {Bombrun}, {Borrachero}, {Bouquillon}, {Bourda},
  {Bouy}, {Bragaglia}, {Breddels}, {Brouillet}, {Br{\"u}semeister},
  {Bucciarelli}, {Budnik}, {Burgess}, {Burgon}, {Burlacu}, {Busonero}, {Buzzi},
  {Caffau}, {Cambras}, {Campbell}, {Cancelliere}, {Cantat-Gaudin}, {Carlucci},
  {Carrasco}, {Castellani}, {Charlot}, {Charnas}, {Charvet}, {Chassat},
  {Chiavassa}, {Clotet}, {Cocozza}, {Collins}, {Collins}, {Costigan}, {Crifo},
  {Cross}, {Crosta}, {Crowley}, {Dafonte}, {Damerdji}, {Dapergolas}, {David},
  {David}, {De Cat}, {de Felice}, {de Laverny}, {De Luise}, {De March}, {de
  Martino}, {de Souza}, {Debosscher}, {del Pozo}, {Delbo}, {Delgado},
  {Delgado}, {di Marco}, {Di Matteo}, {Diakite}, {Distefano}, {Dolding}, {Dos
  Anjos}, {Drazinos}, {Dur{\'a}n}, {Dzigan}, {Ecale}, {Edvardsson}, {Enke},
  {Erdmann}, {Escolar}, {Espina}, {Evans}, {Eynard Bontemps}, {Fabre},
  {Fabrizio}, {Faigler}, {Falc{\~a}o}, {Farr{\`a}s Casas}, {Faye}, {Federici},
  {Fedorets}, {Fern{\'a}ndez-Hern{\'a}ndez}, {Fernique}, {Fienga}, {Figueras},
  {Filippi}, {Findeisen}, {Fonti}, {Fouesneau}, {Fraile}, {Fraser}, {Fuchs},
  {Furnell}, {Gai}, {Galleti}, {Galluccio}, {Garabato}, {Garc{\'\i}a-Sedano},
  {Gar{\'e}}, {Garofalo}, {Garralda}, {Gavras}, {Gerssen}, {Geyer}, {Gilmore},
  {Girona}, {Giuffrida}, {Gomes}, {Gonz{\'a}lez-Marcos},
  {Gonz{\'a}lez-N{\'u}{\~n}ez}, {Gonz{\'a}lez-Vidal}, {Granvik}, {Guerrier},
  {Guillout}, {Guiraud}, {G{\'u}rpide}, {Guti{\'e}rrez-S{\'a}nchez}, {Guy},
  {Haigron}, {Hatzidimitriou}, {Haywood}, {Heiter}, {Helmi}, {Hobbs},
  {Hofmann}, {Holl}, {Holland}, {Hunt}, {Hypki}, {Icardi}, {Irwin}, {Jevardat
  de Fombelle}, {Jofr{\'e}}, {Jonker}, {Jorissen}, {Julbe}, {Karampelas},
  {Kochoska}, {Kohley}, {Kolenberg}, {Kontizas}, {Koposov}, {Kordopatis},
  {Koubsky}, {Kowalczyk}, {Krone-Martins}, {Kudryashova}, {Kull}, {Bachchan},
  {Lacoste-Seris}, {Lanza}, {Lavigne}, {Le Poncin-Lafitte}, {Lebreton},
  {Lebzelter}, {Leccia}, {Leclerc}, {Lecoeur-Taibi}, {Lemaitre}, {Lenhardt},
  {Leroux}, {Liao}, {Licata}, {Lindstr{\o}m}, {Lister}, {Livanou}, {Lobel},
  {L{\"o}ffler}, {L{\'o}pez}, {Lopez-Lozano}, {Lorenz}, {Loureiro},
  {MacDonald}, {Magalh{\~a}es Fernandes}, {Managau}, {Mann}, {Mantelet},
  {Marchal}, {Marchant}, {Marconi}, {Marie}, {Marinoni}, {Marrese},
  {Marschalk{\'o}}, {Marshall}, {Mart{\'\i}n-Fleitas}, {Martino}, {Mary},
  {Matijevi{\v{c}}}, {Mazeh}, {McMillan}, {Messina}, {Mestre}, {Michalik},
  {Millar}, {Miranda}, {Molina}, {Molinaro}, {Molinaro}, {Moln{\'a}r},
  {Moniez}, {Montegriffo}, {Monteiro}, {Mor}, {Mora}, {Morbidelli}, {Morel},
  {Morgenthaler}, {Morley}, {Morris}, {Mulone}, {Muraveva}, {Musella},
  {Narbonne}, {Nelemans}, {Nicastro}, {Noval}, {Ord{\'e}novic},
  {Ordieres-Mer{\'e}}, {Osborne}, {Pagani}, {Pagano}, {Pailler}, {Palacin},
  {Palaversa}, {Parsons}, {Paulsen}, {Pecoraro}, {Pedrosa}, {Pentik{\"a}inen},
  {Pereira}, {Pichon}, {Piersimoni}, {Pineau}, {Plachy}, {Plum}, {Poujoulet},
  {Pr{\v{s}}a}, {Pulone}, {Ragaini}, {Rago}, {Rambaux}, {Ramos-Lerate},
  {Ranalli}, {Rauw}, {Read}, {Regibo}, {Renk}, {Reyl{\'e}}, {Ribeiro},
  {Rimoldini}, {Ripepi}, {Riva}, {Rixon}, {Roelens}, {Romero-G{\'o}mez},
  {Rowell}, {Royer}, {Rudolph}, {Ruiz-Dern}, {Sadowski}, {Sagrist{\`a}
  Sell{\'e}s}, {Sahlmann}, {Salgado}, {Salguero}, {Sarasso}, {Savietto},
  {Schnorhk}, {Schultheis}, {Sciacca}, {Segol}, {Segovia}, {Segransan},
  {Serpell}, {Shih}, {Smareglia}, {Smart}, {Smith}, {Solano}, {Solitro},
  {Sordo}, {Soria Nieto}, {Souchay}, {Spagna}, {Spoto}, {Stampa}, {Steele},
  {Steidelm{\"u}ller}, {Stephenson}, {Stoev}, {Suess}, {S{\"u}veges}, {Surdej},
  {Szabados}, {Szegedi-Elek}, {Tapiador}, {Taris}, {Tauran}, {Taylor},
  {Teixeira}, {Terrett}, {Tingley}, {Trager}, {Turon}, {Ulla}, {Utrilla},
  {Valentini}, {van Elteren}, {Van Hemelryck}, {van Leeuwen}, {Varadi},
  {Vecchiato}, {Veljanoski}, {Via}, {Vicente}, {Vogt}, {Voss}, {Votruba},
  {Voutsinas}, {Walmsley}, {Weiler}, {Weingrill}, {Werner}, {Wevers},
  {Whitehead}, {Wyrzykowski}, {Yoldas}, {{\v{Z}}erjal}, {Zucker}, {Zurbach},
  {Zwitter}, {Alecu}, {Allen}, {Allende Prieto}, {Amorim},
  {Anglada-Escud{\'e}}, {Arsenijevic}, {Azaz}, {Balm}, {Beck}, {Bernstein},
  {Bigot}, {Bijaoui}, {Blasco}, {Bonfigli}, {Bono}, {Boudreault}, {Bressan},
  {Brown}, {Brunet}, {Bunclark}, {Buonanno}, {Butkevich}, {Carret}, {Carrion},
  {Chemin}, {Ch{\'e}reau}, {Corcione}, {Darmigny}, {de Boer}, {de Teodoro}, {de
  Zeeuw}, {Delle Luche}, {Domingues}, {Dubath}, {Fodor}, {Fr{\'e}zouls},
  {Fries}, {Fustes}, {Fyfe}, {Gallardo}, {Gallegos}, {Gardiol}, {Gebran},
  {Gomboc}, {G{\'o}mez}, {Grux}, {Gueguen}, {Heyrovsky}, {Hoar}, {Iannicola},
  {Isasi Parache}, {Janotto}, {Joliet}, {Jonckheere}, {Keil}, {Kim},
  {Klagyivik}, {Klar}, {Knude}, {Kochukhov}, {Kolka}, {Kos}, {Kutka}, {Lainey},
  {LeBouquin}, {Liu}, {Loreggia}, {Makarov}, {Marseille}, {Martayan},
  {Martinez-Rubi}, {Massart}, {Meynadier}, {Mignot}, {Munari}, {Nguyen},
  {Nordlander}, {Ocvirk}, {O'Flaherty}, {Olias Sanz}, {Ortiz}, {Osorio},
  {Oszkiewicz}, {Ouzounis}, {Palmer}, {Park}, {Pasquato}, {Peltzer}, {Peralta},
  {P{\'e}turaud}, {Pieniluoma}, {Pigozzi}, {Poels}, {Prat}, {Prod'homme},
  {Raison}, {Rebordao}, {Risquez}, {Rocca-Volmerange}, {Rosen}, {Ruiz-Fuertes},
  {Russo}, {Sembay}, {Serraller Vizcaino}, {Short}, {Siebert}, {Silva},
  {Sinachopoulos}, {Slezak}, {Soffel}, {Sosnowska}, {Strai{\v{z}}ys}, {ter
  Linden}, {Terrell}, {Theil}, {Tiede}, {Troisi}, {Tsalmantza}, {Tur},
  {Vaccari}, {Vachier}, {Valles}, {Van Hamme}, {Veltz}, {Virtanen}, {Wallut},
  {Wichmann}, {Wilkinson}, {Ziaeepour}, \& {Zschocke}}]{gaia}
{Gaia Collaboration}, {Prusti}, T., {de Bruijne}, J.~H.~J., {et~al.} 2016,
  \aap, 595, A1, \dodoi{10.1051/0004-6361/201629272}

\bibitem[{{Garcia} {et~al.}(2019){Garcia}, {Herrero}, {Najarro}, {Camacho}, \&
  {Lorenzo}}]{garcia19}
{Garcia}, M., {Herrero}, A., {Najarro}, F., {Camacho}, I., \& {Lorenzo}, M.
  2019, \mnras, 484, 422, \dodoi{10.1093/mnras/sty3503}

\bibitem[{{Goswami} {et~al.}(2001){Goswami}, {Rao}, \& {Lambert}}]{goswami01}
{Goswami}, A., {Rao}, N.~K., \& {Lambert}, D.~L. 2001, The Observatory, 121, 97

\bibitem[{{Haisch} {et~al.}(2001){Haisch}, {Lada}, \& {Lada}}]{haischdisc}
{Haisch}, Karl~E., J., {Lada}, E.~A., \& {Lada}, C.~J. 2001, \apjl, 553, L153,
  \dodoi{10.1086/320685}

\bibitem[{{Hamuy} {et~al.}(1994){Hamuy}, {Suntzeff}, {Heathcote}, {Walker},
  {Gigoux}, \& {Phillips}}]{hamuy1994}
{Hamuy}, M., {Suntzeff}, N.~B., {Heathcote}, S.~R., {et~al.} 1994, \pasp, 106,
  566, \dodoi{10.1086/133417}

\bibitem[{{Hauschildt} \& {Baron}(1999)}]{Hauschildt1999}
{Hauschildt}, P.~H., \& {Baron}, E. 1999, Journal of Computational and Applied
  Mathematics, 109, 41, \dodoi{10.48550/arXiv.astro-ph/9808182}

\bibitem[{{Hayes} {et~al.}(2022){Hayes}, {Waller}, {Ireland}, {Nielsen},
  {White}, {Bento}, {Venn}, {Pazder}, {McConnachie}, {Simpson}, \&
  {Labrie}}]{hayes22}
{Hayes}, C.~R., {Waller}, F., {Ireland}, M., {et~al.} 2022, in Society of
  Photo-Optical Instrumentation Engineers (SPIE) Conference Series, Vol. 12184,
  Ground-based and Airborne Instrumentation for Astronomy IX, ed. C.~J.
  {Evans}, J.~J. {Bryant}, \& K.~{Motohara}, 121846H,
  \dodoi{10.1117/12.2642905}

\bibitem[{{Hayes} {et~al.}(2023){Hayes}, {Venn}, {Waller}, {Jensen},
  {McConnachie}, {Pazder}, {Sestito}, {Anthony}, {Baker}, {Bassett}, {Bento},
  {Berg}, {Burley}, {Brzeski}, {Case}, {Chapin}, {Chin}, {Chisholm},
  {Churilov}, {Densmore}, {Diaz}, {Dunn}, {Edgar}, {Farrell}, {Firpo},
  {Fitzsimmons}, {Font-Serra}, {Fuentes}, {Ganton}, {Gomez-Jimenez}, {Hardy},
  {Henderson}, {Hill}, {Hoff}, {Ireland}, {Kalari}, {Kelly}, {Klauser},
  {Kondrat}, {Labrie}, {Lambert}, {Luvaul}, {Lawrence}, {Lothrop}, {Macdonald},
  {Mali}, {Margheim}, {McDermid}, {McGregor}, {Miller}, {Miranda}, {Muller},
  {Nielsen}, {Norbury}, {Oberdorf}, {Pai}, {Perez}, {Prado}, {Price}, {Quiroz},
  {Reshetov}, {Robertson}, {Ruiz-Carmona}, {Salinas}, {Sebo}, {Sheinis},
  {Shetrone}, {Shortridge}, {Silversides}, {Silva}, {Simpson}, {Smith},
  {Szeto}, {Tims}, {Toro}, {Urrutia}, {Venkatesan}, {Waller}, {Wevers},
  {Wierzbicki}, {White}, {Young}, \& {Zhelem}}]{hayes}
{Hayes}, C.~R., {Venn}, K.~A., {Waller}, F., {et~al.} 2023, \apj, 955, 17,
  \dodoi{10.3847/1538-4357/acebc0}

\bibitem[{{Herbig}(1978)}]{herbig78}
{Herbig}, G.~H. 1978, in Problems of Physics and Evolution of the Universe, ed.
  L.~V. {Mirzoyan}, 171

\bibitem[{{Herczeg} {et~al.}(2023){Herczeg}, {Chen}, {Donati}, {Dupree},
  {Walter}, {Hillenbrand}, {Johns-Krull}, {Manara}, {G{\"u}nther}, {Fang},
  {Schneider}, {Valenti}, {Alencar}, {Venuti}, {Alcal{\'a}}, {Frasca},
  {Arulanantham}, {Linsky}, {Bouvier}, {Brickhouse}, {Calvet}, {Espaillat},
  {Campbell-White}, {Carpenter}, {Chang}, {Cruz}, {Dahm}, {Eisl{\"o}ffel},
  {Edwards}, {Fischer}, {Guo}, {Henning}, {Ji}, {Jose}, {Kastner}, {Launhardt},
  {Principe}, {Robinson}, {Serna}, {Siwak}, {Sterzik}, \& {Takasao}}]{herc23}
{Herczeg}, G.~J., {Chen}, Y., {Donati}, J.-F., {et~al.} 2023, \apj, 956, 102,
  \dodoi{10.3847/1538-4357/acf468}

\bibitem[{{Hirst} \& {Cardenes}(2016)}]{archive}
{Hirst}, P., \& {Cardenes}, R. 2016, in Society of Photo-Optical
  Instrumentation Engineers (SPIE) Conference Series, Vol. 9913, Software and
  Cyberinfrastructure for Astronomy IV, ed. G.~{Chiozzi} \& J.~C. {Guzman},
  99131E, \dodoi{10.1117/12.2231833}

\bibitem[{{Howell} {et~al.}(2009){Howell}, {Johnson}, \& {Adamson}}]{howell09}
{Howell}, S.~B., {Johnson}, K.~J., \& {Adamson}, A.~J. 2009, \pasp, 121, 16,
  \dodoi{10.1086/597139}

\bibitem[{{Ireland} {et~al.}(2014){Ireland}, {Anthony}, {Burley}, {Chisholm},
  {Churilov}, {Dunn}, {Frost}, {Lawrence}, {Loop}, {McGregor}, {Martell},
  {McConnachie}, {McDermid}, {Pazder}, {Reshetov}, {Robertson}, {Sheinis},
  {Tims}, {Young}, \& {Zhelem}}]{ireland14}
{Ireland}, M., {Anthony}, A., {Burley}, G., {et~al.} 2014, in Society of
  Photo-Optical Instrumentation Engineers (SPIE) Conference Series, Vol. 9147,
  Ground-based and Airborne Instrumentation for Astronomy V, ed. S.~K.
  {Ramsay}, I.~S. {McLean}, \& H.~{Takami}, 91471J, \dodoi{10.1117/12.2057356}

\bibitem[{{Ireland} {et~al.}(2018){Ireland}, {White}, {Bento}, {Farrell},
  {Labrie}, {Luvaul}, {Nielsen}, \& {Simpson}}]{ireland18}
{Ireland}, M.~J., {White}, M., {Bento}, J.~P., {et~al.} 2018, in Society of
  Photo-Optical Instrumentation Engineers (SPIE) Conference Series, Vol. 10707,
  Software and Cyberinfrastructure for Astronomy V, ed. J.~C. {Guzman} \&
  J.~{Ibsen}, 1070735, \dodoi{10.1117/12.2314418}

\bibitem[{{Ireland} {et~al.}(2012){Ireland}, {Barnes}, {Cochrane}, {Colless},
  {Connor}, {Horton}, {Gibson}, {Lawrence}, {Martell}, {McGregor}, {Nicolle},
  {Nield}, {Orr}, {Robertson}, {Ryder}, {Sheinis}, {Smith}, {Staszak}, {Tims},
  {Xavier}, {Young}, \& {Zheng}}]{ireland12}
{Ireland}, M.~J., {Barnes}, S., {Cochrane}, D., {et~al.} 2012, in Society of
  Photo-Optical Instrumentation Engineers (SPIE) Conference Series, Vol. 8446,
  Ground-based and Airborne Instrumentation for Astronomy IV, ed. I.~S.
  {McLean}, S.~K. {Ramsay}, \& H.~{Takami}, 844629, \dodoi{10.1117/12.925746}

\bibitem[{{Ireland} {et~al.}(2016){Ireland}, {Artigau}, {Burley}, {Edgar},
  {Margheim}, {Robertson}, {Pazder}, {McDermid}, \& {Zhelem}}]{ireland16}
{Ireland}, M.~J., {Artigau}, {\'E}., {Burley}, G., {et~al.} 2016, in Society of
  Photo-Optical Instrumentation Engineers (SPIE) Conference Series, Vol. 9908,
  Ground-based and Airborne Instrumentation for Astronomy VI, ed. C.~J.
  {Evans}, L.~{Simard}, \& H.~{Takami}, 99087A, \dodoi{10.1117/12.2233927}

\bibitem[{{Itoh} {et~al.}(2020){Itoh}, {Misawa}, {Horiuchi}, \&
  {Aoki}}]{itohqso}
{Itoh}, D., {Misawa}, T., {Horiuchi}, T., \& {Aoki}, K. 2020, \mnras, 499,
  3094, \dodoi{10.1093/mnras/staa2793}

\bibitem[{{Jofr{\'e}} {et~al.}(2019){Jofr{\'e}}, {Heiter}, \&
  {Soubiran}}]{abundances}
{Jofr{\'e}}, P., {Heiter}, U., \& {Soubiran}, C. 2019, \araa, 57, 571,
  \dodoi{10.1146/annurev-astro-091918-104509}

\bibitem[{{Kalari} {et~al.}(2024){Kalari}, {Seifahrt}, \& {Diaz}}]{kalari24}
{Kalari}, V., {Seifahrt}, A., \& {Diaz}, R. 2024, in Society of Photo-Optical
  Instrumentation Engineers (SPIE) Conference Series, Vol. 13096, Ground-based
  and Airborne Instrumentation for Astronomy X, ed. J.~J. {Bryant},
  K.~{Motohara}, \& J.~{Vernet}, 1309681

\bibitem[{Labrie {et~al.}(2023)Labrie, Simpson, Cardenes, Turner, Soraisam,
  Quint, Oberdorf, Placco, Berke, Smirnova, Conseil, Vacca, \&
  Thomas-Osip}]{dragons}
Labrie, K., Simpson, C., Cardenes, R., {et~al.} 2023, Research Notes of the
  AAS, 7, 214, \dodoi{10.3847/2515-5172/ad0044}

\bibitem[{{Lothrop} {et~al.}(2020){Lothrop}, {Hoff}, {MacDonald}, {Dunn},
  {Densmore}, {Pazder}, {Chapin}, {Burley}, {Anthony}, {Lambert}, \&
  {Reshetov}}]{lothrop}
{Lothrop}, J., {Hoff}, B., {MacDonald}, S., {et~al.} 2020, in Society of
  Photo-Optical Instrumentation Engineers (SPIE) Conference Series, Vol. 11447,
  Ground-based and Airborne Instrumentation for Astronomy VIII, ed. C.~J.
  {Evans}, J.~J. {Bryant}, \& K.~{Motohara}, 114473R,
  \dodoi{10.1117/12.2561629}

\bibitem[{{Lyke} {et~al.}(2020){Lyke}, {Higley}, {McLane}, {Schurhammer},
  {Myers}, {Ross}, {Dawson}, {Chabanier}, {Martini}, {Busca}, {Mas des
  Bourboux}, {Salvato}, {Streblyanska}, {Zarrouk}, {Burtin}, {Anderson},
  {Bautista}, {Bizyaev}, {Brandt}, {Brinkmann}, {Brownstein}, {Comparat},
  {Green}, {de la Macorra}, {Mu{\~n}oz Guti{\'e}rrez}, {Hou}, {Newman},
  {Palanque-Delabrouille}, {P{\^a}ris}, {Percival}, {Petitjean}, {Rich},
  {Rossi}, {Schneider}, {Smith}, {Vivek}, \& {Weaver}}]{sdss}
{Lyke}, B.~W., {Higley}, A.~N., {McLane}, J.~N., {et~al.} 2020, \apjs, 250, 8,
  \dodoi{10.3847/1538-4365/aba623}

\bibitem[{{Madhusudhan}(2019)}]{transmission}
{Madhusudhan}, N. 2019, \araa, 57, 617,
  \dodoi{10.1146/annurev-astro-081817-051846}

\bibitem[{{Martioli} {et~al.}(2022){Martioli}, {H{\'e}brard}, {Fouqu{\'e}},
  {Artigau}, {Donati}, {Cadieux}, {Bellotti}, {Lecavelier des Etangs}, {Doyon},
  {do Nascimento}, {Arnold}, {Carmona}, {Cook}, {Cortes-Zuleta}, {de Almeida},
  {Delfosse}, {Folsom}, {K{\"o}nig}, {Moutou}, {Ould-Elhkim}, {Petit},
  {Stassun}, {Vidotto}, {Vandal}, {Benneke}, {Boisse}, {Bonfils}, {Boyd},
  {Brasseur}, {Charbonneau}, {Cloutier}, {Collins}, {Cristofari}, {Crossfield},
  {D{\'\i}az}, {Fausnaugh}, {Figueira}, {Forveille}, {Furlan}, {Girardin},
  {Gnilka}, {Gomes da Silva}, {Gu}, {Guerra}, {Howell}, {Hussain}, {Jenkins},
  {Kiefer}, {Latham}, {Matson}, {Matthews}, {Morin}, {Naves}, {Ricker},
  {Seager}, {Takami}, {Twicken}, {Vanderburg}, {Vanderspek}, \&
  {Winn}}]{Martioli2022}
{Martioli}, E., {H{\'e}brard}, G., {Fouqu{\'e}}, P., {et~al.} 2022, \aap, 660,
  A86, \dodoi{10.1051/0004-6361/202142540}

\bibitem[{{Mayor} \& {Queloz}(1995{\natexlab{a}})}]{peg51}
{Mayor}, M., \& {Queloz}, D. 1995{\natexlab{a}}, \nat, 378, 355,
  \dodoi{10.1038/378355a0}

\bibitem[{{Mayor} \& {Queloz}(1995{\natexlab{b}})}]{Mayor1995}
---. 1995{\natexlab{b}}, \nat, 378, 355, \dodoi{10.1038/378355a0}

\bibitem[{{McConnachie} {et~al.}(2022){McConnachie}, {Hayes}, {Ireland},
  {Waller}, {Berg}, {Pazder}, {Margheim}, {Kalari}, {Farrell}, \&
  {Robertson}}]{mccon22}
{McConnachie}, A.~W., {Hayes}, C., {Ireland}, M., {et~al.} 2022, in Society of
  Photo-Optical Instrumentation Engineers (SPIE) Conference Series, Vol. 12184,
  Ground-based and Airborne Instrumentation for Astronomy IX, ed. C.~J.
  {Evans}, J.~J. {Bryant}, \& K.~{Motohara}, 121841E,
  \dodoi{10.1117/12.2630407}

\bibitem[{{McConnachie} {et~al.}(2024){McConnachie}, {Hayes}, {Robertson},
  {Pazder}, {Ireland}, {Burley}, {Churilov}, {Lothrop}, {Zhelem}, {Kalari},
  {Anthony}, {Baker}, {Berg}, {Chapin}, {Chin}, {Densmore}, {Diaz}, {Dunn},
  {Edgar}, {Farrell}, {Firpo}, {Fuentes}, {Gomez-Jimenez}, {Hardy},
  {Henderson}, {Hill}, {Labrie}, {Jensen}, {Lambert}, {Lawrence}, {Macdonald},
  {Margheim}, {Millar}, {Muller}, {Nielsen}, {P{\'e}rez}, {Quiroz},
  {Ruiz-Carmona}, {Sebo}, {Sestito}, {Silva}, {Simpson}, {Smith}, {Venkatesan},
  {Waller}, {Waller}, {Wevers}, {Venn}, \& {Young}}]{mcconn24}
{McConnachie}, A.~W., {Hayes}, C.~R., {Robertson}, J.~G., {et~al.} 2024, arXiv
  e-prints, arXiv:2401.07452, \dodoi{10.48550/arXiv.2401.07452}

\bibitem[{{Merrill}(1924)}]{merrill24}
{Merrill}, P.~W. 1924, \pasp, 36, 225

\bibitem[{{Merrill}(1951)}]{merrill51}
---. 1951, \apj, 114, 37, \dodoi{10.1086/145449}

\bibitem[{{Miller} \& {Norris}(2008)}]{bryan}
{Miller}, B.~W., \& {Norris}, R. 2008, in Society of Photo-Optical
  Instrumentation Engineers (SPIE) Conference Series, Vol. 7016, Observatory
  Operations: Strategies, Processes, and Systems II, ed. R.~J. {Brissenden} \&
  D.~R. {Silva}, 70160V, \dodoi{10.1117/12.790169}

\bibitem[{{Miller} {et~al.}(2020){Miller}, {Stephens}, \& {Nunez}}]{miller20}
{Miller}, B.~W., {Stephens}, A.~W., \& {Nunez}, A. 2020, in Astronomical
  Society of the Pacific Conference Series, Vol. 522, Astronomical Data
  Analysis Software and Systems XXVII, ed. P.~{Ballester}, J.~{Ibsen},
  M.~{Solar}, \& K.~{Shortridge}, 635

\bibitem[{{Napolitano} {et~al.}(2023){Napolitano}, {Pandey}, {Myers}, {Lan},
  {Anand}, {Aguilar}, {Ahlen}, {Alexander}, {Brooks}, {Canning}, {Circosta},
  {De La Macorra}, {Doel}, {Eftekharzadeh}, {Fawcett}, {Font-Ribera},
  {Garcia-Bellido}, {Gontcho A Gontcho}, {Le Guillou}, {Guy}, {Honscheid},
  {Juneau}, {Kisner}, {Landriau}, {Meisner}, {Miquel}, {Moustakas}, {Percival},
  {Prochaska}, {Schubnell}, {Tarl{\'e}}, {Weaver}, {Weiner}, {Zhou}, {Zou}, \&
  {Zou}}]{napolitano}
{Napolitano}, L., {Pandey}, A., {Myers}, A.~D., {et~al.} 2023, \aj, 166, 99,
  \dodoi{10.3847/1538-3881/ace62c}

\bibitem[{{Nielsen} {et~al.}(2018){Nielsen}, {Price}, {Young}, \&
  {Ireland}}]{nielsen}
{Nielsen}, J.~G., {Price}, I.~A., {Young}, P.~J., \& {Ireland}, M.~J. 2018, in
  Society of Photo-Optical Instrumentation Engineers (SPIE) Conference Series,
  Vol. 10707, Software and Cyberinfrastructure for Astronomy V, ed. J.~C.
  {Guzman} \& J.~{Ibsen}, 1070706, \dodoi{10.1117/12.2313068}

\bibitem[{{N{\'u}{\~n}ez} \& {Walker}(2008)}]{nunez}
{N{\'u}{\~n}ez}, A., \& {Walker}, S. 2008, in Society of Photo-Optical
  Instrumentation Engineers (SPIE) Conference Series, Vol. 7019, Advanced
  Software and Control for Astronomy II, ed. A.~{Bridger} \& N.~M. {Radziwill},
  70190S, \dodoi{10.1117/12.787712}

\bibitem[{{Pazder} {et~al.}(2016){Pazder}, {Burley}, {Ireland}, {Robertson},
  {Sheinis}, \& {Zhelem}}]{pazder16}
{Pazder}, J., {Burley}, G., {Ireland}, M.~J., {et~al.} 2016, in Society of
  Photo-Optical Instrumentation Engineers (SPIE) Conference Series, Vol. 9908,
  Ground-based and Airborne Instrumentation for Astronomy VI, ed. C.~J.
  {Evans}, L.~{Simard}, \& H.~{Takami}, 99087F, \dodoi{10.1117/12.2234366}

\bibitem[{{Pazder} {et~al.}(2020){Pazder}, {McConnachie}, {Ireland}, {Anthony},
  {Bassett}, {Burley}, {Chapin}, {Churilov}, {Densmore}, {Dunn}, {Farrell},
  {Henderson}, {Hoff}, {Lambert}, {Lothrop}, {MacDonald}, {Margheim},
  {Reshetov}, {Wevers}, {Waller}, {Young}, \& {Zhelem}}]{pazder20}
{Pazder}, J., {McConnachie}, A., {Ireland}, M., {et~al.} 2020, in Society of
  Photo-Optical Instrumentation Engineers (SPIE) Conference Series, Vol. 11447,
  Society of Photo-Optical Instrumentation Engineers (SPIE) Conference Series,
  1144743, \dodoi{10.1117/12.2561985}

\bibitem[{{Pazder} {et~al.}(2022){Pazder}, {McConnachie}, {Ireland}, {Anthony},
  {Bassett}, {Burley}, {Chapin}, {Churilov}, {Densmore}, {Dunn}, {Farrell},
  {Henderson}, {Hoff}, {Lothrop}, {Macdonald}, {Margheim}, {Reshetov},
  {Wevers}, {Waller}, {Young}, \& {Zhelem}}]{pazder22}
{Pazder}, J., {McConnachie}, A., {Ireland}, M., {et~al.} 2022, in Society of
  Photo-Optical Instrumentation Engineers (SPIE) Conference Series, Vol. 12184,
  Ground-based and Airborne Instrumentation for Astronomy IX, ed. C.~J.
  {Evans}, J.~J. {Bryant}, \& K.~{Motohara}, 121841D,
  \dodoi{10.1117/12.2630715}

\bibitem[{{Pence} {et~al.}(2010){Pence}, {Chiappetti}, {Page}, {Shaw}, \&
  {Stobie}}]{fits}
{Pence}, W.~D., {Chiappetti}, L., {Page}, C.~G., {Shaw}, R.~A., \& {Stobie}, E.
  2010, \aap, 524, A42, \dodoi{10.1051/0004-6361/201015362}

\bibitem[{{Pepe} {et~al.}(2002){Pepe}, {Mayor}, {Galland}, {Naef}, {Queloz},
  {Santos}, {Udry}, \& {Burnet}}]{pepe2002}
{Pepe}, F., {Mayor}, M., {Galland}, F., {et~al.} 2002, \aap, 388, 632,
  \dodoi{10.1051/0004-6361:20020433}

\bibitem[{{Placco} {et~al.}(2023){Placco}, {Almeida-Fernandes}, {Holmbeck},
  {Roederer}, {Mardini}, {Hayes}, {Venn}, {Chiboucas}, {Deibert}, {Gamen},
  {Heo}, {Jeong}, {Kalari}, {Martioli}, {Xu}, {Diaz}, {Gomez-Jimenez},
  {Henderson}, {Prado}, {Quiroz}, {Ruiz-Carmona}, {Simpson}, {Urrutia},
  {McConnachie}, {Pazder}, {Burley}, {Ireland}, {Waller}, {Berg}, {Robertson},
  {Hartman}, {Jones}, {Labrie}, {Perez}, {Ridgway}, \& {Thomas-Osip}}]{sv2}
{Placco}, V.~M., {Almeida-Fernandes}, F., {Holmbeck}, E.~M., {et~al.} 2023,
  \apj, 959, 60, \dodoi{10.3847/1538-4357/ad077e}

\bibitem[{{Puxley} \& {J{\o}rgensen}(2006)}]{puxley}
{Puxley}, P., \& {J{\o}rgensen}, I. 2006, in Society of Photo-Optical
  Instrumentation Engineers (SPIE) Conference Series, Vol. 6270, Society of
  Photo-Optical Instrumentation Engineers (SPIE) Conference Series, ed. D.~R.
  {Silva} \& R.~E. {Doxsey}, 62700V, \dodoi{10.1117/12.668934}

\bibitem[{{Robertson}(2017)}]{robertson}
{Robertson}, J.~G. 2017, \pasa, 34, e035, \dodoi{10.1017/pasa.2017.29}

\bibitem[{{Sestito} {et~al.}(2024){Sestito}, {Hayes}, {Venn}, {Jensen},
  {McConnachie}, {Pazder}, {Waller}, {Ardern-Arentsen}, {Jablonka}, {Martin},
  {Matsuno}, {Navarro}, {Starkenburg}, {Vitali}, {Bassett}, {Berg}, {Diaz},
  {Edgar}, {Firpo}, {Gomez-Jimenez}, {Kalari}, {Lambert}, {Lawrence},
  {Robertson}, {Ruiz-Carmona}, {Salinas}, {Sebo}, \& {Venkatesan}}]{sestito}
{Sestito}, F., {Hayes}, C.~R., {Venn}, K.~A., {et~al.} 2024, \mnras, 528, 4838,
  \dodoi{10.1093/mnras/stae244}

\bibitem[{Sheinis {et~al.}(2016)Sheinis, Anthony, Baker, Burley, Churilov,
  Edgar, Ireland, Kondrat, Pazder, Robertson, Young, \& Zhelem}]{sheinis16}
Sheinis, A.~I., Anthony, A., Baker, G., {et~al.} 2016, SPIE Proceedings, 9908,
  990817, \dodoi{10.1117/12.2232231}

\bibitem[{{Simon}(2019)}]{dwarfs}
{Simon}, J.~D. 2019, \araa, 57, 375,
  \dodoi{10.1146/annurev-astro-091918-104453}

\bibitem[{{Skarka}(2014)}]{skarka14}
{Skarka}, M. 2014, \aap, 562, A90, \dodoi{10.1051/0004-6361/201322491}

\bibitem[{{Sokal} {et~al.}(2018){Sokal}, {Deen}, {Mace}, {Lee}, {Oh}, {Kim},
  {Kidder}, \& {Jaffe}}]{sokal18}
{Sokal}, K.~R., {Deen}, C.~P., {Mace}, G.~N., {et~al.} 2018, \apj, 853, 120,
  \dodoi{10.3847/1538-4357/aaa1e4}

\bibitem[{{Tollestrup} {et~al.}(2010){Tollestrup}, {Kleinman}, {Goodsell},
  {Arriagada}, {Lazo}, {Rogers}, {Galvez}, \& {White}}]{toll10}
{Tollestrup}, E.~V., {Kleinman}, S.~J., {Goodsell}, S.~J., {et~al.} 2010, in
  Society of Photo-Optical Instrumentation Engineers (SPIE) Conference Series,
  Vol. 7735, Ground-based and Airborne Instrumentation for Astronomy III, ed.
  I.~S. {McLean}, S.~K. {Ramsay}, \& H.~{Takami}, 773505,
  \dodoi{10.1117/12.857563}

\bibitem[{{Tolstoy} {et~al.}(2009){Tolstoy}, {Hill}, \& {Tosi}}]{dwarfs2}
{Tolstoy}, E., {Hill}, V., \& {Tosi}, M. 2009, \araa, 47, 371,
  \dodoi{10.1146/annurev-astro-082708-101650}

\bibitem[{{Tomasella} {et~al.}(2010){Tomasella}, {Munari}, \&
  {Zwitter}}]{tomasella10}
{Tomasella}, L., {Munari}, U., \& {Zwitter}, T. 2010, \aj, 140, 1758,
  \dodoi{10.1088/0004-6256/140/6/1758}

\bibitem[{{Triaud}(2018)}]{Triaud2018}
{Triaud}, A. H.~M.~J. 2018, in Handbook of Exoplanets, ed. H.~J. {Deeg} \&
  J.~A. {Belmonte}, 2, \dodoi{10.1007/978-3-319-55333-7_2}

\bibitem[{{Vacca} \& {Sandell}(2011)}]{vacca11}
{Vacca}, W.~D., \& {Sandell}, G. 2011, \apj, 732, 8,
  \dodoi{10.1088/0004-637X/732/1/8}

\bibitem[{Webb {et~al.}(1999)Webb, Flambaum, Churchill, Drinkwater, \&
  Barrow}]{qso}
Webb, J.~K., Flambaum, V.~V., Churchill, C.~W., Drinkwater, M.~J., \& Barrow,
  J.~D. 1999, Physical review letters, 82, 884

\bibitem[{{Webb} {et~al.}(1999){Webb}, {Flambaum}, {Churchill}, {Drinkwater},
  \& {Barrow}}]{quasarfund}
{Webb}, J.~K., {Flambaum}, V.~V., {Churchill}, C.~W., {Drinkwater}, M.~J., \&
  {Barrow}, J.~D. 1999, \prl, 82, 884, \dodoi{10.1103/PhysRevLett.82.884}

\bibitem[{{Yang} {et~al.}(2005){Yang}, {Johns-Krull}, \& {Valenti}}]{yang05}
{Yang}, H., {Johns-Krull}, C.~M., \& {Valenti}, J.~A. 2005, \apj, 635, 466,
  \dodoi{10.1086/497070}

\bibitem[{{Young} \& {Nielsen}(2016)}]{young16}
{Young}, P.~J., \& {Nielsen}, J.~G. 2016, in Society of Photo-Optical
  Instrumentation Engineers (SPIE) Conference Series, Vol. 9913, Software and
  Cyberinfrastructure for Astronomy IV, ed. G.~{Chiozzi} \& J.~C. {Guzman},
  99132Q, \dodoi{10.1117/12.2232235}

\bibitem[{{Zapatero Osorio} {et~al.}(2002){Zapatero Osorio}, {B{\'e}jar},
  {Pavlenko}, {Rebolo}, {Allende Prieto}, {Mart{\'\i}n}, \& {Garc{\'\i}a
  L{\'o}pez}}]{zapatero}
{Zapatero Osorio}, M.~R., {B{\'e}jar}, V.~J.~S., {Pavlenko}, Y., {et~al.} 2002,
  \aap, 384, 937, \dodoi{10.1051/0004-6361:20020046}

\bibitem[{{Zhelem} {et~al.}(2018){Zhelem}, {Churilov}, {Kondrat}, {Waller},
  {Lawrence}, {Edgar}, {Baker}, {Farrel}, {Young}, {Nielsen}, {Mali}, {Muller},
  {Klauser}, {Pai}, {Ireland}, {Sheinis}, {Robertson}, {Case}, \&
  {McDermid}}]{zhelem18}
{Zhelem}, R., {Churilov}, V., {Kondrat}, Y., {et~al.} 2018, in Society of
  Photo-Optical Instrumentation Engineers (SPIE) Conference Series, Vol. 10702,
  Ground-based and Airborne Instrumentation for Astronomy VII, ed. C.~J.
  {Evans}, L.~{Simard}, \& H.~{Takami}, 107026H, \dodoi{10.1117/12.2313059}

\bibitem[{{Zhelem} {et~al.}(2020){Zhelem}, {Churilov}, {Case}, {Kondrat},
  {Fiegert}, {Waller}, {Lawrence}, {Farrell}, \& {Ireland}}]{zhelem20}
{Zhelem}, R., {Churilov}, V., {Case}, S., {et~al.} 2020, in Society of
  Photo-Optical Instrumentation Engineers (SPIE) Conference Series, Vol. 11203,
  Advances in Optical Astronomical Instrumentation 2019, ed. S.~C. {Ellis} \&
  C.~{d'Orgeville}, 1120318, \dodoi{10.1117/12.2539924}

\end{thebibliography}
\bibliographystyle{aasjournal}



\end{document}